\documentclass[11pt]{article}
\usepackage{amsmath,amssymb,amsthm,amsxtra,overpic,bbm,bm,epsfig,subfigure,multirow}
\usepackage{color}

\usepackage{comment}

\def\thefootnote{\fnsymbol{footnote}}

\begin{document}

\vspace{0.2cm}

\begin{center}
{\Large\bf Viability of exact tri-bimaximal, golden-ratio and bimaximal \\ mixing patterns and renormalization-group running effects}
\end{center}

\vspace{0.2cm}

\begin{center}
{\bf Jue Zhang $^{a,~b}$} \footnote{E-mail: zhangjue@ihep.ac.cn} \quad {\bf Shun Zhou $^{b,~a}$} \footnote{E-mail: zhoush@ihep.ac.cn}
\\
{$^a$Center for High Energy Physics, Peking University, Beijing 100871, China \\
$^b$Institute of High Energy Physics, Chinese Academy of
Sciences, Beijing 100049, China}
\end{center}

\vspace{1.5cm}

\begin{abstract}
In light of the latest neutrino oscillation data, we examine whether the leptonic flavor mixing matrix can take on an \emph{exact} form of tri-bimaximal (TBM), golden-ratio (GR) or bimaximal (BM) mixing pattern at a superhigh-energy scale, where such a mixing pattern could be realized by a flavor symmetry, and become compatible with experimental data at the low-energy scale. Within the framework of the Minimal Supersymmetric Standard Model (MSSM), the only hope for realizing such a possibility is to count on the corrections from the renomalization-group (RG) running. In this work we focus on these radiative corrections, and fully explore the allowed parameter space for each of these mixing patterns. We find that when the upper bound on the sum of neutrino masses $\Sigma^{}_\nu \equiv m^{}_1 + m^{}_2 + m^{}_3 < 0.23~\text{eV}$ at the $95\%$ confidence level from Planck 2015 is taken into account, none of these mixing patterns can be identified as the leptonic mixing matrix below the seesaw threshold. If this cosmological upper bound on the sum of neutrino masses were relaxed, the TBM and GR mixing patterns would still be compatible with the latest neutrino oscillation data at the $3\sigma$ level, but not at the $1\sigma$ level. Even in this case, no such a possibility exists for the BM mixing.
\end{abstract}

\begin{flushleft}
\hspace{0.8cm} PACS number(s): 12.38.Bx, 14.60.Pq
\end{flushleft}

\def\thefootnote{\arabic{footnote}}
\setcounter{footnote}{0}

\newpage
\section{Introduction}

In retrospect, it is remarkable that experimentalists were able to identify the existence of two large (i.e., $\theta^{}_{12} \approx 34^\circ$ and $\theta^{}_{23} \approx 45^\circ$) and one small (i.e., $\theta^{}_{13} < 10^\circ$) flavor mixing angles in the lepton sector about one decade ago~\cite{Agashe:2014kda}. Around the same time, several appealing constant mixing matrices were postulated by theorists so as to explain the observed flavor mixing pattern. The most interesting ones include the TBM, GR and BM mixing matrices, which can be collectively parametrized as
\begin{equation}
U = \left( \begin{array}{ccc}
             c^{}_{12} & s^{}_{12} & 0 \\
              \displaystyle - \frac{s^{}_{12}}{\sqrt{2}} & \displaystyle + \frac{c^{}_{12}}{\sqrt{2}} & \displaystyle \frac{1}{\sqrt{2}} \\
             \displaystyle + \frac{s^{}_{12}}{\sqrt{2}} & \displaystyle -\frac{c^{}_{12}}{\sqrt{2}} & \displaystyle \frac{1}{\sqrt{2}}
           \end{array}
\right) \; ,
\end{equation}
where $s^{}_{12} \equiv \sin \theta^{}_{12}$ and $c^{}_{12} \equiv \cos \theta^{}_{12}$ have been defined. We have $\theta^{}_{23} = 45^\circ$ and $\theta^{}_{13} = 0$ for all three mixing patterns in question, while $\theta^{}_{12} = \arctan(1/\sqrt{2}) \approx 35.3^\circ$ for TBM~\cite{TBM}, $\theta^{}_{12} = 45^\circ$ for BM~\cite{BM} and $\theta^{}_{12} = \arctan[2/(1+\sqrt{5})] \approx 31.7^\circ$ for GR~\cite{GR}.\footnote{There exists an alternative mixing pattern involving the golden ratio~\cite{GR_B}, which assumes $\theta^{}_{12} = \cos^{-1}\left[(\sqrt{5} + 1)/4\right] = 36.0^\circ$, quite similar to that for the TBM mixing, so we concentrate on $\theta^{}_{12} = \tan^{-1}\left[2/(\sqrt{5} + 1)\right] \approx 31.7^\circ$ in this work.} For quite a long time, these constant mixing patterns had been regarded as good candidates for the leptonic flavor mixing matrix, as they all agreed with the experimental data fairly well. Motivated by the simplicity of these mixing patterns and the excellent agreement with data, a great number of works have been carried out to pursue a deep connection between these mixing matrices and discrete flavor symmetries. See, e.g., Refs.~\cite{flavor}, for recent reviews on this topic.

Recently, a relatively-large value of $\theta_{13}^{}$ (i.e., $\theta^{}_{13} \approx 9^\circ$) has been discovered in the Daya Bay, RENO, and Double Chooz reactor neutrino experiments~\cite{dayabay, reno,chooz}. As an immediate consequence of this great discovery, none of these mixing matrices is able to account for the oscillation data at the low-energy scale. Two different approaches have been implemented to solve this problem:
\begin{itemize}
\item First, one can take the constant mixing patterns with $\theta^{}_{13} = 0$ as the leading-order approximation, which will be modified by significant corrections from various sources~\cite{modifications}.

\item Second, one can simply abandon these mixing patterns and search for new ones, which may also be derived from flavor symmetries but with a non-vanishing value of $\theta_{13}^{}$ \cite{group_scan,Zhang:2011aw}.
\end{itemize}
In this paper, we follow the first approach and assume that one of the three constant mixing patterns is valid at a superhigh-energy scale, where a suitable flavor symmetry is at work and responsible for that simple mixing matrix. Then, after the RG running effects on the mixing parameters are taken into account, the mixing matrix at the low-energy scale becomes compatible with neutrino oscillation data.

In fact, this idea has been studied in the literature but in different contexts. In Ref.~\cite{seesaw_threshold}, it has been found that the seesaw threshold effects can actually be quite significant, rendering some of the constant mixing patterns exactly viable above the seesaw scale. However, this analysis depends very much on the seesaw models themselves. With the ignorance of high-energy physics, we consider the RG running effects below the seesaw scale, where the heavy particles introduced to realize seesaw mechanisms and flavor symmetries have been integrated out. To be specific, we choose to work in the framework of MSSM.\footnote{The case of the Standard Model (SM) is not considered here, as the RG running effects are known to be negligible. Therefore, none of three mixing patterns can be exact below the seesaw threshold.} In such a set-up, the only way that one can rescue these mixing matrices is to count on the corrections from RG running, and therefore a detailed RG study would manifest their fates below the seesaw threshold. There are also extensive RG studies on the above mixing matrices even in this case. Before the emergence of a large value of $\theta_{13}^{}$, one often made discussions on a general ground by considering both possibilities, small or large radiative corrections to $\theta_{13}^{}$ \cite{Antusch:2003kp,Luo:2005fc,Dighe:2006sr,Dighe:2007ksa,Dighe:2008wn}. The conditions of obtaining a large value of $\theta_{13}^{}$ at low energies have been partially identified. Later, shortly after the first hint of a large value of $\theta_{13}^{}$ from the global fit of neutrino oscillation data, a more detailed analysis focusing on the large correction side and the case of TBM was performed in \cite{Goswami:2009yy}. With the advent of the actual discovery of $\theta_{13}^{}$, one put the above matrices into a further scrutiny, and in \cite{Luo:2012ce} it was found that generating $\theta_{13}^{} \approx 9^\circ$ at low energies from a zero value at the high-energy scale is quite challenging.
%%%%%%%%%%%%%%%%%%%%%%%%%% Table 1 %%%%%%%%%%%%%%%%%%%%%%%%%%%%
%%%%%%%%%%%%%%%%%%%%%%%%%%%%%%%%%%%%%%%%%%%%%%%%%%%%%%%%%%%%%%%%%
\begin{table}[t]
\begin{center}
\vspace{-0.25cm} \caption{The best-fit values, together with the
1$\sigma$, 2$\sigma$ and 3$\sigma$ intervals, for three neutrino
mixing angles $\{\theta^{}_{12}, \theta^{}_{13}, \theta^{}_{23}\}$, two mass-squared differences $\{\Delta m^2_{21} \equiv m^2_2 - m^2_1, \Delta m^2_{31}\equiv m^2_3 - m^2_1~{\rm or}~\Delta m^2_{32} \equiv m^2_3 - m^2_2\}$ and the Dirac CP-violating phase $\delta$ from a global analysis of current experimental data~\cite{Gonzalez-Garcia:2014bfa}. Here we quote the global-fit results of the NuFIT2.0 version (available from www.nu-fit.org), which will be used in our numerical calculations.} \vspace{0.2cm}
\label{tb:fit}
\begin{tabular}{c|c|c|c|c}
\hline
\hline
Parameter & Best fit & 1$\sigma$ range & 2$\sigma$ range & 3$\sigma$ range \\
\hline
\multicolumn{5}{c}{Normal neutrino mass hierarchy (NH)} \\ \hline
%------------------------------------------------------------
$\theta_{12}/^\circ$
& $33.48$ & 32.73 --- 34.26 & 31.98 --- 35.04 & 31.29 --- 35.91 \\
%------------------------------------------------------------
$\theta_{13}/^\circ$
& $8.50$ & 8.29 --- 8.70 & 8.08 --- 8.90 & 7.85 --- 9.10 \\
%------------------------------------------------------------
$\theta_{23}/^\circ$
& $42.3$  & 40.7 --- 45.3 & 39.1 --- 48.3 & 38.2 --- 53.3 \\
%------------------------------------------------------------
$\delta/^\circ$ &  $306$ & 236 --- 345 & 0 --- 24
$\oplus$ 166 --- 360 & 0 --- 360 \\
%------------------------------------------------------------
$\Delta m^2_{21}/[10^{-5}~{\rm eV}^2]$ &  $7.50$ & 7.33 --- 7.69 & 7.16 --- 7.88 & 7.02 --- 8.09 \\
%------------------------------------------------------------
$\Delta m^2_{31}/[10^{-3}~{\rm eV}^2]$ &  $+2.457$ & +2.410 --- +2.504 & +2.363 --- +2.551 & +2.317 --- +2.607 \\\hline
%%%%%%%%%%%%%%%%%%%%%%%%%%%%%%%%%%%%%%%%%%%%%%%%%%%%%%%%%%%%%
\multicolumn{5}{c}{Inverted neutrino mass hierarchy (IH)} \\ \hline
%------------------------------------------------------------
$\theta_{12}/^\circ$
& $33.48$ & 32.73 --- 34.26 & 31.98 --- 35.04 & 31.29 --- 35.91 \\
%------------------------------------------------------------
$\theta_{13}/^\circ$
& $8.51$ & 8.30 --- 8.71 & 8.09 --- 8.91 & 7.87 --- 9.11 \\
%------------------------------------------------------------
$\theta_{23}/^\circ$
& $49.5$  & 47.3 --- 51.0 & 45.1 --- 52.5 & 38.6 --- 53.3 \\
%------------------------------------------------------------
$\delta/^\circ$ &  $254$ & 192 --- 317 & 0 --- 20
$\oplus$ 130 --- 360 & 0 --- 360 \\
%------------------------------------------------------------
$\Delta m^2_{21}/[10^{-5}~{\rm eV}^2]$ &  $7.50$ & 7.33 --- 7.69 & 7.16 --- 7.88 & 7.02 --- 8.09 \\
%------------------------------------------------------------
$\Delta m^2_{32}/[10^{-3}~{\rm eV}^2]$ &  $-2.449$ & $-2.496$ --- $-2.401$ & $-2.543$ --- $-2.355$ & $-2.590$ --- $-2.307$ \\ \hline\hline
%%%%%%%%%%%%%%%%%%%%%%%%%%%%%%%%%%%%%%%%%%%%%%%%%%%%%%%%%%%%%
\end{tabular}
\end{center}
\end{table}
%%%%%%%%%%%%%%%%%%%%%%%%%%%%%%%%%%%%%%%%%%%%%%%%%%%%%%%%%%%%%%%%%
%%%%%%%%%%%%%%%%%%%%%%%%%%%%%%%%%%%%%%%%%%%%%%%%%%%%%%%%%%%%%%%%%

Our work is different from previous studies in several aspects. First, the latest results from the global-fit analysis of neutrino oscillation data will be used. Second, we aim at providing a more precise and definite answer to the fates of those mixing patterns, namely, to what extent they still agree with current data. To this end, we vary all the parameters that may have significant impact on the RG running, including two Majorana CP-violating phases and one parameter that characterizes the supersymmetry (SUSY) threshold corrections. The allowed parameter space for each mixing pattern is then obtained. Third, the impact of the cosmological bound on the sum of neutrino masses $\Sigma^{}_\nu < 0.23~\text{eV}$ from Planck 2015~\cite{Ade:2015xua} is examined. If this bound is taken into account, none of those three mixing patterns can be identified as the lepton mixing matrix below the seesaw threshold. However, if the cosmological bound were relaxed, the TBM and GR mixing matrices would be compatible with the latest neutrino oscillation data~\cite{Gonzalez-Garcia:2014bfa} at the $3\sigma$ level (see, e.g., Table 1), but not at the $1\sigma$ level. No such a possibility exists for the BM mixing matrix.

The remaining part of the paper is organized as follows. In Section 2, we provide an analytical study on the required corrections from the RG running to the constant mixing patterns, and identify the necessary conditions on the parameters at the superhigh-energy scale. Such an analytical study is later confirmed by a detailed numerical calculation given in Section 3, where we first introduce our numerical set-up, and then present the allowed parameter space for each of these mixing matrices. Finally, we summarize our main results in Section 4.

\section{Analytical Results}

Since we are interested in the radiative corrections below the seesaw threshold, the RG running of neutrino parameters is dictated by the evolution of the coefficient $\kappa$ in the dimension-five Weinberg operator, assuming that neutrinos are Majorana particles. In the MSSM, the Weinberg operator is given by
\begin{eqnarray}
\mathcal{W}_{\text{d=5}} = -\frac{1}{2} ~ \kappa \left(\mathbb{L} \cdot \mathbb{H}^{}_{\rm u}\right)\left(\mathbb{L} \cdot \mathbb{H}^{}_{\rm u}\right) \; ,
\end{eqnarray}
where $\mathbb{L}$ and $\mathbb{H}^{}_{\rm u}$ are chiral superfields that contain the left-handed lepton doublets $\ell_{\rm L}^{}$ and the Higgs doublet $H_u^{}$ of hypercharge $+1/2$, respectively. After the spontaneous breaking of the electroweak gauge symmetry, neutrinos obtain an effective Majorana mass term via $M_\nu^{} = \kappa v^2 \tan^2\beta/(1+\tan^2\beta)$, where $v \approx 174~{\rm GeV}$ is the electroweak vacuum expectation value (vev), and $\tan\beta$ is the ratio of the vev's of two MSSM Higgs doublets.

At the one-loop level, the RG equation (RGE) of $\kappa$ reads~\cite{kapparunning}
\begin{eqnarray} \label{eq:kappa}
16\pi^2 \frac{\mathrm{d} \kappa}{\mathrm{d} t} = \alpha(t) \kappa + \left[ (Y_l^{} Y_l^\dagger)\kappa + \kappa (Y_l^{} Y_l^\dagger)^T \right],
\end{eqnarray}
where $t(\mu) = \ln(\mu/\mu_0)$ with $\mu^{}_0$ being a reference mass scale, $Y_l^{}$ is the charged-lepton Yukawa coupling, and $\alpha(t) \approx -6 g_1^2/5 -6 g_2^2 + 6 y_t^2$. In the $\alpha(t)$ function, $g_{1}^{}$ and $g^{}_2$ denote respectively the $U(1)_{\rm Y}^{}$ and $SU(2)_{\rm L}^{}$ gauge couplings, while $y_t^{}$ is the Yukawa coupling of the top quark. Note that the contributions from two generations of light quarks have been safely neglected.

There are two distinct ways to study the running behavior of neutrino masses and flavor mixing parameters. In the first way, starting with the RGE of $\kappa$~\cite{RGE}, one can derive a set of RGEs for neutrino masses and flavor mixing parameters. By inspecting the structure of those RGEs, some general running behavior of neutrino masses and lepton mixing angles can be observed~\cite{RGE,Ohlsson:2013xva}. However, a further more quantitative estimation on the size of these radiative corrections becomes formidable, as the RGEs are non-linear in nature and thus difficult to solve analytically. In the second way, one instead focuses on the evolution of $\kappa$ itself. The RG running effects are viewed as small perturbations to the initial value of $\kappa$, and then a further study on those perturbations yields the corrections to neutrino masses and mixing angles~\cite{Dighe:2006zk,Dighe:2007ksa}. In this connection, such a approach surpasses the previous one, as it can provide useful relations between the high-energy and low-energy values. But for the phase parameters this approach is still quite clumsy. As these two approaches are complementary to each other, we shall adopt both of them in this paper, i.e., discussing the radiative corrections to neutrino masses and mixing angles in the second approach, while employing the first approach to study the running of phase parameters. The RGEs of three CP-violating phases are summarized in Appendix A.

In the flavor basis where $Y_l^{}$ is diagonal, the evolution of the neutrino mass matrix $M_\nu^{}$ can be found according to Eq.~(\ref{eq:kappa}), namely,
\begin{eqnarray}
M^{}_\nu = I^{}_0 \begin{pmatrix}
I^{}_e & 0 & 0\\
0 & I^{}_\mu & 0\\
0 & 0 & I^{}_\tau
\end{pmatrix} M_\nu^\Lambda
\begin{pmatrix}
I^{}_e & 0 & 0\\
0 & I^{}_\mu & 0\\
0 & 0 & I^{}_\tau
\end{pmatrix} \; ,
\end{eqnarray}
where the evolution functions $I_0^{}$ and $I_\alpha^{}$ (for $\alpha = e, \mu, \tau$) are defined as
\begin{eqnarray}
I^{}_0 &=& \text{exp} \left[-\frac{1}{16\pi^2} \int_{t(\lambda)}^{t(\Lambda)} \alpha(t) ~ {\rm d} t \right] \; , \\
I_\alpha &=& \text{exp} \left[-\frac{1}{16\pi^2} \int_{t(\lambda)}^{t(\Lambda)} y_\alpha^{}(t)^2 ~ {\rm d} t \right]\; ,
\end{eqnarray}
and $M_\nu^\Lambda$ is the neutrino mass matrix at the high-energy scale $\Lambda$. Note that a superscript ``$\Lambda$" will be attached to the quantity at the high-energy scale $\Lambda$, so as to distinguish it from its counterpart at the low-energy scale $\lambda$.

Since the electron and muon masses are negligibly small compared to the tau-lepton mass (i.e., $y^{}_e \ll y^{}_\mu \ll y_\tau^{}$), only the contributions from $y_\tau^{}$ will be kept, which should be an excellent approximation. Furthermore, we expand the evolution function $I_\tau^{} = 1 - \epsilon + \mathcal{O}(\epsilon^2)$ in terms of a small parameter \begin{eqnarray}
\epsilon &=& \frac{1}{16\pi^2} \int_{t(\lambda)}^{t(\Lambda)} y_\tau^{}(t)^2 ~ {\rm d} t  \approx \frac{1}{16\pi^2} y_\tau^2 \ln(\Lambda/\lambda) \; .
\end{eqnarray}
In the MSSM, although $y_\tau^{}$ can be as large as of order one for large values of $\tan\beta$, the parameter $\epsilon$ remains small enough. For instance, taking $\tan\beta =50$, $\Lambda = 10^{14}_{}~\text{GeV}$ and $\lambda = 10^{3}_{}~\text{GeV}$, one finds $\epsilon \approx 0.01$. With these approximations, $M_\nu$ is obtained by perturbing $M_\nu^\Lambda$ in the following way
\begin{eqnarray} \label{eq:mnulambda}
M^{}_\nu /I^{}_0 = M_\nu^\Lambda -\epsilon \begin{pmatrix}
0 & 0 & (M_\nu^\Lambda)_{13}^{}\\
0 & 0 & (M_\nu^\Lambda)_{23} \\
(M_\nu^\Lambda)_{31}^{} & (M_\nu^\Lambda)_{32} & 2(M_\nu^\Lambda)_{33}
\end{pmatrix} + \mathcal{O}(\epsilon^2) \; .
\end{eqnarray}
To obtain the corrections to neutrino parameters, we first reconstruct $M_\nu^\Lambda$ from the neutrino masses, mixing angles and CP-violating phases at the scale $\Lambda$, namely,
\begin{eqnarray}
M_\nu^\Lambda &=& (U_\nu^{\Lambda})^*~D_\nu^\Lambda~(U_\nu^\Lambda)^\dagger \; ,
\end{eqnarray}
where $D_\nu^\Lambda \equiv \text{Diag}\{m_1^\Lambda, m_2^\Lambda, m_3^\Lambda\}$, and $U_\nu^\Lambda$ is just the Maki-Nakagawa-Sakata-Pontecorvo (MNSP) matrix at the scale $\Lambda$, as $Y_l^{}$ is always kept diagonal. Since $\theta_{23}^\Lambda = \pi/4$ and $\theta_{13}^\Lambda =0$ hold for all the three mixing patterns, we simply parametrize $U_\nu^\Lambda$ as
\begin{eqnarray}
U_\nu^\Lambda = V \left(\theta_{12}^\Lambda,~ \theta_{13}^\Lambda=0,~ \theta_{23}^\Lambda = \frac{\pi}{4},~ \delta \right) \cdot \text{Diag}\{e^{-{\rm i} \varphi_1^\Lambda/2}, e^{-{\rm i} \varphi_2^\Lambda/2}, 1\} \;,
\end{eqnarray}
using the convention introduced in Appendix A. Given $M^\Lambda_\nu$ in Eq.~(9), we can diagonalize neutrino mass matrix in Eq.~(\ref{eq:mnulambda}) and arrive at
\begin{eqnarray}
\label{eq:t12}
\theta_{12}^{} &\approx & \theta_{12}^\Lambda + \frac{\epsilon}{4}  \mathrm{Re}[Z_{21}^{}] \sin 2\theta_{12}^\Lambda \; , \\
\label{eq:t13}
\theta_{13}^{} &\approx & \frac{\epsilon}{4} |Z_{31} - Z_{32}| \sin 2\theta_{12}^\Lambda \; , \\
\label{eq:t23}
\theta_{23}^{} &\approx & \frac{\pi}{4} + \frac{\epsilon}{2} \left( \mathrm{Re}[Z_{32}] \cos^2\theta_{12}^\Lambda + \mathrm{Re}[Z_{31}] \sin^2 \theta_{12}^\Lambda \right) \; ,
\end{eqnarray}
where $Z_{ij}$ is defined as
\begin{eqnarray}
Z_{ij} \equiv \frac{(m_i^\Lambda)^2 + 2 m_i^\Lambda m_j^\Lambda e^{\mathrm{i}(\varphi_i^\Lambda-\varphi_j^\Lambda)} + (m_j^\Lambda)^2 }{(m_i^\Lambda)^2 - (m_j^\Lambda)^2} \; ,
\end{eqnarray}
with $\varphi_3^\Lambda = 0$ understood. Note that the approximate formulas in Eqs.~(\ref{eq:t12}), (\ref{eq:t13}) and (\ref{eq:t23}) are valid as long as the factors $\epsilon |Z^{}_{ij}|$ are small. As we will show later, this is really the case for the ranges of parameters in our calculations. The real part of $Z_{ij}^{}$ can be further simplified to
\begin{eqnarray}
\mathrm{Re}[Z_{ij}^{}] = \frac{| m_i^\Lambda e^{\mathrm{i} \varphi_i^\Lambda} + m_j^\Lambda e^{\mathrm{i} \varphi_j^\Lambda}|^2}{(m_i^\Lambda)^2 - (m_j^\Lambda)^2} \; .
\end{eqnarray}
The above formulas for $\theta_{12}^{}$ and $\theta_{23}^{}$ coincide with those given in Ref.~\cite{Dighe:2007ksa}.\footnote{However, we disagree with \cite{Dighe:2007ksa} on the formula for $\theta_{13}^{}$.}
The details of obtaining the above corrections can be found in Appendix B, where we also discuss the conditions under which these formulas are valid. In addition, one also obtains the corrections to three neutrino masses as
\begin{eqnarray}
m_1^2 &\approx & (m_1^\Lambda)^2 (1 - 2 \epsilon \sin^2\theta_{12}^\Lambda) I_0^2 \; , \\
m_2^2 &\approx & (m_2^\Lambda)^2 (1 - 2 \epsilon \cos^2\theta_{12}^\Lambda) I_0^2 \; , \\
m_3^2 &\approx & (m_3^\Lambda)^2 (1 - 2 \epsilon) I_0^2 \; ,
\end{eqnarray}
and two neutrino mass-squared differences
\begin{eqnarray} \label{eq:sol}
\Delta m_{21}^2 &\approx & I_0^2 \left \{(\Delta m_{21}^2)^\Lambda + 2 \epsilon [ \sin^2\theta_{12}^\Lambda (m_1^\Lambda)^2 -  \cos^2\theta_{12}^\Lambda (m_2^\Lambda)^2] \right \} \; , \\ \label{eq:atm}
\Delta m_{32}^2 &\approx & I_0^2 \left \{(\Delta m_{32}^2)^\Lambda + 2 \epsilon [\cos^2\theta_{12}^\Lambda (m_2^\Lambda)^2 -  (m_3^\Lambda)^2] \right \} \; .
\end{eqnarray}

Given the relations between the boundary values of neutrino parameters, we are ready to discuss the requirements for the high-energy parameters so that their low-energy counterparts are within the regions allowed by current experiments. In our discussions, we assume $I^{}_0 \approx 1$ and $\epsilon \lesssim 0.01$, which are proved to hold perfectly via exact numerical calculations. Some comments are in order:
\begin{itemize}
\item Since the running of three neutrino masses is mild, $m_i^\Lambda$'s are then subject to the same constraints derived from low-energy experiments. Currently, the most stringent bound on neutrino masses comes from cosmology, i.e., $\Sigma^{}_\nu < 0.23~\text{eV}$ from the Planck 2015 data~\cite{Ade:2015xua}, implying that each individual neutrino mass has to satisfy $m_i^{} \lesssim 0.07~\text{eV}$. Considering that such a bound is only at $2\sigma$ level and depends on the data sets used in the statistical analysis, we conservatively choose a looser bound on $m_i^{}$ in this paper, namely, $m_i^{} \lesssim 0.2~\text{eV}$.

\item Even for $m_i^\Lambda \sim 0.2~\text{eV}$, one can verify that the correction term to $|\Delta m_{3i}^2|^\Lambda$ in Eq.~(\ref{eq:atm}) is on the order of $10^{-4}~\text{eV}^2$. Therefore, to be consistent with the observed value of $|\Delta m^2_{3i}|$ in neutrino oscillation experiments, the mass-squared difference $|\Delta m_{3i}^2|^\Lambda$ needs to be around $10^{-3}~\text{eV}^2$.

\item Achieving $\theta_{13} \sim 0.16$ at low energies requires a nearly-degenerate mass spectrum of neutrinos. To see this point, we first notice from Eq.~(\ref{eq:t13}) that $|Z_{31}^{}|$ and $Z^{}_{32}$ have to be of order $\mathcal{O}(10)$, as $\epsilon \approx 0.01$. Then, due to $|\Delta m_{3i}^2|^\Lambda \sim 10^{-3}~\text{eV}^2$, the absolute masses $m_i^\Lambda$ should be all around $\sim 0.1~\text{eV}$, resulting in the quasi-degenerate mass spectrum. Because of such a requirement for mass degeneracy, $(\Delta m_{21}^2)^\Lambda$ needs to be around $10^{-4}~\text{eV}^2$ in the cases of TBM and GR, so as to offset the correction term in Eq.~(\ref{eq:sol}), which is of order $10^{-4}~\text{eV}^2$ for $m_i^\Lambda \sim 0.1~\text{eV}$. In the case of BM, however, $(\Delta m_{21}^2)^\Lambda$ is around $10^{-5}~\text{eV}^2$, since the correction terms are now negligible due to a cancellation between $\sin^2\theta_{12}^\Lambda$ and $\cos^2\theta_{12}^\Lambda$ for $\theta^\Lambda_{12} = 45^\circ$.

\item In the TBM and GR cases, $\theta^\Lambda_{12}$ is already close to the best-fit value of $\theta^{}_{12}$, so only a small correction is allowed, indicating that the phase difference between $\varphi_1^{\Lambda}$ and $\varphi_2^{\Lambda}$ should be around $\pi$, ensuring that $\mathrm{Re}[Z_{21}^{}]$ is vanishingly small. In the BM case, we do need a large but negative correction to $\theta_{12}^\Lambda$. However, because of $\mathrm{Re}[Z_{21}^{}] > 0$, the leading-order contribution is always positive, pushing $\theta_{12}^{}$ further away from its desired best-fit value at low energies. Therefore, before resorting to high-order corrections, one needs to suppress the leading-order contribution in the first place. In summary, we have $\varphi_2^\Lambda \approx \varphi_1^\Lambda + \pi$ for all three mixing patterns.
\end{itemize}

The above conditions are necessary for three constant mixing patterns to satisfy the experimental data at low energies. However, even with those conditions fulfilled, those patterns may still be in tension with low-energy data due to the following reasons:
\begin{itemize}
\item As mentioned above, the leading order correction to $\theta_{12}^{}$ is always positive, indicating that $\theta_{12}^{}$ tends to increase when running towards low energy.\footnote{If one goes beyond the leading-order perturbation, $\theta_{12}^{}$ becomes decreasing when running towards low energies. Such an observation is verified by our numerical calculations, and has also been mentioned in Ref.~\cite{Antusch:2003kp}.} Such an increase would be welcomed by GR, while a severe tension with the low-energy data would be generated for TBM and BM.
\item From Eq.~(\ref{eq:t23}), one notices that the sign of the correction to $\theta^{}_{23}$ depends on the neutrino mass hierarchy. More explicitly, the first or second octant of $\theta^{}_{23}$ corresponds to the IH or NH case. Such a correlation, however, is exactly opposite to what we have observed in the latest global-fit results~\cite{Gonzalez-Garcia:2014bfa}.
\item There also exists a correlation between $\theta_{13}$ and the correction to $\theta_{23}$, namely, $\delta \theta_{23} \equiv \theta_{23}^\Lambda - \theta_{23}$. Taking the NH as an example, we have $\varphi_2^{\Lambda} = \varphi_1^{\Lambda} +\pi$ and find that $Z_{31}^{}$ and $Z_{32}^{}$ are approximately given by
\begin{eqnarray}
Z_{31}^{} \approx 2(1 + e^{-\mathrm{i} \varphi_1^{\Lambda}}) \frac{|m_1^\Lambda|^2}{(\Delta  m_{31}^2)^\Lambda} \; , \quad  Z_{32}^{} \approx 2(1 - e^{-\mathrm{i} \varphi_1^{\Lambda}}) \frac{|m_1^\Lambda|^2}{(\Delta  m_{31}^2)^\Lambda} \; ,
\end{eqnarray}
in the limit of a nearly-degenerate mass spectrum. The ratio of $\delta \theta_{23}$ to $\theta_{13}$ is then given by
\begin{eqnarray}
\label{eq:t13_d23_bound}
\frac{\delta \theta^{}_{23}}{\theta^{}_{13}} \approx \frac{1-\cos 2\theta_{12}^\Lambda \cos\varphi_1^\Lambda} {\sin 2 \theta_{12}^\Lambda } \geq \tan\theta_{12}^\Lambda \; .
\end{eqnarray}
The above inequality indicates that the correction to $\theta^{}_{23}$ has to be on the order of $(\tan\theta_{12}^\Lambda) \theta_{13}$, which is around $6^\circ$ (or $9^\circ$) in the case of TBM and GR (or BM) with $\theta_{13} \sim 9^\circ$. Such a large correction would drive $\theta_{23}$ to be outside the allowed $1\sigma$ range at low energies, according to Ref.~\cite{Gonzalez-Garcia:2014bfa}.

\end{itemize}

All the above features have been observed previously in the literature \cite{Antusch:2003kp,Luo:2005fc,Dighe:2006sr,Dighe:2007ksa,Dighe:2008wn,Goswami:2009yy,Luo:2012ce}, except for the last one about the bound on $\delta \theta_{23}$. In the next section, we will confirm those analytical findings through a detailed numerical study.

\section{Numerical Analysis}

\subsection{General Approach}

In the MSSM, we adopt the one-loop RGEs of neutrino parameters. Although in the analytical study, we have neglected the contributions from two generations of lighter quarks and leptons, they are now included in the numerical analysis. The actual running is performed in two steps. First, we run the various gauge and Yukawa couplings from low energies to high energies in order to determine their high-energy boundary values. Then, we start with $\kappa$, which can be reconstructed from the constant mixing patterns and neutrino masses, and run the neutrino parameters to the low-energy scale.

In the above first step, we fix the low-energy scale to be $\lambda = 10^3~\text{GeV}$, and the matching of the Yukawa couplings in the SM with those in the MSSM is performed at the same scale. The values of the running SM quantities are taken from \cite{Antusch:2013jca}. Moreover, since large values of $\tan\beta$ are favored in the following running, it is then necessary to include the so-called SUSY threshold corrections~\cite{SUSY} as well, which would result in modifications to the down-type quark and charged-lepton Yukawa couplings at the matching scale. Following \cite{Antusch:2013jca}, in the basis where both $Y_u^{}$ and $Y_l^{}$ are diagonal, we have the matching conditions
\begin{eqnarray}
Y_u^{\text{MSSM}} &\simeq & Y_u^{\text{SM}} \frac{1}{\sin\overline{\beta}} \; , \\
Y_d^{\text{MSSM}} &\simeq & \text{Diag}\left\{ \frac{1}{1+\overline{\eta}_q}, \frac{1}{1+\overline{\eta}_q}, \frac{1}{1+\overline{\eta}_b} \right\} ~Y_d^{\text{SM}} \frac{1}{\cos\overline{\beta}} \; , \\
Y_l^{\text{MSSM}} &\simeq & \text{Diag}\left\{ \frac{1}{1+\overline{\eta}_l} , \frac{1}{1+\overline{\eta}_l}, 1 \right\} ~Y_l^{\text{SM}} \frac{1}{\cos\overline{\beta}} \; ,
\end{eqnarray}
where $\overline{\eta}^{}_q$ and $\overline{\eta}^{}_l$ are the parameters that describe the SUSY threshold corrections, and we have absorbed the corrections to $y_\tau^{}$ by a redefinition of $\beta \rightarrow \overline{\beta}$, namely, $\cos\overline{\beta} = (1+\eta_l^\prime) \cos\beta$, where $\eta_l^\prime$ also denotes the SUSY threshold correction and its exact definition can be found in \cite{Antusch:2013jca}. Hence, by considering the SUSY threshold corrections, three additional parameters $\overline{\eta}$'s, together with a redefined $\tan\overline{\beta}$, are introduced. However, since $\overline{\eta}_q^{}$ and $\overline{\eta}_l^{}$ only correct the Yukawa couplings of the first two generations of down-type quarks and charged leptons, whose contributions to the RG running of neutrino parameters are already quite small, we set $\overline{\eta}_{q,l}^{}=0$. Only $\overline{\eta}^{}_b$ and $\tan\overline{\beta}$ are relevant.

Next, we run all the obtained MSSM quantities to the high-energy scale $\Lambda$, which can vary in a wide range. At the scale $\Lambda$, we further verify the resultant Yukawa couplings to see if they are still in the perturbative region, i.e., ${\cal O}(Y^{}_f) < 4\pi$. Having found the high-energy boundary values for the gauge and Yukawa couplings, we finally perform the running towards the low-energy scale $\lambda$ with a boundary value of $\kappa$. When reconstructing $\kappa$, we consider various different values of two Majorana CP-violating phases and neutrino masses, in addition to the mixing angles implied by three constant mixing patterns. A list of free parameters and their chosen ranges in our numerical study can be found in Table~\ref{tb:parameters}, where the values of $\overline{\eta}_b$ and $\tan\overline{\beta}$ are consistent with those from Ref.~\cite{Antusch:2013jca}. Two cases of neutrino mass hierarchy are also distinguished, in the assumption that the hierarchy patterns at the low- and high-energy scales are matched, i.e., they are both in NH, or both in IH.\footnote{The case with mismatched neutrino mass hierarchies at low and high energies is seldom discussed in the literature (except for a vague mention in~\cite{Antusch:2003kp}). However, a preliminary study shows that it can revive some mixing scenarios that would be eliminated in the assumption of identical mass hierarchy.}
%%%%%%%%%%%%%%%%%%%%%%%%%%%%%% Table 1 %%%%%%%%%%%%%%%%%%%%%%%%%%%%%%%%
\begin{table}
\centering
\begin{tabular}{c | c}
\hline \hline
Free Parameters & Range \\
\hline
$\overline{\eta}_b^{}$ & [-0.6, 0.6] \\
$\tan\overline{\beta}$ & [10, 50] \\
$\Lambda/\text{GeV}$ & $[10^6, 10^{14}]$ \\
$\varphi_{1,2}^\Lambda / \text{deg}$ & $[-180, 180]$ \\
$m_{\text{lightest}}^\Lambda /\text{eV}$ & $[0.001, 0.2]$ \\
$(\Delta m_{21}^2)^\Lambda /10^{-5}~\text{eV}^2$ & $[0.1, 100]$ \\
$|\Delta m_{32}^2|^\Lambda /10^{-3}~\text{eV}^2$ & $[0.1, 10]$ \\
\hline \hline
\end{tabular}
\caption{The ranges of free parameters chosen in the numerical calculations.}
\label{tb:parameters}
\end{table}
%%%%%%%%%%%%%%%%%%%%%%%%%%%%%%%%%%%%%%%%%%%%%%%%%%%%%%%%%%%%%%%

The neutrino parameters at the low-energy scale $\lambda$ are then confronted to the latest global-fit results from Ref.~\cite{Gonzalez-Garcia:2014bfa}. In our computations, in order to improve the sampling efficiency, we have employed the program $\texttt{MULTINEST}$~\cite{multinest} for the parameter scan. The computational details and the allowed parameter space for each mixing scenario will be presented in the next subsection.

\subsection{Parameter Space}

\subsubsection{Tri-bimaximal mixing}

Let us begin with the case of NH, for which the best-fit points are given in Table~\ref{tb:TBM}. The $\chi^2_{}$ function in the fitting procedure is constructed as follows
\begin{eqnarray}
\chi^2 = \sum_{i=1}^n \left( \frac{y_i^\text{th}(x) - y_i^\text{ex}}{\sigma_i^{\text{ex}}} \right)^2,
\end{eqnarray}
where $y_i^{\text{th}}$ is the theoretical prediction obtained by sampling from the parameter space $x = (x_1^{}, ..., x_m^{})$, and $y_i^\text{ex}$ is the measured value with an uncertainty $\sigma_i^\text{ex}$. In our numerical study, we take $\sin^2\theta_{ij}$ and two mass-squared differences $(\Delta m^2_{21}, \Delta m^2_{31})$ as the five observables $y_i^{}$'s. Their experimental values $y_i^{\text{ex}}$'s are taken to be the best-fit ones from~\cite{Gonzalez-Garcia:2014bfa}, and the uncertainties $\sigma_i^\text{ex}$'s are obtained by symmetrizing the $1\sigma$ errors. In Table~\ref{tb:TBM}, we also list the pull for each observable $y_i^{}$, with its definition given by
\begin{eqnarray}
\text{pull}(y_i) = \frac{y_i^\text{th}(x) - y_i^\text{ex}}{\sigma_i^{\text{ex}}} \; ,
\end{eqnarray}
where the theoretical prediction is referred to that at the best-fit point. We then see that only a fair $\chi^2$ fit is reached for TBM in the NH case, and $\theta_{12}^{}$ and $\theta_{23}^{}$ indeed have large pulls, confirming our previous analytical findings. As shown in Table~\ref{tb:TBM}, it is not difficult to achieve the observed value of $\theta_{13}^{}$.
%%%%%%%%%%%%%%%%%%%%%%%%%%%%%%%%%% Table 2 %%%%%%%%%%%%%%%%%%%%%%%%%%%%
\begin{table}[t]
\centering
\begin{tabular}{c | c c c c}
\hline \hline
\multirow{2}{*}{Parameter} & \multicolumn{2}{c}{TBM, NH} & \multicolumn{2}{c}{TBM, IH} \\
\cline{2-3} \cline{4-5}
& best-fit & pull & best-fit & pull\\
\hline
$\overline{\eta}_b$ & -0.53 & - & 0.54 & - \\
$\tan\overline{\beta}$ & 41.8 & - & 30.1 & - \\
$\Lambda/ \text{GeV}$ & $1.94\times 10^{6}$ & - & $2.01\times 10^{12}$ & - \\
$\varphi_{1}^\Lambda / \text{deg}$ & -56.3 & - & 0.8 & - \\
$\varphi_{2}^\Lambda / \text{deg}$ & 134.5 & - & 181.3 & - \\
$m_{\text{lightest}}^\Lambda /\text{eV}$ & 0.198 & - & 0.188 & - \\
$(\Delta m_{21}^2)^\Lambda /10^{-5}~\text{eV}^2$ & 32.9 & - & 43.9 & - \\
$|\Delta m_{32}^2|^\Lambda /10^{-3}~\text{eV}^2$ & 2.96 & - & 2.33 & - \\
\hline
$\sin^2\theta_{12}$ & 0.333 & 2.32 & 0.341 & 2.93 \\
$\sin^2\theta_{13}$ & 0.0214 & -0.37 & 0.0213 & -0.54 \\
$\sin^2\theta_{23}$ & 0.613 & 4.02 & 0.396 & -5.9 \\
$\Delta m_{21}^2 /10^{-5}~\text{eV}^2$ & 7.50 & -0.008 & 7.50 & -0.003 \\
$|\Delta m_{3i}^2| /10^{-3}~\text{eV}^2$ & 2.458 & 0.017 & 2.450 & 0.024 \\
$m_{\text{lightest}} /\text{eV}$ & 0.183 & - & 0.185 & - \\
$\delta / \text{deg}$ & 137.6 & - & 0.45 & - \\
$\varphi_{1} / \text{deg}$ & -43.0 & - & -0.03 & - \\
$\varphi_{2} / \text{deg}$ & 140.3 & - & 180.9 & - \\
$m^{}_{ee} / \text{eV}$ & 0.063 & - & 0.062 & - \\
$m^{}_\beta / \text{eV}$ & 0.18 & - & 0.19 & - \\
\hline
$\chi^2_{\rm min}$ & & $\mathbf{21.7}$ & & $\mathbf{43.7}$ \\
\hline \hline
\end{tabular}
\caption{The best-fit points of low-energy observables for the TBM mixing pattern in both cases of NH and IH, where the input parameters at the high-energy scale are also given. }
\label{tb:TBM}
\end{table}
%%%%%%%%%%%%%%%%%%%%%%%%%%%%%%%%%%%%%%%%%%%%%%%%%%%%%%%%%%%%%%%%%%%

\begin{figure}[!t]
\begin{center}
\subfigure{%
\includegraphics[width=0.47\textwidth]{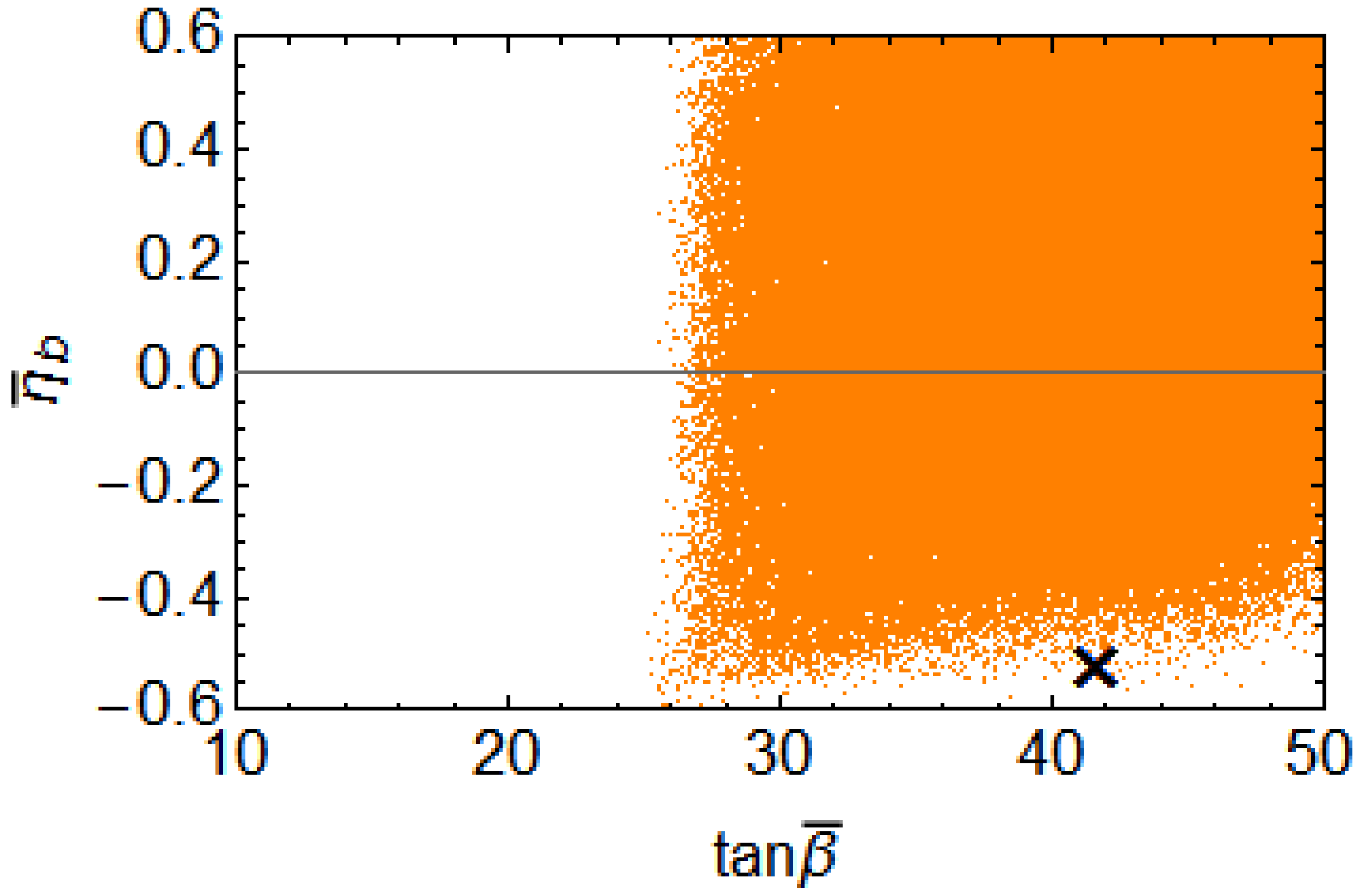}
}%
\subfigure{%
\hspace{0.1cm}
\includegraphics[width=0.5\textwidth, height=0.3\textwidth]{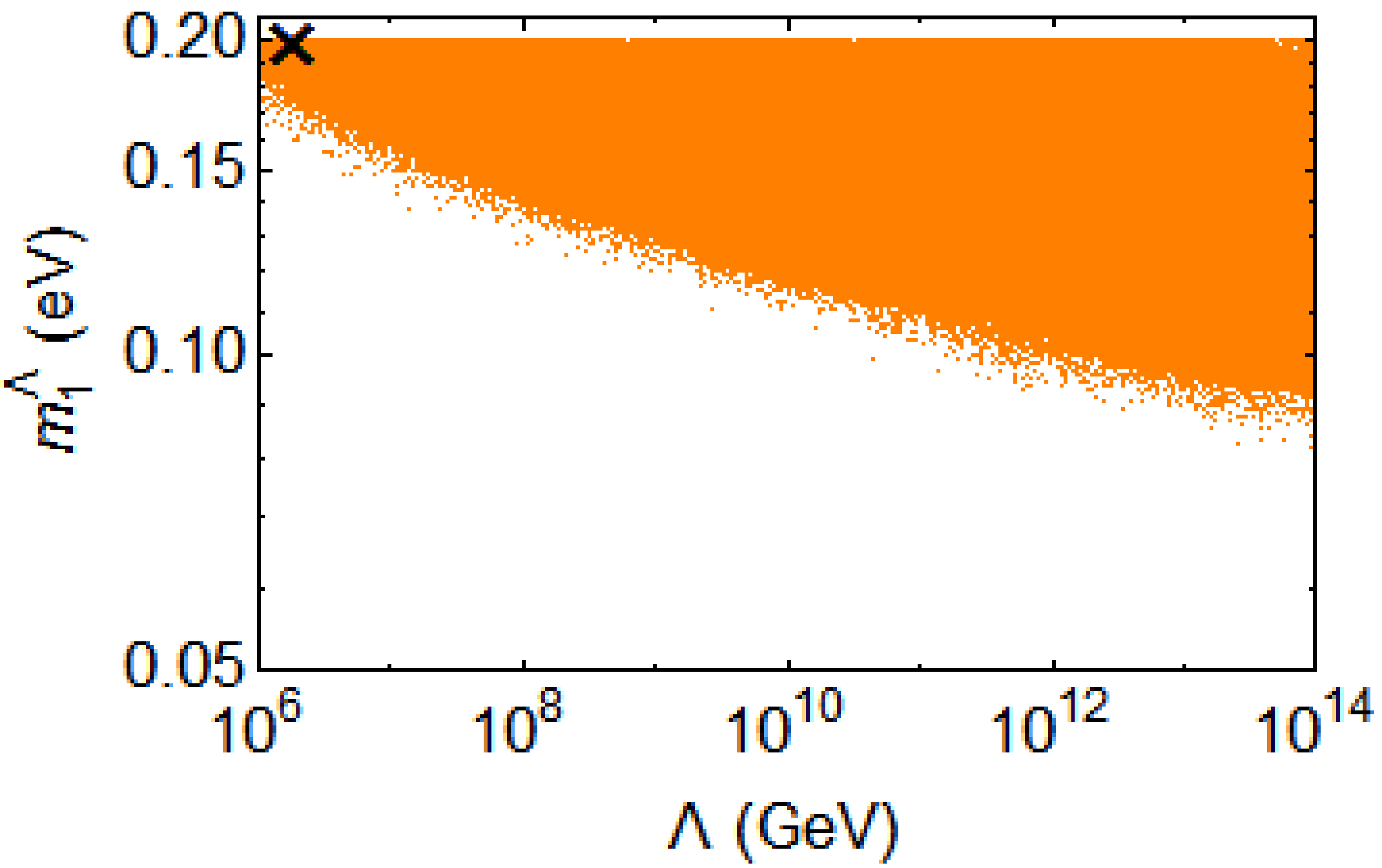}
}%
\vspace{-0.3cm}
\subfigure{%
\includegraphics[width=0.47\textwidth]{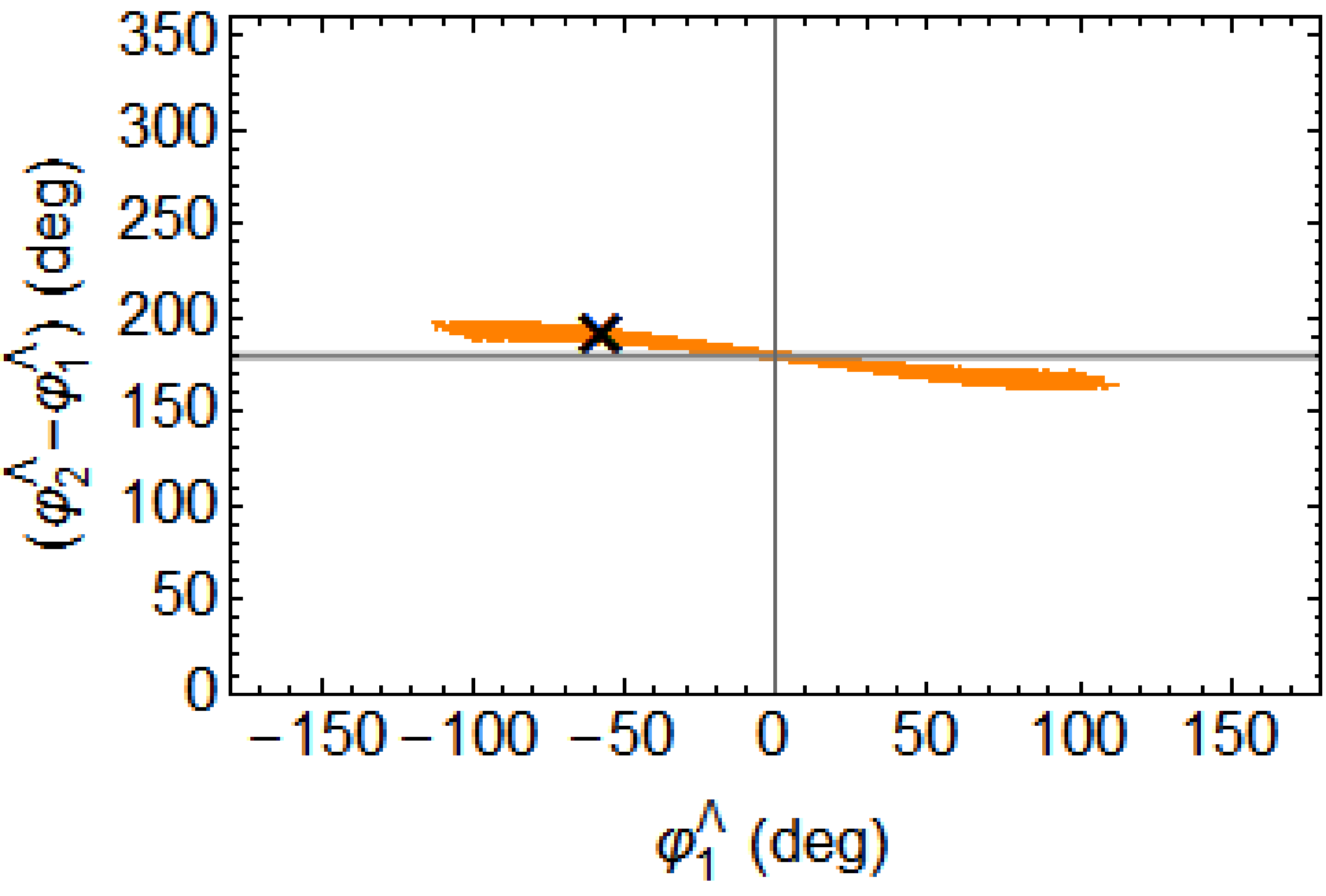}
}%
\subfigure{%
\hspace{0.4cm}
\includegraphics[width=0.47\textwidth]{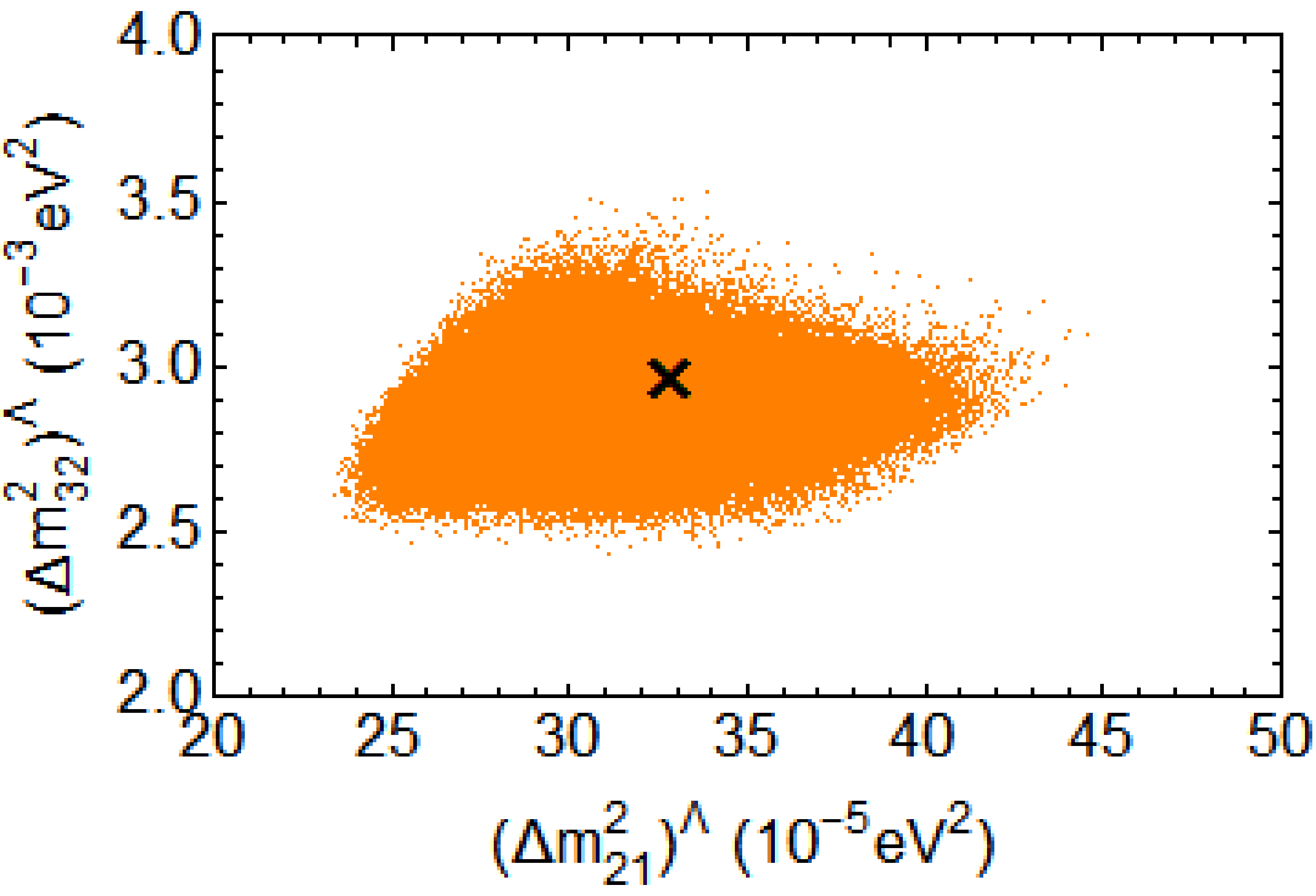}
}%
\end{center}
\vspace{-0.5cm}
\caption{Allowed parameter space for the free parameters given in Table~\ref{tb:parameters} for TBM in the case of NH, where the black crosses stand for the best-fit points.}
\label{fg:TBM_NH_HE}
\end{figure}

We next present the allowed parameter space for the free parameters in Fig.~\ref{fg:TBM_NH_HE}, where for each yellow point their predictions on three mixing angles and two mass-squared differences are within the $3\sigma$ ranges from Ref.~\cite{Gonzalez-Garcia:2014bfa}. As one can see, large values of $\tan\overline{\beta}$ and $m_1^\Lambda$ are favored, and such a feature is expected from the consideration of large corrections to $\theta^{}_{13}$. The allowed ranges for $\Lambda$ and $\overline{\eta}_b$, however, are vast. The reason may be that $\Lambda$ and $\overline{\eta}_b$ are helping each other when trying to obtain a relatively large value of $\epsilon$. In fact, $\overline{\eta}_b$ also helps $\tan\overline{\beta}$ to reach a lower value.
%%%%%%%%%%%%%%%%%%%%%%%%%%%%%%%%%%% Fig. 1 %%%%%%%%%%%%%%%%%%%%%%%%%%%%%%
%%%%%%%%%%%%%%%%%%%%%%%%%%%%%%%%%%%%%%%%%%%%%%%%%%%%%%%%%%%%%%%%%%%

Regarding the Majorana CP-violating phases at high energies, we indeed see from Fig.~\ref{fg:TBM_NH_HE} that the region with a near $\pi$ difference between them is highly favored. Moreover, it also shows that $\varphi_1^\Lambda$ tends to be in the first and fourth quadrants. This can be understood by the correlation between $\theta_{13}^{}$ and $\delta \theta_{23}^{}$. According to Eq.~(\ref{eq:t13_d23_bound}), $\delta\theta_{23}^{}$ is bounded from below, and this lower bound, which gives a better fit to the low-energy data, is reached when $\cos\varphi_1^\Lambda = 1$. Finally, the allowed parameter space of two neutrino mass-squared differences also agrees with our previous analytical study, namely, $(\Delta m_{21}^2)^\Lambda$ in the $10^{-4}~\text{eV}^2$ range while $(\Delta m_{32}^2)^\Lambda$ in the $10^{-3}~\text{eV}^2$ range.
%%%%%%%%%%%%%%%%%%%%%%%%%%%%%%%%%%%%%% Fig. 2 %%%%%%%%%%%%%%%%%%%%%%%%
\begin{figure}[!t]
\begin{center}
\subfigure{%
\hspace{0.5cm}
\includegraphics[width=0.45\textwidth , height=0.3\textwidth]{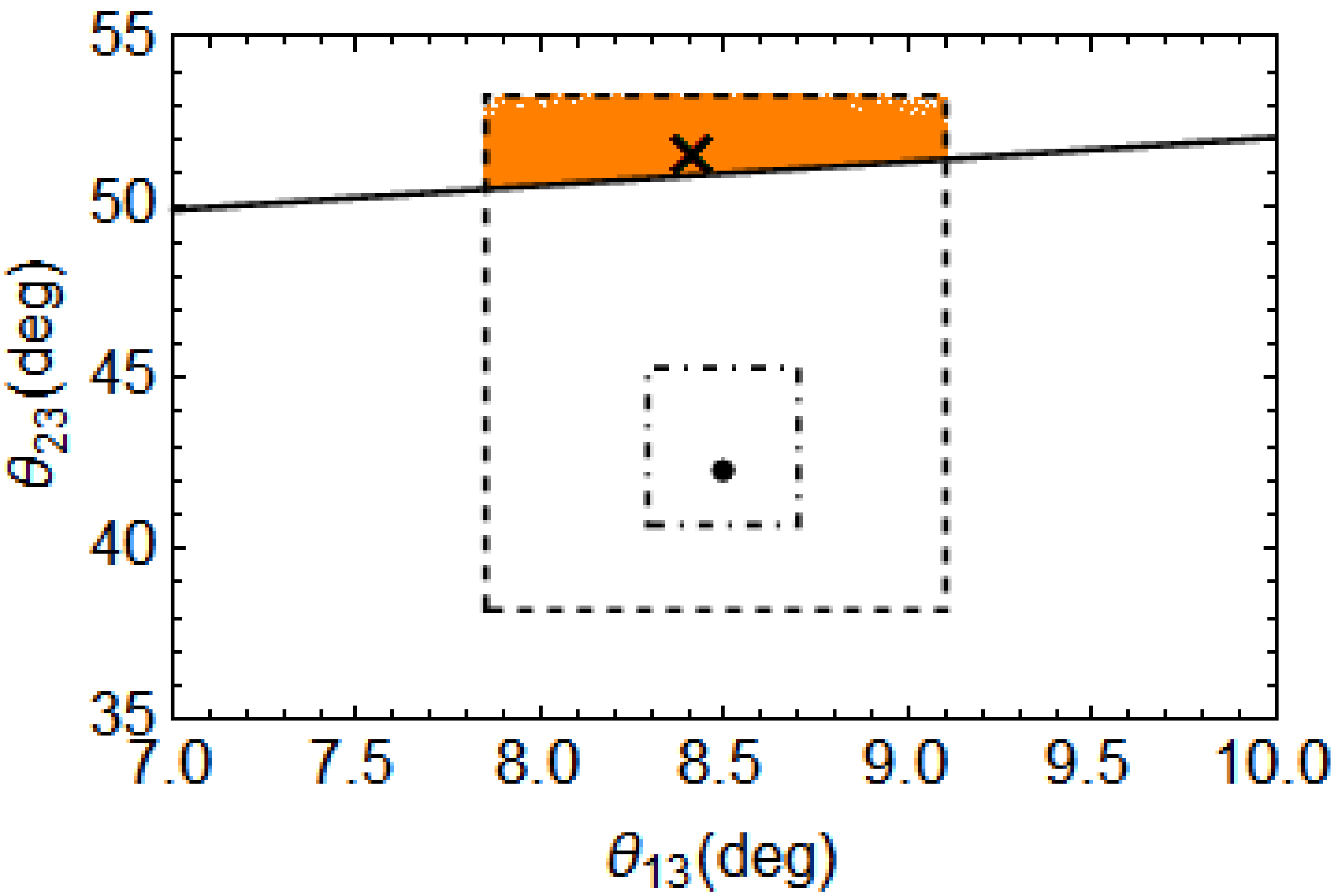}
}%
\subfigure{%
\hspace{0.1cm}
\includegraphics[width=0.48\textwidth, height=0.29\textwidth]{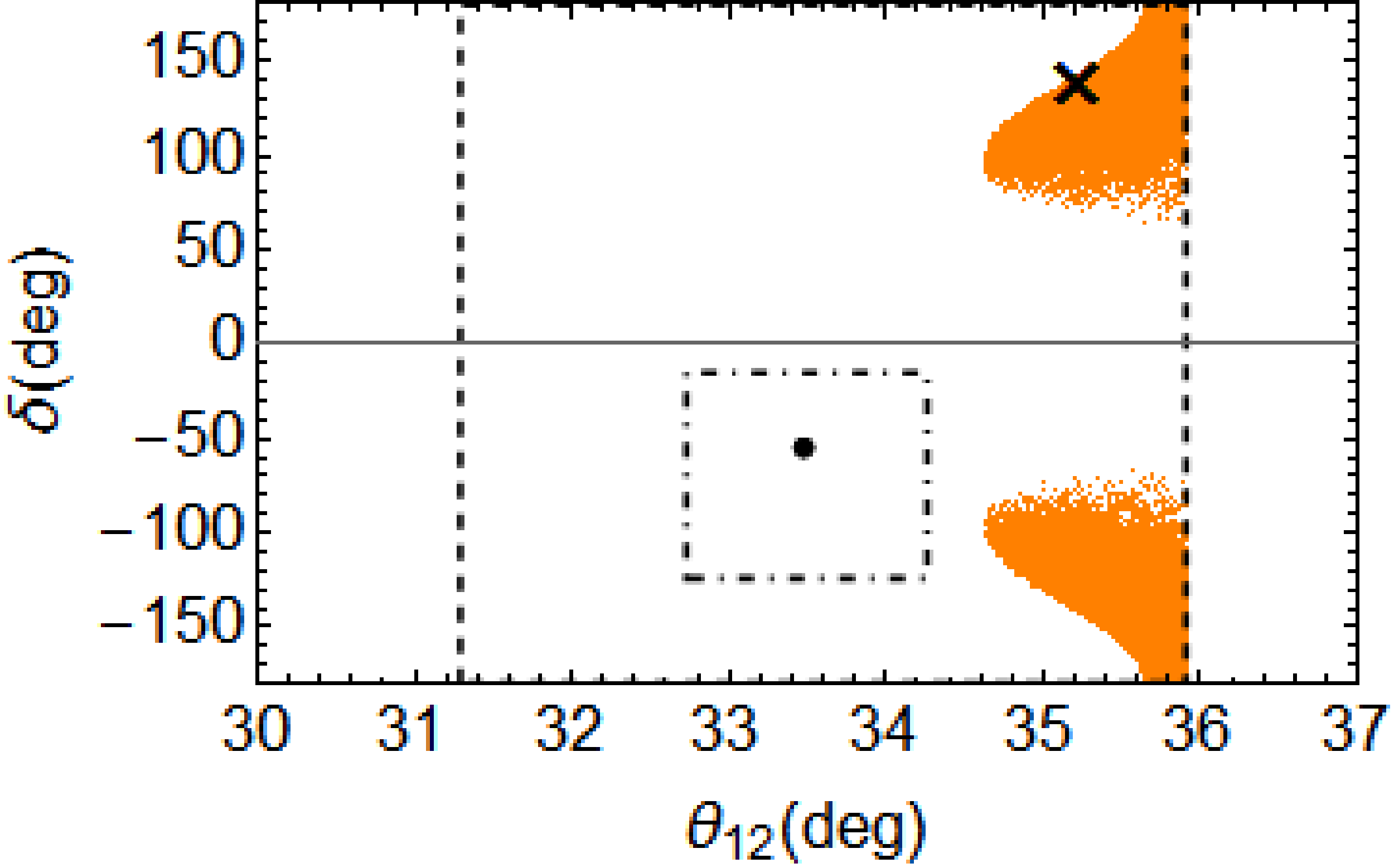}
}%
\vspace{-0.13cm}
\subfigure{%
\hspace{0.2cm}
\includegraphics[width=0.465\textwidth, height=0.31\textwidth]{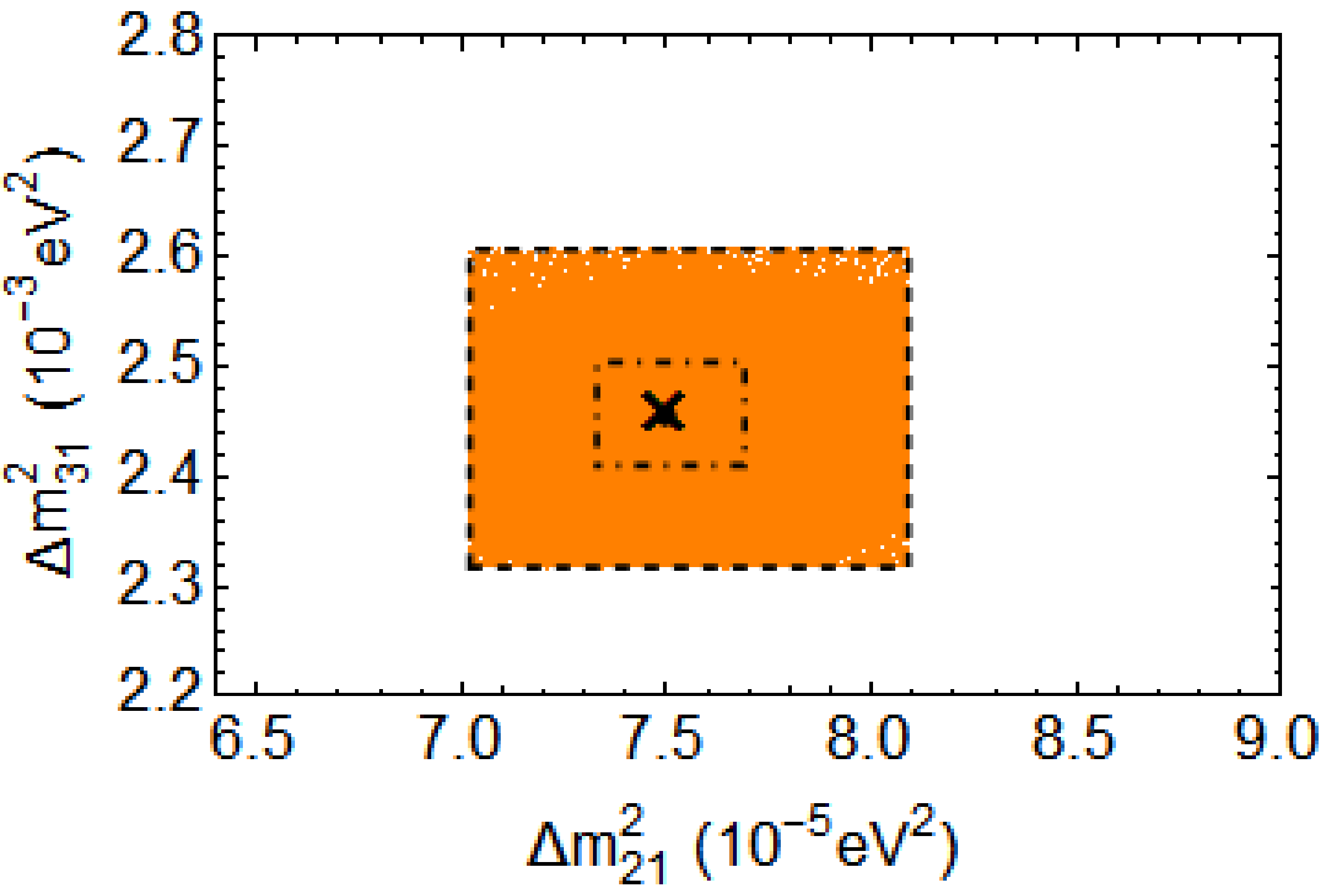}
}%
\subfigure{%
\hspace{0.4cm}
\includegraphics[width=0.455\textwidth, height=0.295\textwidth]{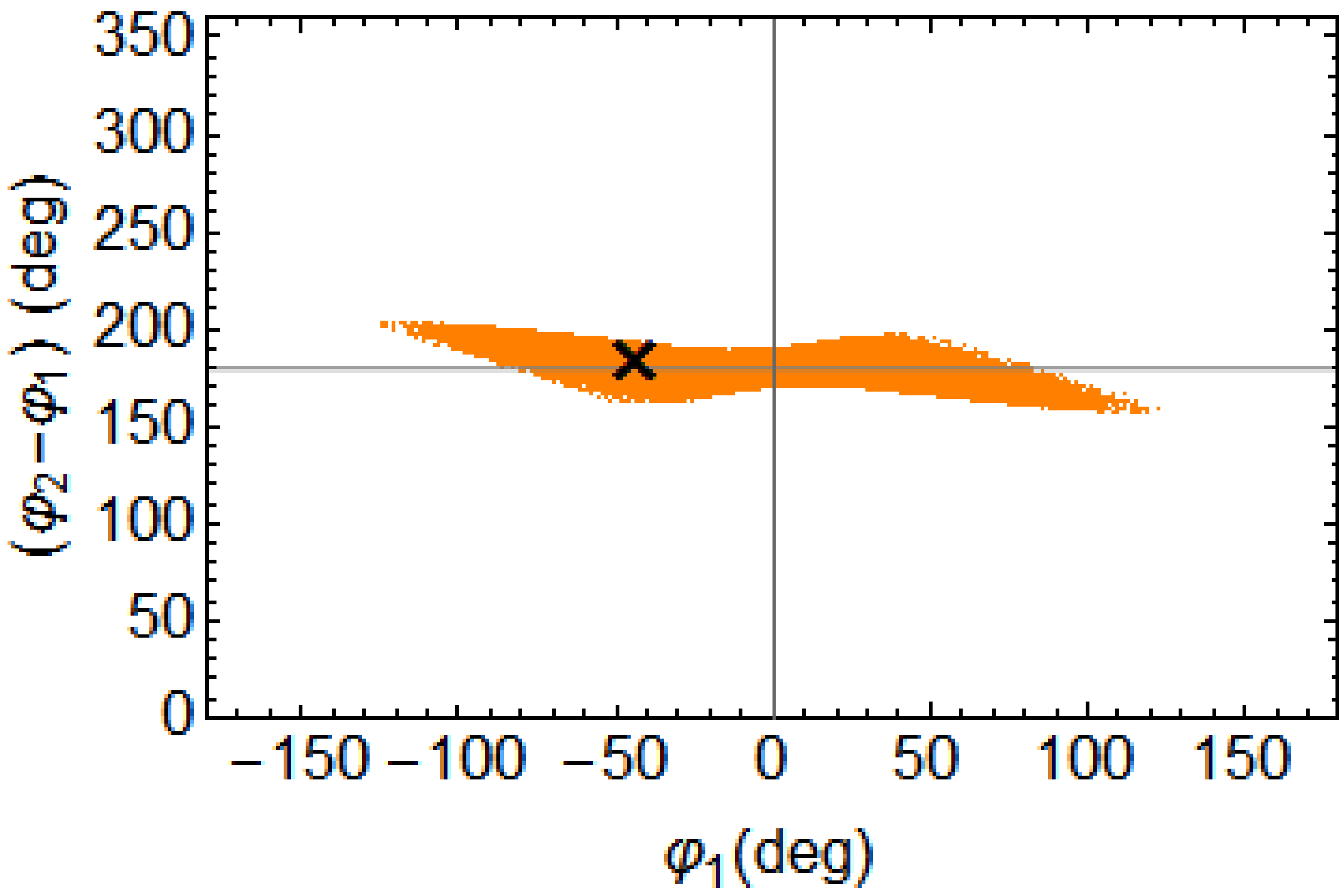}
}%
\vspace{-0.3cm}
\subfigure{%
\includegraphics[width=0.49\textwidth]{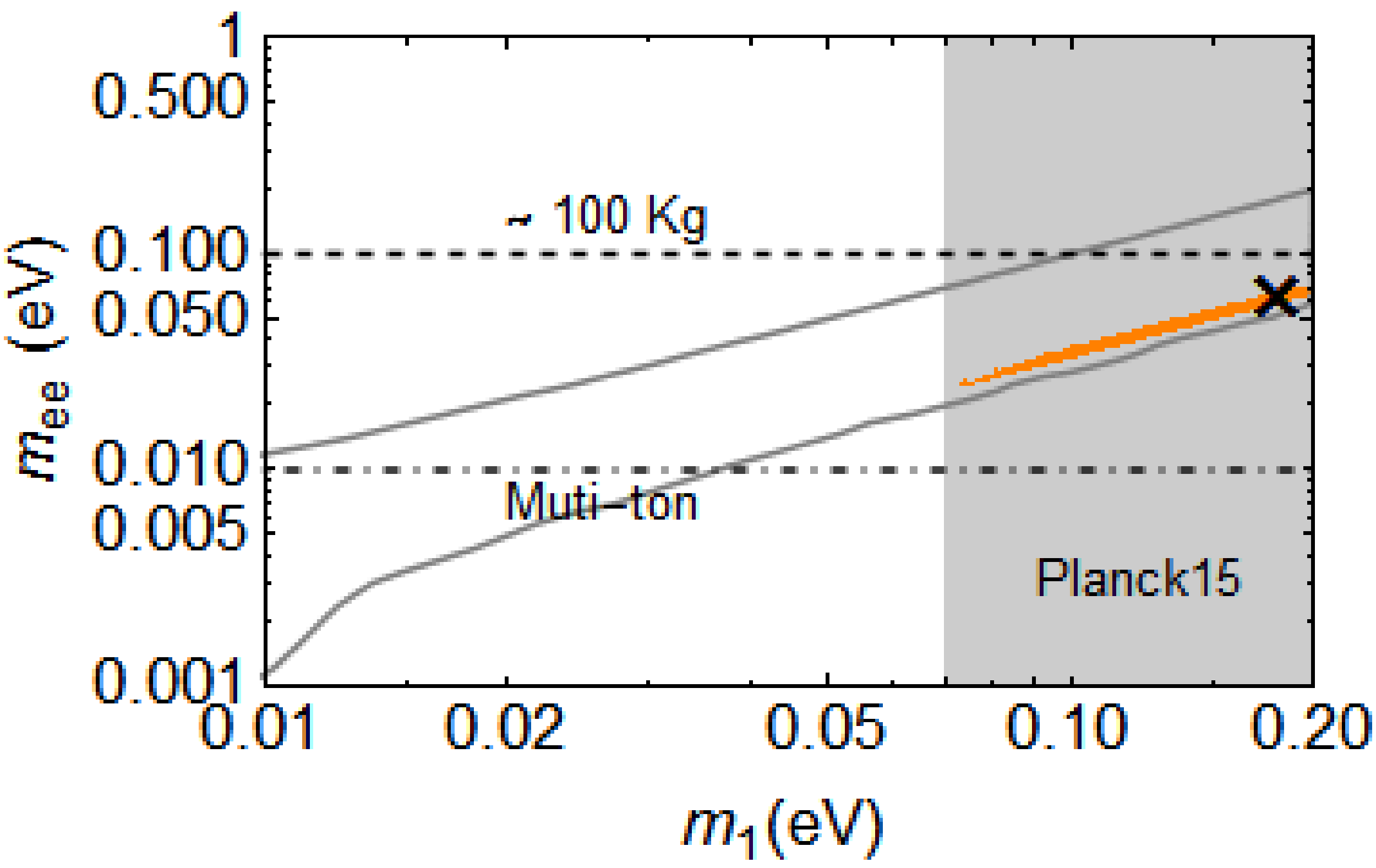}
}%
\subfigure{%
\hspace{0cm}
\includegraphics[width=0.5\textwidth, height=0.31\textwidth]{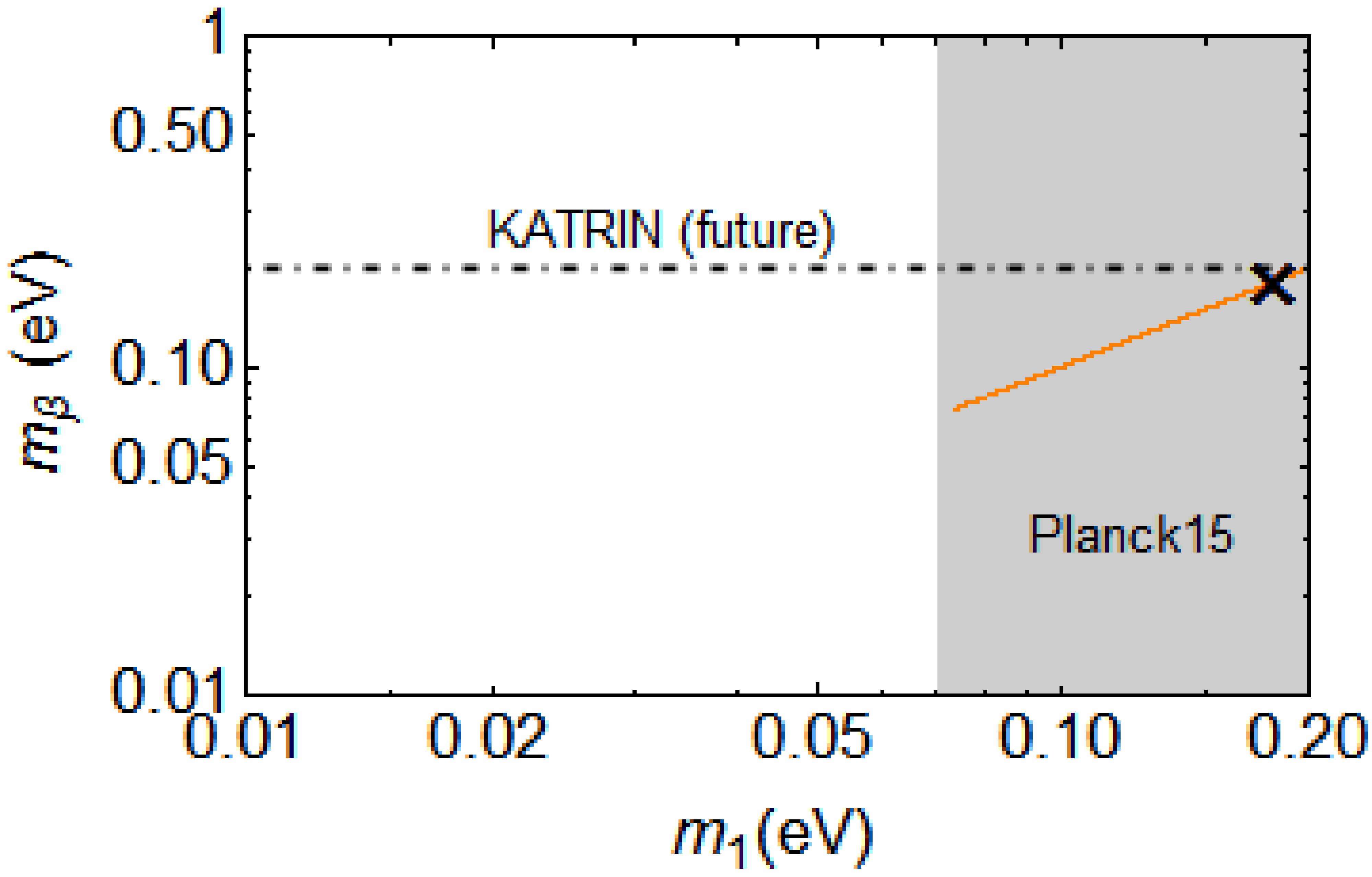}
}%
\end{center}
\vspace{-0.5cm}
\caption{Predictions for the low-energy neutrino parameters in the case of TBM for NH. The dashed contours in the plots of mixing angles and mass-squared differences refer to the $1\sigma$ and $3\sigma$ ranges taken from Ref.~\cite{Gonzalez-Garcia:2014bfa}, while the best-fit values from Ref.~\cite{Gonzalez-Garcia:2014bfa} are indicated by the black solid dots. The black solid line in the first plot denotes the lower bound in Eq.~(22). For the two plots in the last row, the shaded areas indicate the exclusion regions by the Planck 2015 data. The grey curves in the plot of ($m_{ee}^{}$, $m_1^{}$) stand for the allowed region for $m_{ee}^{}$, when three mixing angles vary within their $3\sigma$ ranges and the two Majorana phases are free.}
\label{fg:TBM_NH_LE}
\end{figure}
%%%%%%%%%%%%%%%%%%%%%%%%%%%%%%%%%%%%%%%%%%%%%%%%%%%%%%%%%%%%%%%%%%%%%%%%%%

Th predictions for the low-energy neutrino parameters are shown in Fig.~\ref{fg:TBM_NH_LE}, together with experimental constraints, and some comments on them will be instructive.
\begin{itemize}
\item The correlation between $\theta_{13}^{}$ and $\delta \theta_{23}^{}$ is indeed confirmed, as can be seen in the sub-figure of $\theta_{13}^{}$ versus $\theta_{23}^{}$. The black solid line corresponds to the lower bound derived in Eq.~(\ref{eq:t13_d23_bound}), and the allowed values for $\theta_{23}^{}$ are indeed above that line. For this reason, one can also observe that TBM in the NH case is excluded at $1\sigma$ level. However, there remains a portion of parameter space at the $3\sigma$ level.
\item In the sub-figure of $\theta_{12}$ versus $\delta$, we confirm the previous observation that $\theta_{12}^{}$ tends to become larger at low energies. Such an increase also renders the TBM pattern disfavored at the $1\sigma$ level. In addition, the predicted value of $\delta$ turns out to be in the second and third quadrants.
    To understand this feature, we recall that at the scale $\Lambda$, because of $\theta_{13}^{}=0$, the Dirac CP-violating phase $\delta$ is ill-defined. However, when the downward running takes place, $\theta_{13}^{}$ becomes non-zero, and $\delta$ restores its definition. The limiting value of $\delta$ at $\Lambda$ can then be determined by requiring its derivative ${\rm d}\delta/{\rm d}t$ to be finite \cite{Antusch:2003kp,Dighe:2008wn,Luo:2006tb}. From the RGE of $\delta$ listed in Appendix A, such a requirement yields
\begin{eqnarray}
\tan\delta^\Lambda = \frac{m_1^\Lambda \sin\varphi_1^\Lambda - (1+\zeta) m_2^\Lambda \sin\varphi_2^\Lambda}{m_1^\Lambda \cos\varphi_1^\Lambda - (1+\zeta) m_2 \cos\varphi_2^\Lambda - \zeta m_3^\Lambda} \; ,
\end{eqnarray}
where $\zeta \equiv \Delta m_{21}^2/\Delta m_{32}^2$. In the approximation of $\zeta \approx 0$ and $\varphi_2^\Lambda \approx \varphi_1^\Lambda + \pi$, we find that $\delta^\Lambda$ is close to $\varphi_2^\Lambda$ or $\varphi_2^\Lambda+\pi$. Because $\delta$ and the sign of $\sin\theta_{13}$ are closely related in the MNSP matrix, such a two-fold ambiguity on $\delta$ would be further removed if imposing the fact that we always keep $\theta_{13}$ to be in the first quadrant during the RG running. This can be seen by inspecting the RGE of $\theta_{13}^{}$. According to \cite{Antusch:2013jca}, we have
\begin{eqnarray}
\frac{\text{d}\theta_{13}}{\text{d} t} \propto \frac{m_1 \cos(\varphi_1^{} - \delta) - (1+\zeta) m_2 \cos(\varphi_2^{} - \delta) - \zeta m_3 \cos\delta}{\Delta m_{32}^2} \; ,
\end{eqnarray}
at the leading order. Given $\zeta \ll 1$, $\varphi_2^\Lambda \approx \varphi_1^\Lambda + \pi$ and $\Delta m_{31}^2 >0$ for NH, we find that ${\rm d}\theta_{13}^{}/{\rm d}t$ can only be negative when $\delta^\Lambda_{} \approx \varphi_2^\Lambda$. Such a negative value of ${\rm d}\theta^{}_{13}/{\rm d}t$ would then lead $\theta_{13}$ to enter the desired first quadrant from a zero boundary value. Hence, we need to have $\delta^\Lambda_{} \approx \varphi_2^\Lambda$ at the scale $\Lambda$.

During the running from $\Lambda$ to the low-energy scale, because of $\varphi_2^\Lambda - \varphi_1^\Lambda \approx \pi$, all the three phases $\delta$, $\varphi_1^{}$, and $\varphi_2^{}$ do not change much, according to their RGEs given in Appendix A. Therefore, the equality of $\delta$ and $\varphi_2^{}$ also holds approximately at low energies, as one can see from the best-fit points given in Table \ref{tb:TBM}.

\item From the left plot in the middle row of Fig.~\ref{fg:TBM_NH_LE}, we can observe that two neutrino mass-squared differences $\Delta m_{21}^2$ and $\Delta m_{31}^2$ can be easily reproduced. Comparing the phases $\{\varphi^{}_1, \varphi^{}_2\}$ in the right plot in the middle row of Fig.~\ref{fg:TBM_NH_LE} with those in~Fig.~\ref{fg:TBM_NH_HE}, one can see that the shape of allowed regions remains nearly unchanged, but becomes slightly broader. The reason is the mild running effect under the condition $\varphi_2 \approx \varphi_1 + \pi$, as we explain above.

\item As one can observe from Fig.~\ref{fg:TBM_NH_HE}, only a large value $m^\Lambda_1 \gtrsim 0.1~{\rm eV}$ is allowed for a wide range of the high-energy scale $\Lambda$. Since $m^{}_1$ does not run significantly, it still sits in the quasi-degenerate region at low energies. Now we explore the implications for the neutrinoless double-beta decay and the tritium beta-decay experiments. The effective neutrino mass for neutrinoless double-beta decays is
\begin{eqnarray}
m^{}_{ee} \equiv \left||U^{}_{e1}|^2 e^{-{\rm i}\varphi^{}_1} m^{}_1 + |U^{}_{e2}|^2 e^{-{\rm i}\varphi^{}_2}m^{}_2 + |U^{}_{e3}|^2 e^{-2{\rm i}\delta} m^{}_3\right| \; ,
%    (30)
\label{eq:mbb}
\end{eqnarray}
where $\varphi^{}_1$ and $\varphi^{}_2$ are two Majorana CP-violating phases. In the standard parametrization, we have $|U^{}_{e1}|^2 = \cos^2\theta^{}_{13} \cos^2 \theta^{}_{12}$, $|U^{}_{e2}|^2 = \cos^2\theta^{}_{13} \sin^2 \theta^{}_{12}$ and $|U^{}_{e3}|^2 = \sin^2 \theta^{}_{13}$. Given $\varphi^{}_2 - \varphi^{}_1 \approx \pi$ and $\sin^2 \theta^{}_{13} \ll 1$, one can obtain $m^{}_{ee} \approx m^{}_1 \cos 2\theta^{}_{12}$ in the limit of nearly-degenerate neutrino masses. As shown in Fig.~\ref{fg:TBM_NH_LE}, such a large effective neutrino mass is almost reached by current 100 kilogram-scale experiments, will be definitely accessible by the future multi-ton scale experiments~\cite{deGouvea:2013onf,Rodejohann:2012xd}. On the other hand, the effective neutrino mass relevant for beta decays is defined as
\begin{equation}
m^{}_\beta \equiv \sqrt{|U^{}_{e1}|^2 m^2_1 + |U^{}_{e2}|^2 m^2_2 + |U^{}_{e3}|^2 m^2_3} \; ,
\end{equation}
which approximates to $m^{}_\beta \approx m^{}_1$ in the nearly-degenerate mass limit. Therefore, as shown in Fig.~\ref{fg:TBM_NH_LE}, $m^{}_\beta$ is approximately given by $m^{}_1$, which is beyond the reach of the KATRIN experiment~\cite{Osipowicz:2001sq,fortheKATRIN:2013saa}.

Moreover, from the last row of Fig.\ref{fg:TBM_NH_LE}, it is evident that the allowed range of $m_1^{}$ has already been excluded by the latest cosmological upper bound on the sum of neutrino masses $\Sigma^{}_\nu < 0.23~{\rm eV}$ from Planck 2015 data~\cite{Ade:2015xua}. Therefore, if this cosmological bound is taken into account, the exact TBM at a high-energy scale is not allowed in the NH case.
\end{itemize}

%%%%%%%%%%%%%%%%%%%%%%%%%%%%%%%%%%%% Fig. 3 %%%%%%%%%%%%%%%%%%%%%%%%%%%%%%%
\begin{figure}[!t]
\begin{center}
\subfigure{%
\hspace{0.1cm}
\includegraphics[width=0.47\textwidth]{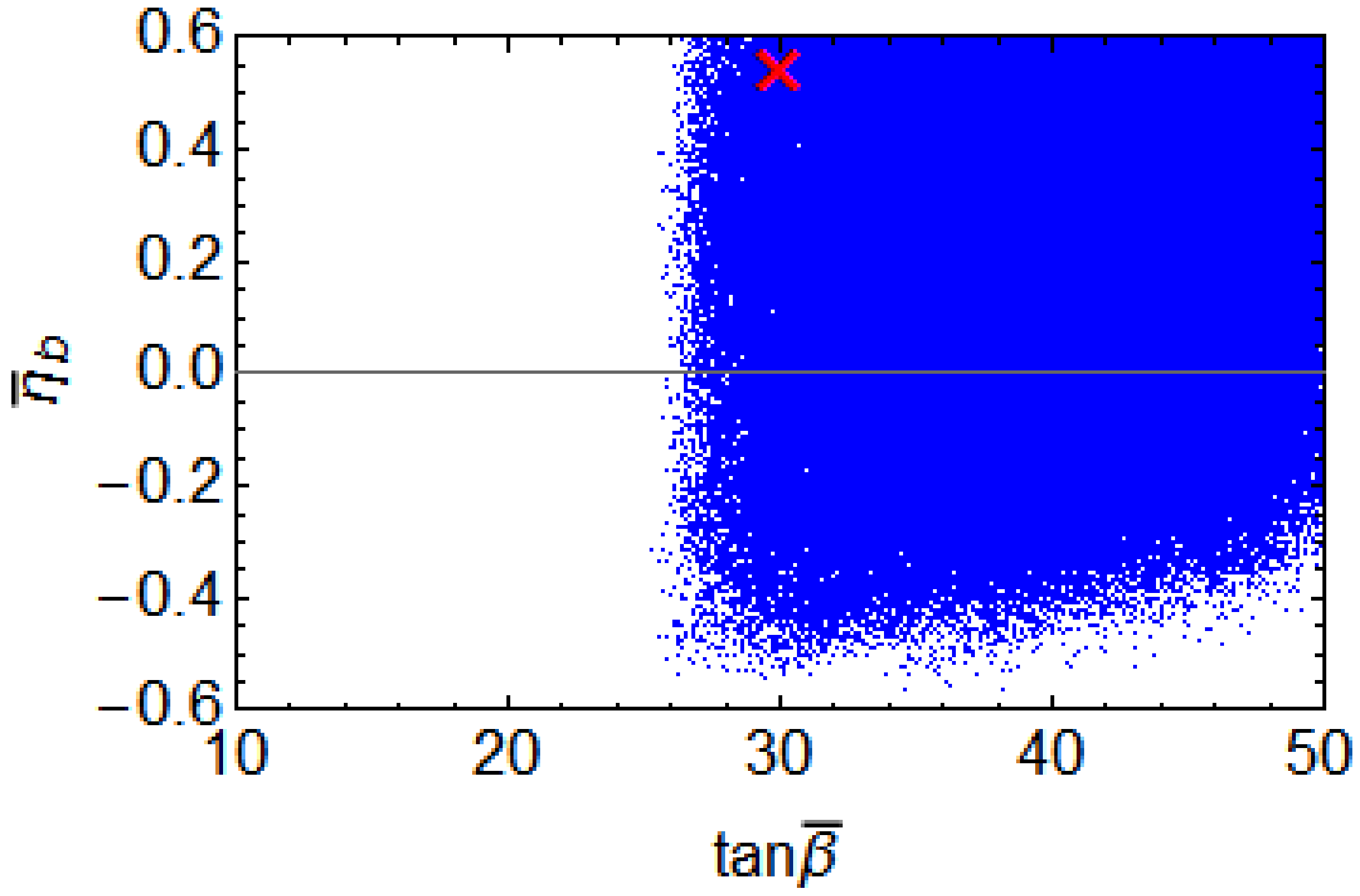}
}%
\subfigure{%
\hspace{0.5cm}
\includegraphics[width=0.48\textwidth, height=0.3\textwidth]{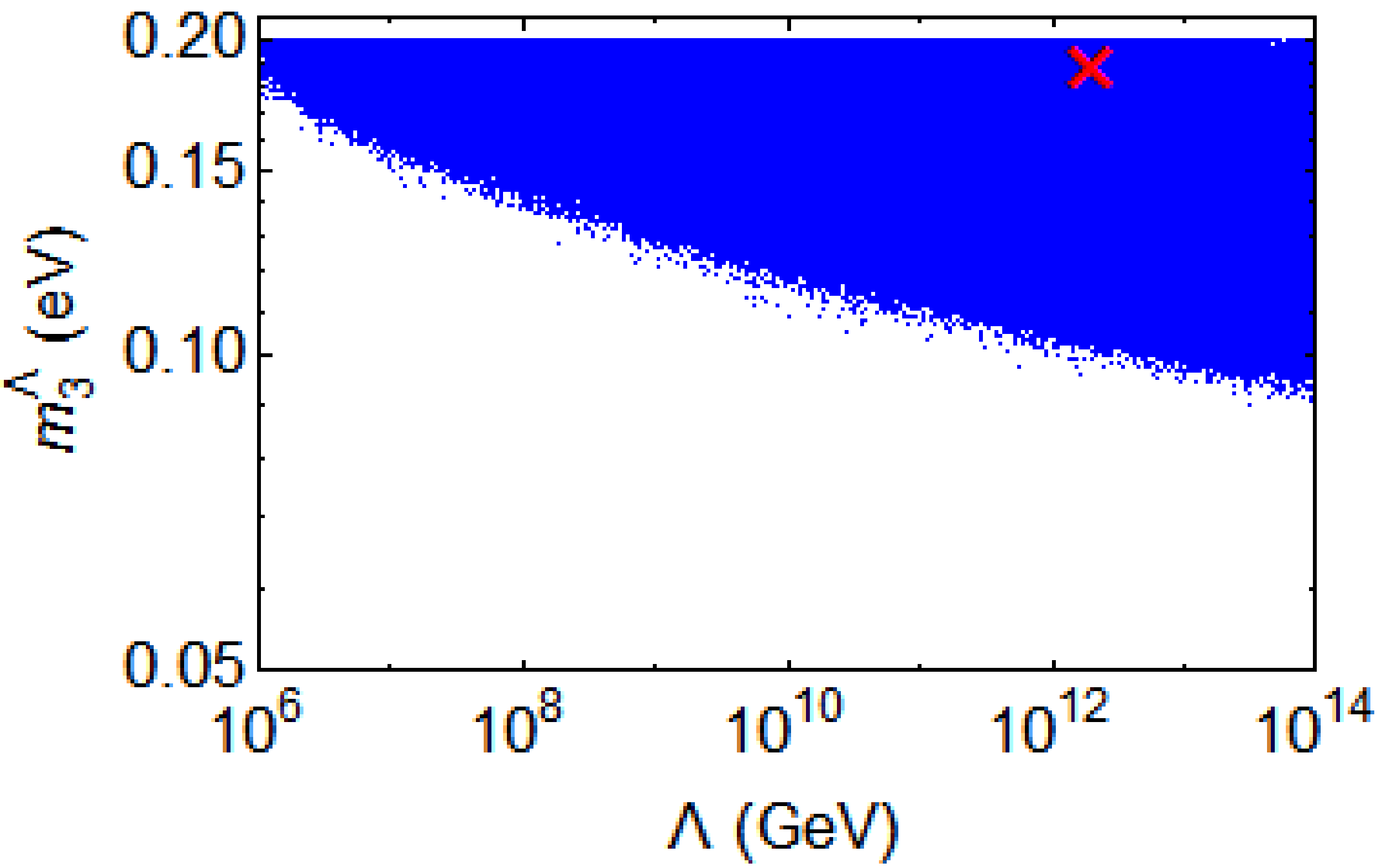}
}%
\vspace{-0.3cm}
\subfigure{%
\hspace{0.1cm}
\includegraphics[width=0.46\textwidth, height=0.3\textwidth]{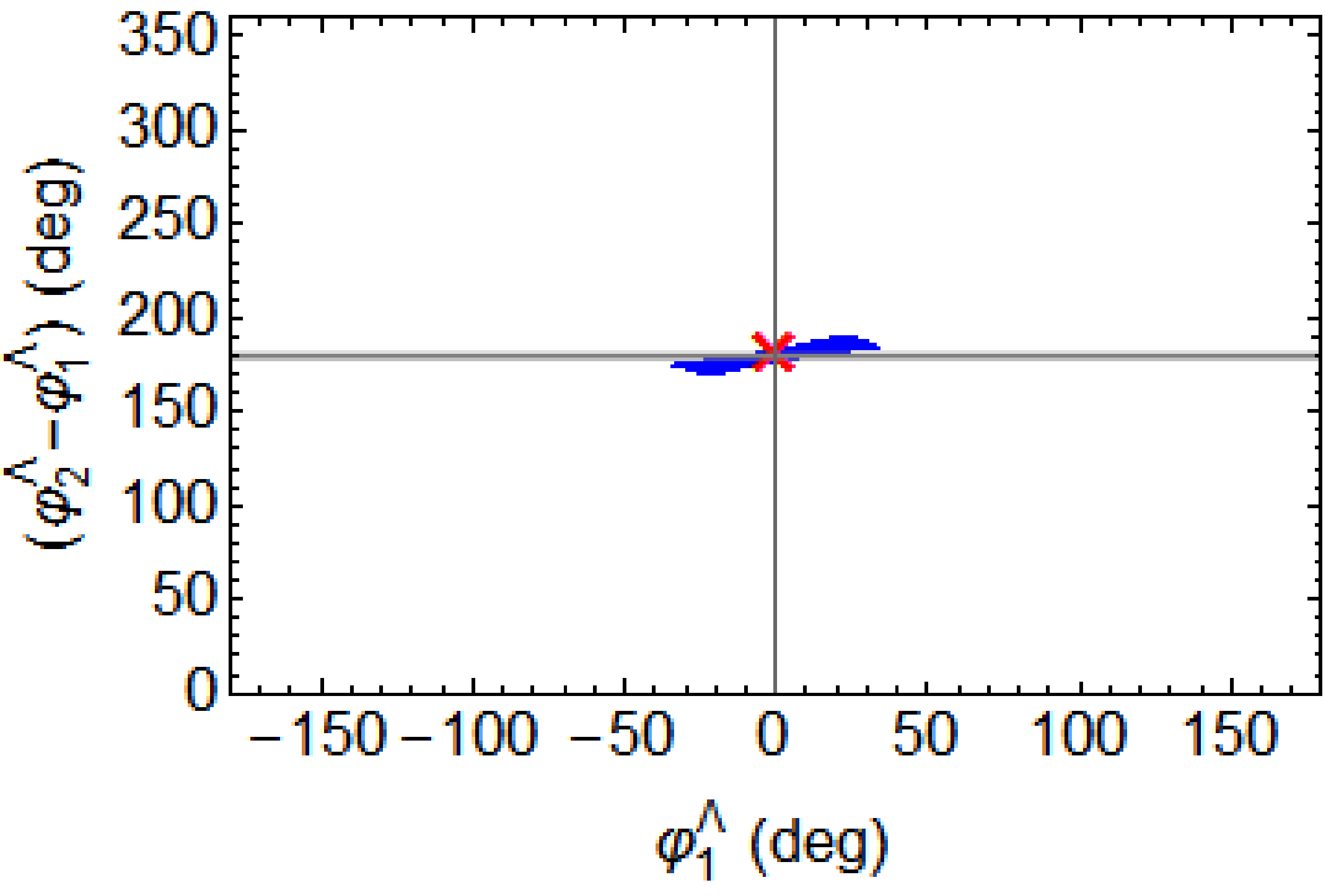}
}%
\subfigure{%
\hspace{1.0cm}
\includegraphics[width=0.45\textwidth]{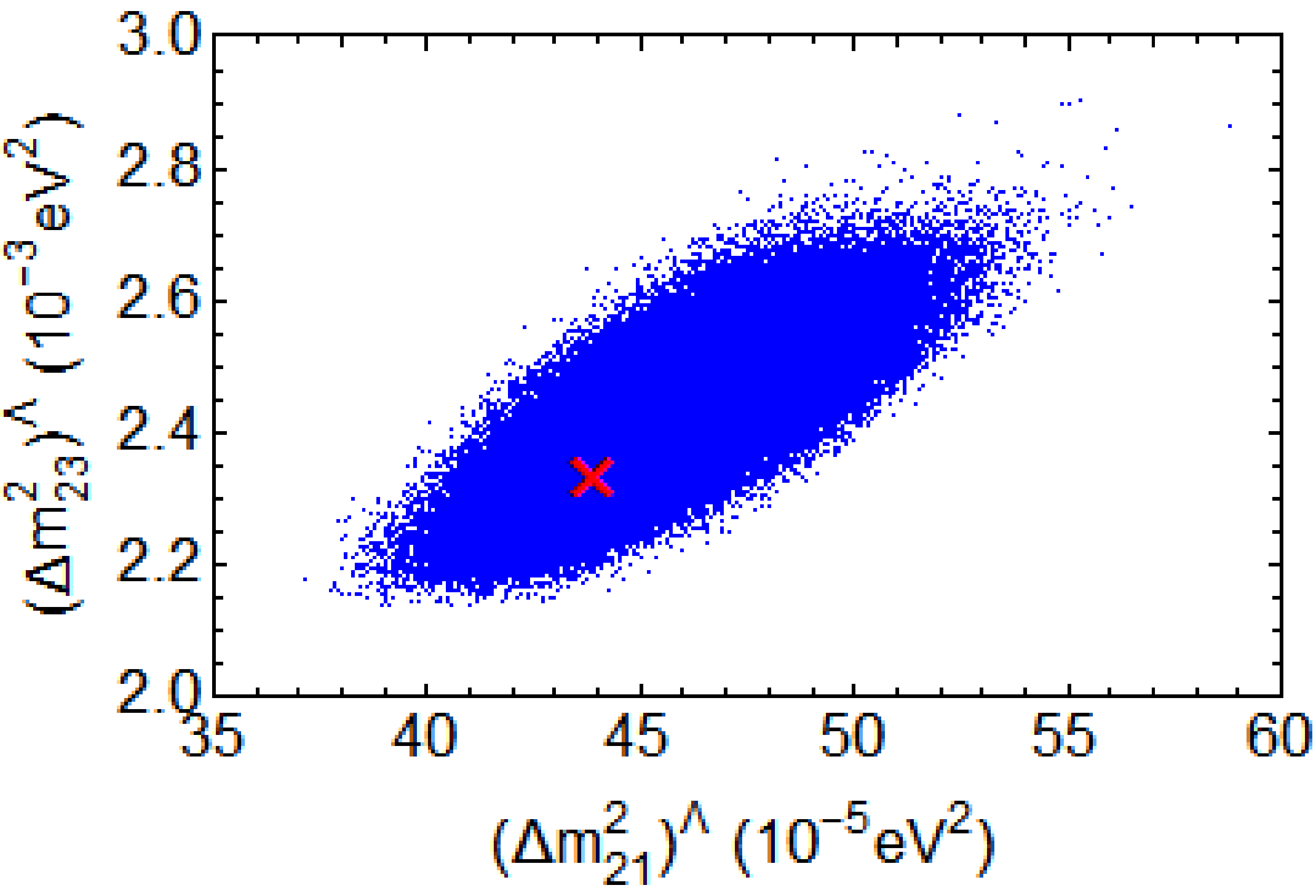}
}%
\end{center}
\vspace{-0.3cm}
\caption{Allowed parameter space for the free parameters given in Table \ref{tb:parameters} for the case of TBM in IH, where the red crosses stand for the best-fit points.}
\label{fg:TBM_IH_HE}
\end{figure}
%%%%%%%%%%%%%%%%%%%%%%%%%%%%%%%%%%%%%%%%%%%%%%%%%%%%%%%%%%%%%%%%%%%%%%%%%%%%
%%%%%%%%%%%%%%%%%%%%%%%%%%%%% Fig. 4 %%%%%%%%%%%%%%%%%%%%%%%%%%%%%%%%%%%%%%%
\begin{figure}[!t]
\begin{center}
\subfigure{%
\hspace{0.5cm}
\includegraphics[width=0.45\textwidth , height=0.3\textwidth]{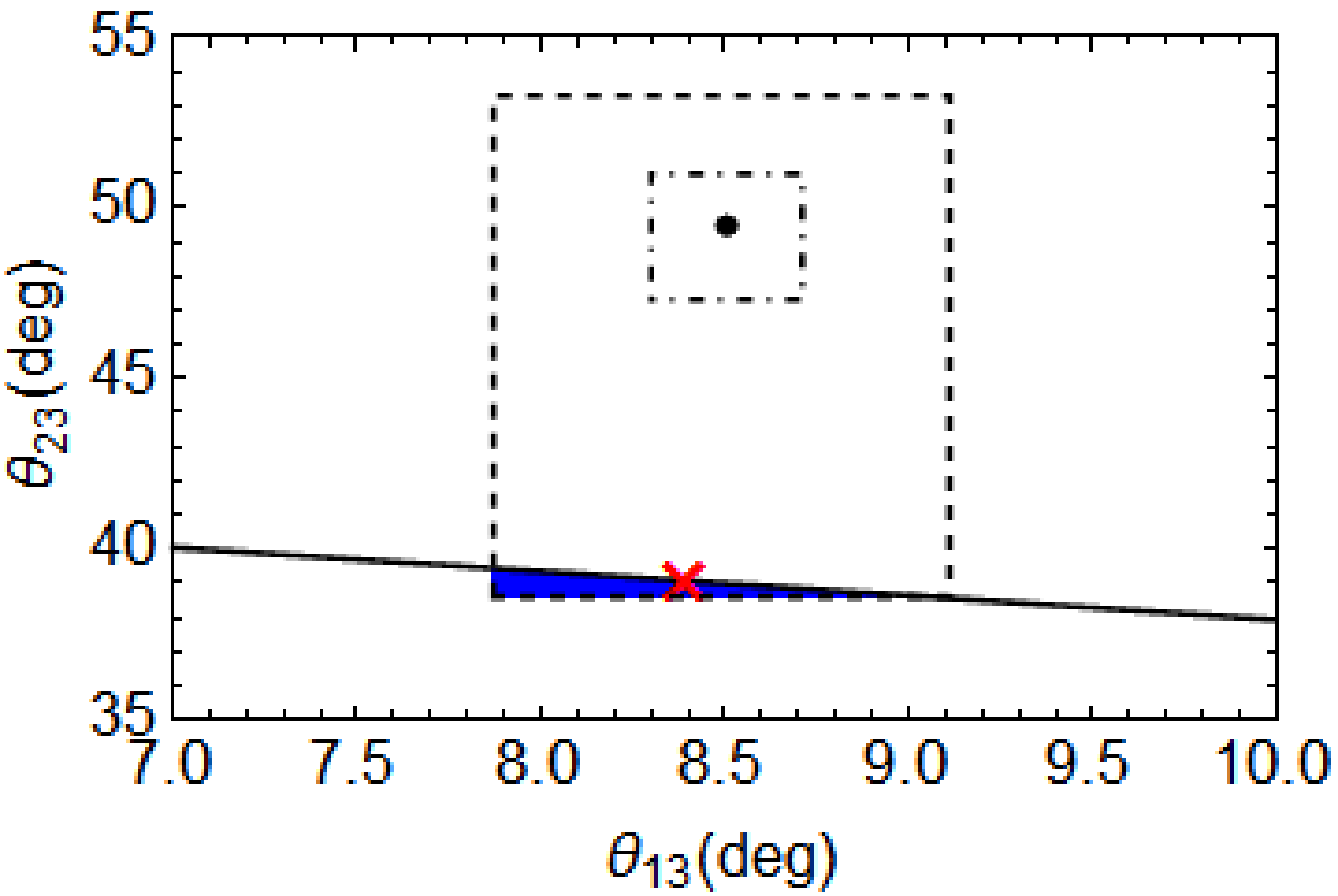}
}%
\subfigure{%
\hspace{0.1cm}
\includegraphics[width=0.48\textwidth, height=0.29\textwidth]{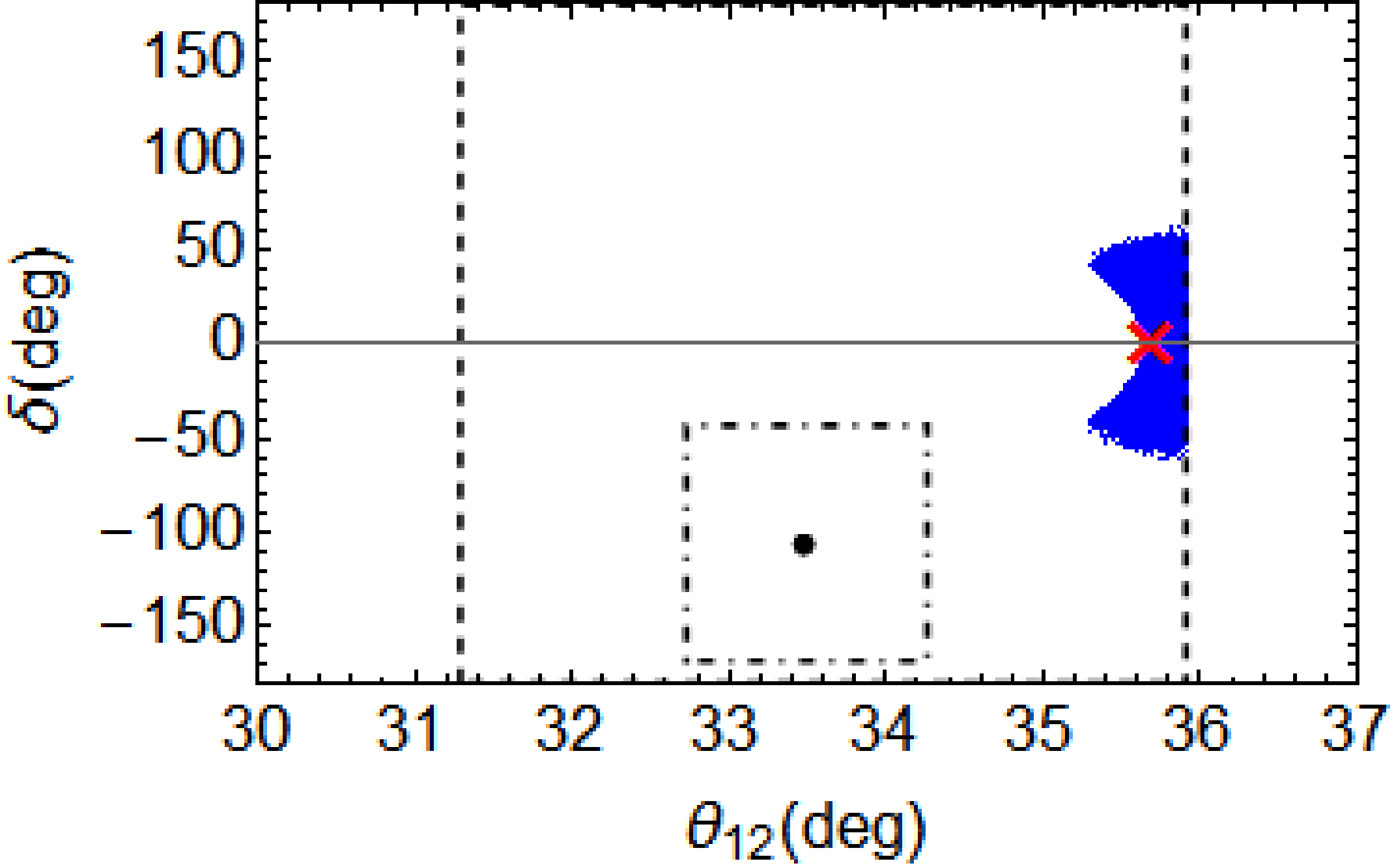}
}%
\vspace{-0.3cm}
\subfigure{%
\hspace{0.2cm}
\includegraphics[width=0.465\textwidth, height=0.30\textwidth]{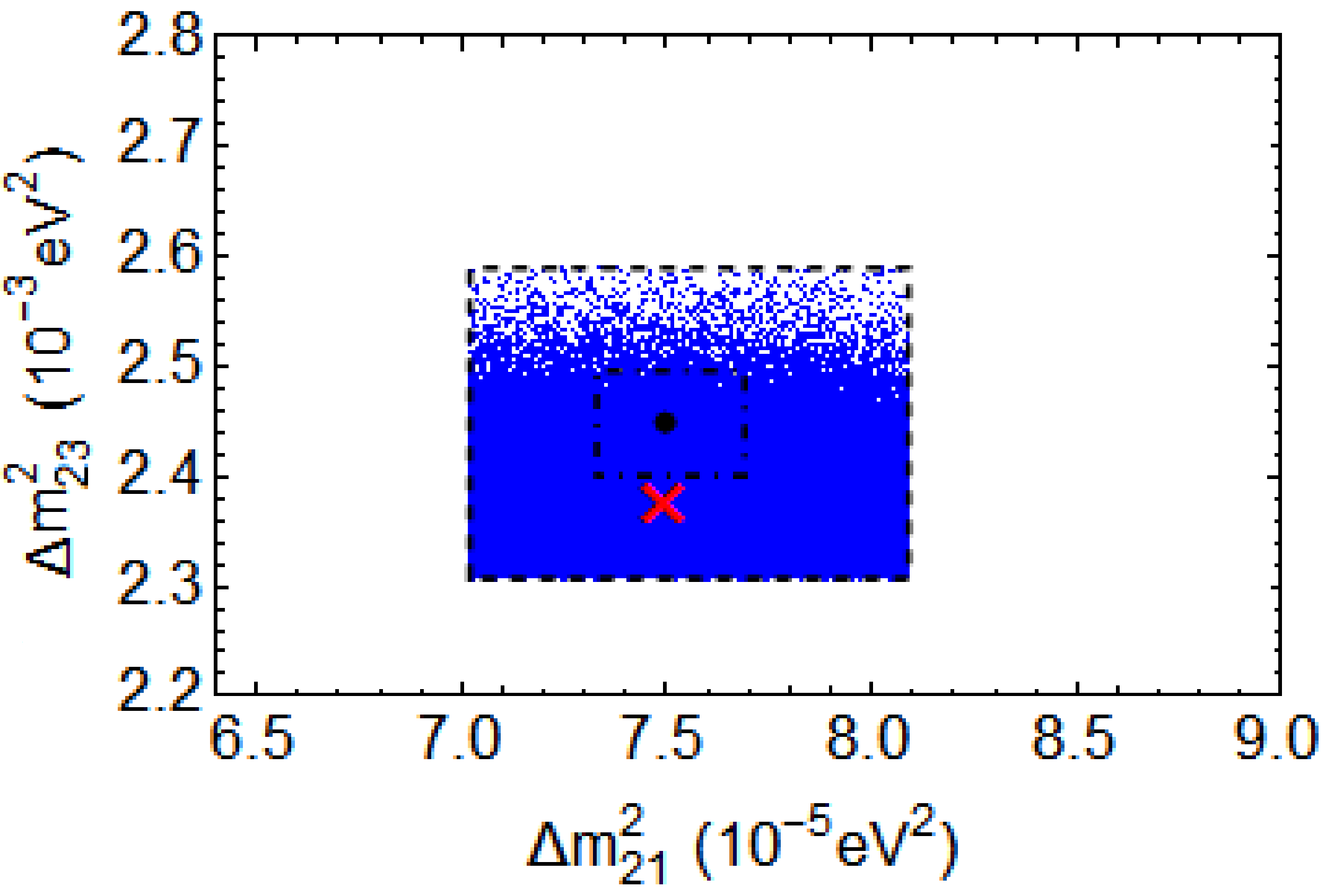}
}%
\subfigure{%
\hspace{0.4cm}
\includegraphics[width=0.455\textwidth, height=0.295\textwidth]{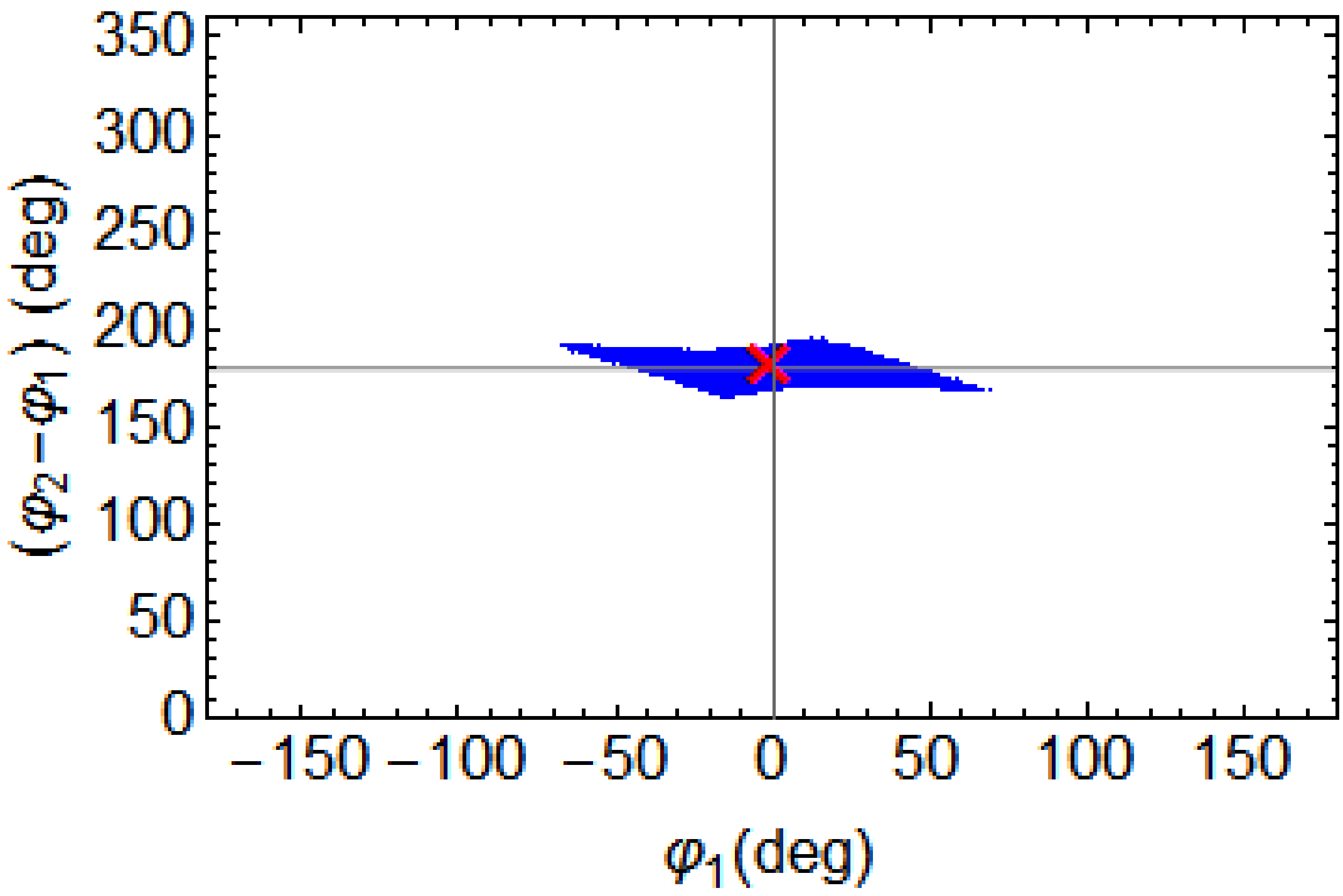}
}%
\vspace{-0.3cm}
\subfigure{%
\includegraphics[width=0.49\textwidth]{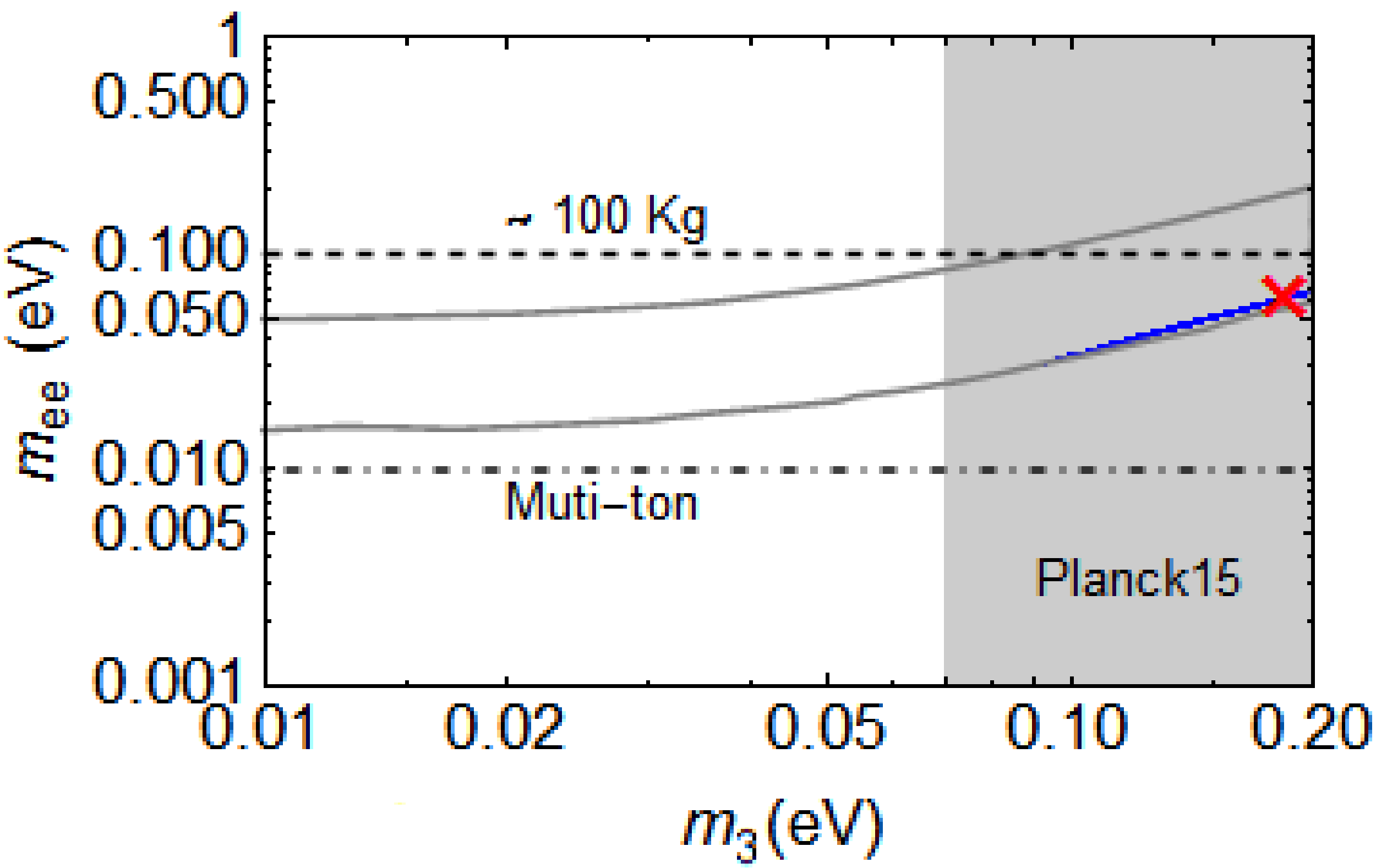}
}%
\subfigure{%
\hspace{0cm}
\includegraphics[width=0.5\textwidth, height=0.31\textwidth]{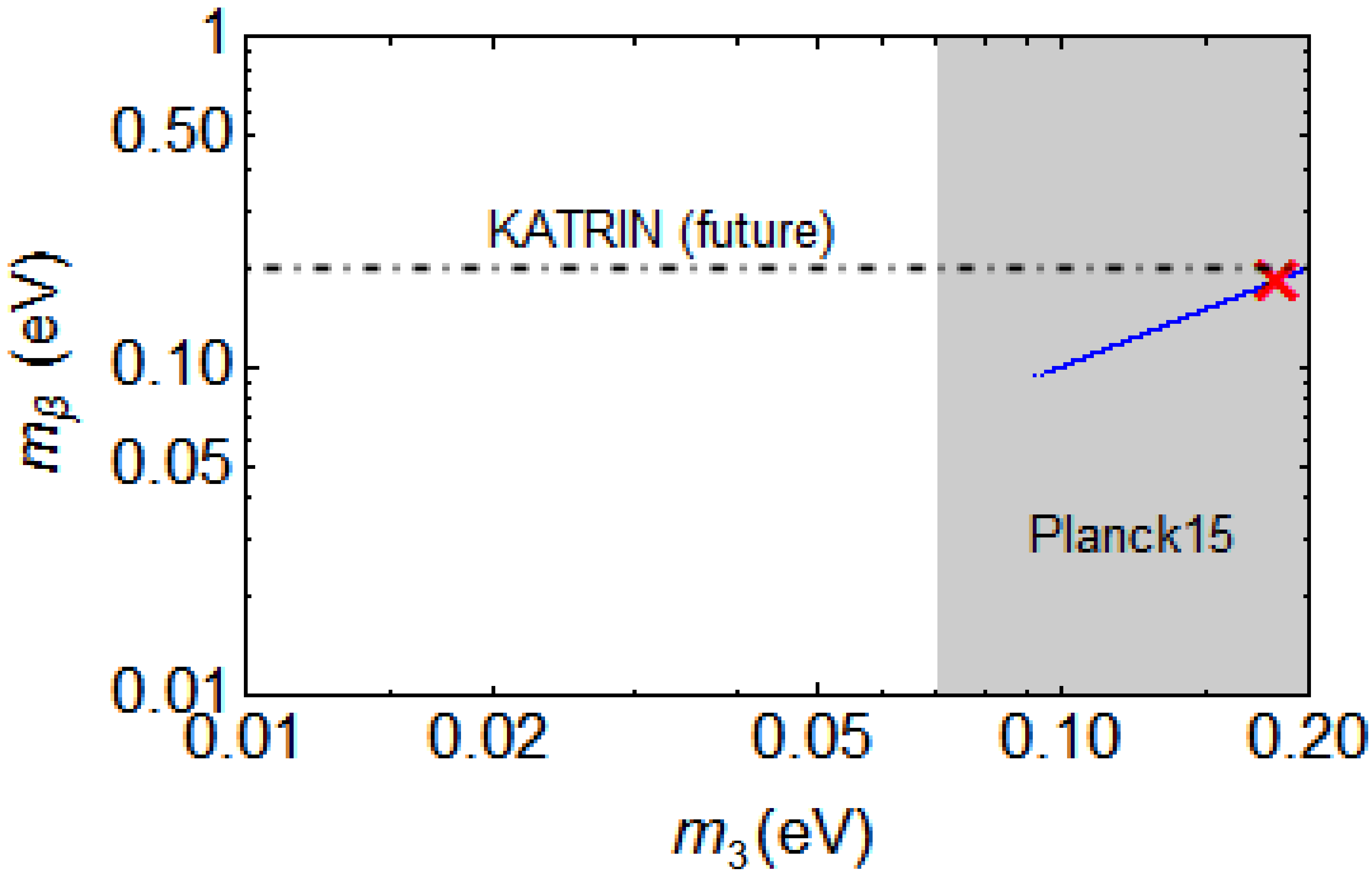}
}%
\end{center}
\vspace{-0.5cm}
\caption{Predictions for the low energy neutrino parameters in the case of TBM for IH. Detailed descriptions of each plot follow those in the caption of Fig.~\ref{fg:TBM_NH_LE}.}
\label{fg:TBM_IH_LE}
\end{figure}
%%%%%%%%%%%%%%%%%%%%%%%%%%%%%%%%%%%%%%%%%%%%%%%%%%%%%%%%%%%%%%%%%%%%%%%%%%%%%%
Next, we proceed with the IH case. The best-fit points in this case have also been given in Table~\ref{tb:TBM}, while the allowed parameter space for free parameters and the predictions for the low-energy observables are summarized in Fig.~\ref{fg:TBM_IH_HE} and Fig.~\ref{fg:TBM_IH_LE}, respectively. Although the results in the IH case are quite similar to those in the NH case, we can observe the following two different features. First, the $\chi^2$ fit to neutrino oscillation data in IH becomes worse, and this can be traced to the larger pull of $\theta_{23}^{}$, which originates from the fact that the best-fit value of $\theta_{23}^{}$ in Ref.~\cite{Gonzalez-Garcia:2014bfa} is farther away from $\pi/4$ in IH than that in NH. Second, from Fig.~\ref{fg:TBM_IH_LE}, we notice that the prediction of $\delta$ at low energies favors the first and fourth quadrants, instead of the second and third quadrants as in the previous NH case. The difference between the predictions of $\delta$ in two cases is approximately $\pi$, which can be understood by revisiting previous discussions on $\delta$. In the IH case, one has to apply $\Delta m_{31}^2 < 0$ to the RGE of $\delta$. As a consequence, $\delta$ tends to be close to $\varphi_1^{}$, instead of $\varphi_2^{}$. Furthermore, one can notice that the octant of $\theta^{}_{23}$ and the value of $\delta$ are correlated with the neutrino mass hierarchy, as previously observed in the context of $\mu$-$\tau$ symmetry~\cite{Luo:2014upa,Xing:2015fdg} and leptonic mixing sum rule~\cite{Zhang:2016djh}.

Finally, we briefly summarize the consequences of requiring both an exact TBM mixing pattern at high energies and the correct phenomenology at low energies. First, the large radiative correction needed for $\theta_{13}^{}$ forces neutrino masses to lie in the quasi-degenerate region, which in fact causes the TBM case to be in danger with the Planck data. Second, within the quasi-degenerate mass region, two Majorana phases have to be different by about $180^\circ$ so as to protect $\theta_{12}^{}$ from too large corrections. Given the above two facts, the Dirac CP-violating phase $\delta$ is found to be aligned with one Majorana phase, and there exists a correlation between $\theta^{}_{13}$ and $\theta^{}_{23}$ at low energies. It is such a correlation, together with a sizable value of $\theta_{12}$, that renders the TBM mixing pattern to be incompatible with the latest global-fit result at the $1\sigma$ level. Allowed parameter space of mixing parameters at the $3\sigma$ level are available if the cosmological bound on neutrino masses is relaxed.

\subsubsection{Golden-ratio and bimaximal mixing}
%%%%%%%%%%%%%%%%%%%%%%%%%%%%%%% Table 3 %%%%%%%%%%%%%%%%%%%%%%%%%%%%%%%%%%%%
\begin{table}[!t]
\centering
\begin{tabular}{c c c c c}
\hline
\multirow{2}{*}{Parameter} & \multicolumn{2}{c}{GR, NH} & \multicolumn{2}{c}{GR, IH} \\
\cline{2-3} \cline{4-5}
& best-fit & pull & best-fit & pull\\
\hline
$\overline{\eta}_b$ & -0.32 & - & 0.39 & - \\
$\tan\overline{\beta}$ & 32.5 & - & 40.6 & - \\
$\Lambda/ \text{GeV}$ & $3.59\times 10^{11}$ & - & $1.25\times 10^{8}$ & - \\
$\varphi_{1}^\Lambda / \text{deg}$ & 8.9 & - & -6.63 & - \\
$\varphi_{2}^\Lambda / \text{deg}$ & -176.8 & - & 177.1 & - \\
$m_{\text{lightest}}^\Lambda /\text{eV}$ & 0.18 & - & 0.19 & - \\
$(\Delta m_{21}^2)^\Lambda /10^{-5}~\text{eV}^2$ & 39.4 & - & 57.7 & - \\
$|\Delta m_{32}^2|^\Lambda /10^{-3}~\text{eV}^2$ & 2.93 & - & 2.52 & - \\
\hline
$\sin^2\theta_{12}$ & 0.305 & 0.116 & 0.305 & 0.065 \\
$\sin^2\theta_{13}$ & 0.0216 & -0.23 & 0.0214 & -0.43 \\
$\sin^2\theta_{23}$ & 0.591 & 3.48 & 0.409 & -5.49 \\
$\Delta m_{21}^2 /10^{-5}~\text{eV}^2$ & 7.46 & 0.21 & 7.49 & -0.017 \\
$|\Delta m_{3i}^2| /10^{-3}~\text{eV}^2$ & 2.46 & 0.08 & 2.377 & -1.52 \\
$m_{\text{lightest}} /\text{eV}$ & 0.17 & - & 0.18 & - \\
$\delta / \text{deg}$ & -150.4 & - & -29.3 & - \\
$\varphi_{1} / \text{deg}$ & 41.8 & - & -41.7 & - \\
$\varphi_{2} / \text{deg}$ & -163.1 & - & 162.6 & - \\
$m^{}_{ee} / \text{eV}$ & 0.076 & - & 0.081 & - \\
$m^{}_\beta / \text{eV}$ & 0.17 & - & 0.18 & - \\
\hline
$\chi^2_{\rm min}$ & & $\mathbf{12.2}$ & & $\mathbf{30.3}$ \\
\hline
\end{tabular}
\caption{The best-fit points of low-energy observables for the GR mixing pattern in both cases of NH and IH, where the input parameters at the high-energy scale are also given. }
\label{tb:GR}
\end{table}
%%%%%%%%%%%%%%%%%%%%%%%%%%%%%%%%%%%%%%%%%%%%%%%%%%%%%%%%%%%%%%%%%%%%%%%%%%%%%
For GR, as expected, the main results are quite similar to those for TBM, since the predicted value of $\theta_{12}^\Lambda$ in both cases are not far from current best-fit value of $\theta_{12}^{}$ at low energies. Therefore, our previous discussions on TBM can be readily applied to GR, and we only focus on the differences between these two cases. Although we have carried out a full numerical analysis of GR, only the $\chi^2$ fits to neutrino oscillation data in both NH and IH are presented in Table~\ref{tb:GR}, where one can see slightly better agreement with experimental data.

The differences between these GR and TBM cases arise from the fact that the predicted $\theta_{12}^\Lambda$ in GR is below the current best-fit value of $\theta_{12}^{}$ at low energies, while the opposite for TBM. Such a difference of $\theta_{12}^\Lambda$ will make the fitting to $\theta_{12}$ much easier for the GR case, as $\theta^{}_{12}$ increases when running downward from a high-energy scale. This better fit to $\theta_{12}$ is indeed observed in Table~\ref{tb:GR}, and because of this, the final minimal $\chi^2$ values in GR are smaller than their counterparts in TBM. Although a better fit to $\theta_{12}^{}$ is now obtained, the fate of GR remains the same as that of TBM. The main reason is the previously identified correlation between $\theta_{13}^{}$ and $\theta_{23}^{}$, which leads to a disagreement with the latest global-fit results at $1\sigma$ level.

Then, we come to the BM case. Our numerical study indicates that no allowed parameter space of neutrino mixing angles and mass-squared differences can be found even at the $3\sigma$ level. In Table~\ref{tb:BM}, we list the best-fit points for both NH and IH. As one can see, the minima of $\chi^2$ values for NH and IH are both over 100, because of the poor fit to both $\theta_{12}^{}$ and $\theta_{23}^{}$.

\begin{figure}[!t]
\begin{center}
\subfigure{%
\hspace{0.65cm}
\includegraphics[width=0.45\textwidth]{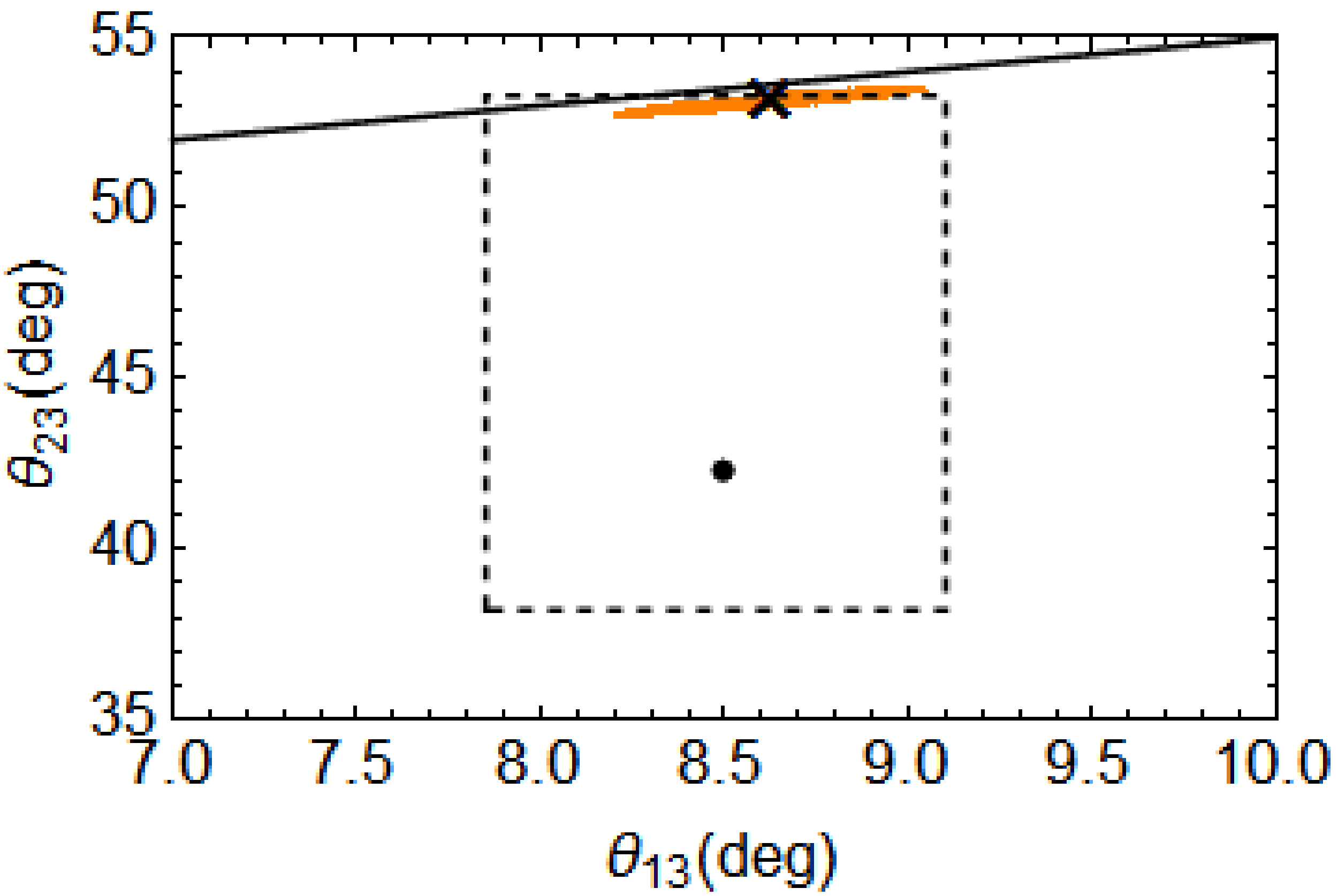}
}%
\subfigure{%
\hspace{0.5cm}
\includegraphics[width=0.46\textwidth, height=0.3\textwidth]{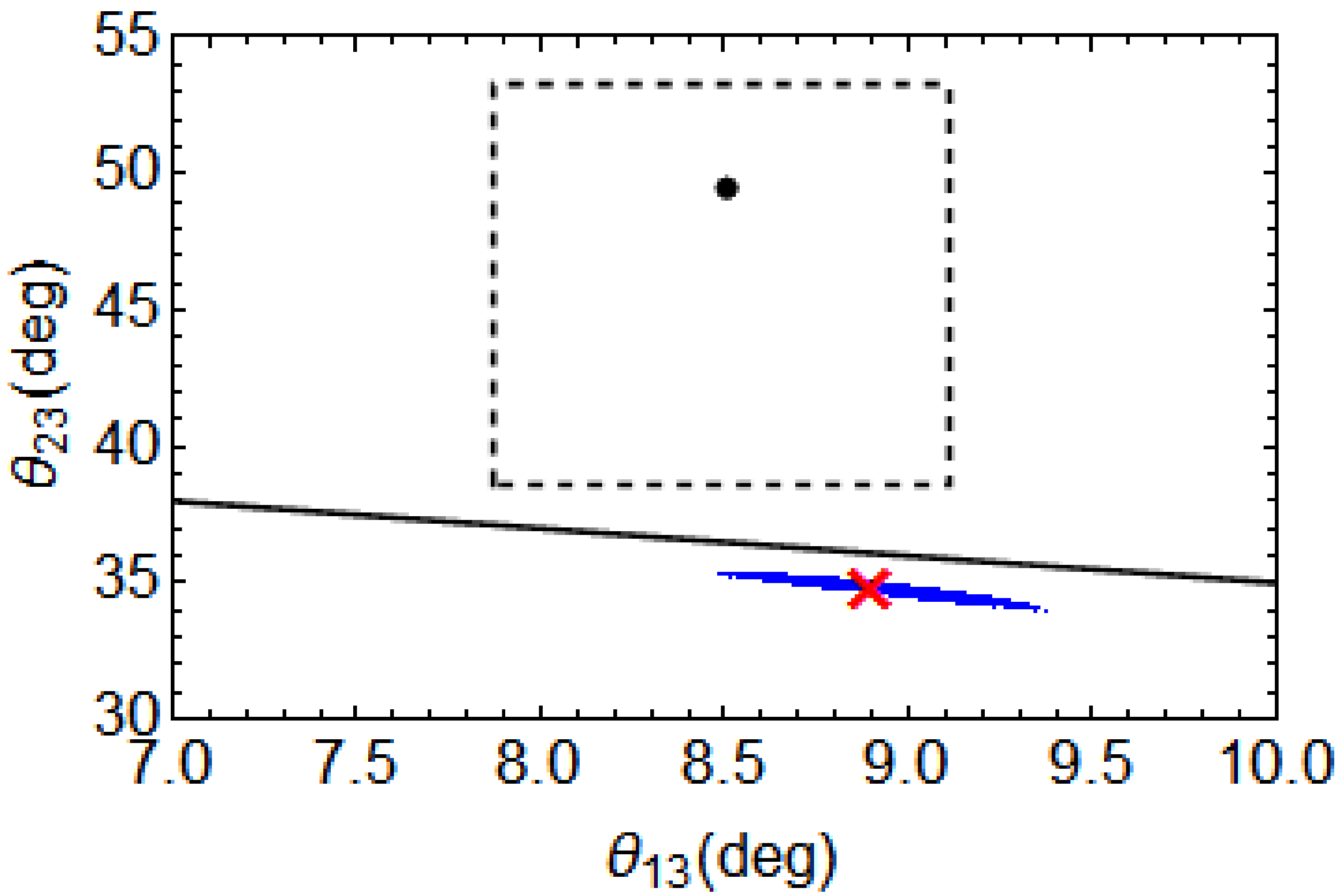}
}%
\vspace{-0.3cm}
\subfigure{%
\includegraphics[width=0.47\textwidth]{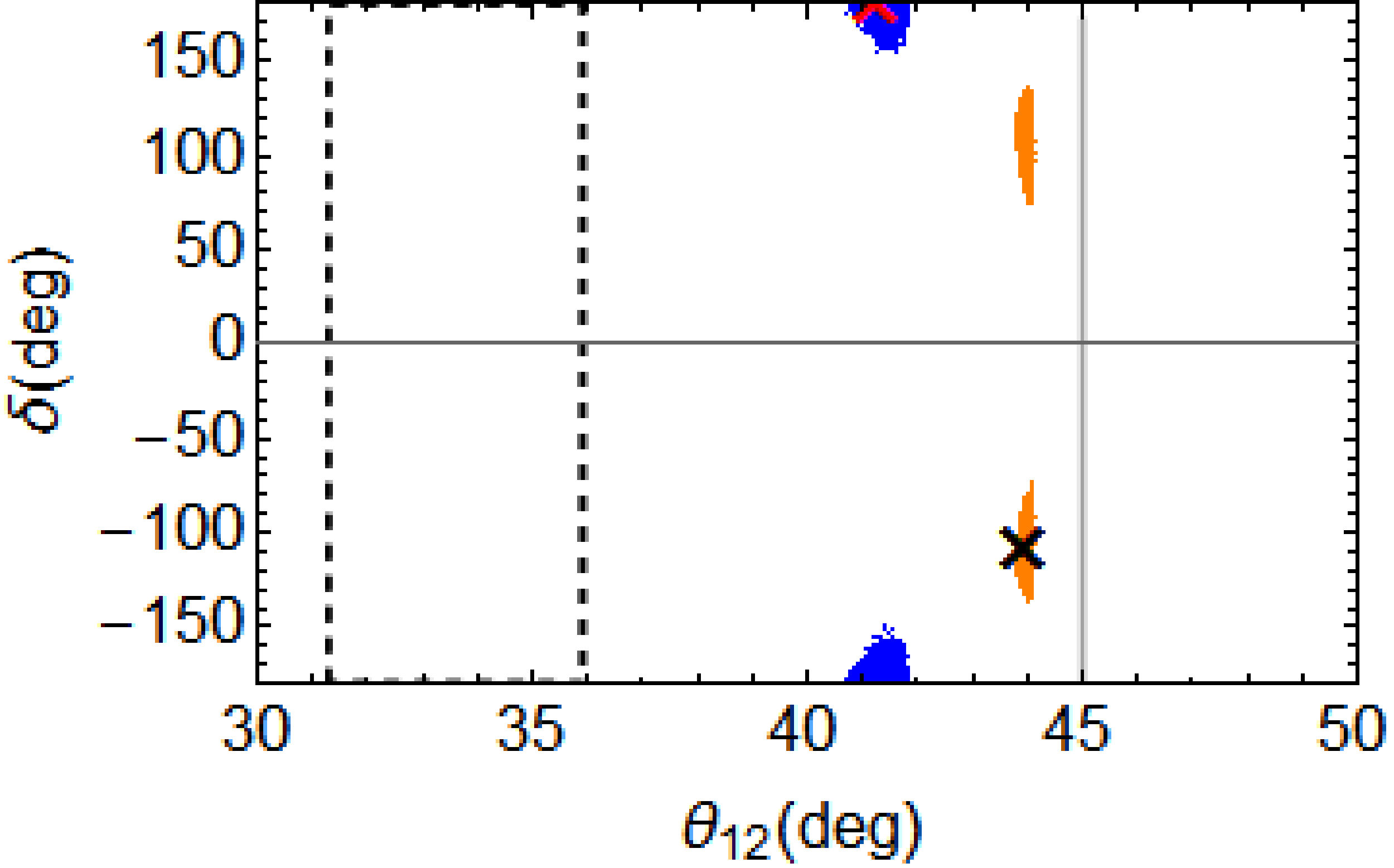}
}%
\subfigure{%
\hspace{0.6cm}
\includegraphics[width=0.45\textwidth, height=0.3\textwidth]{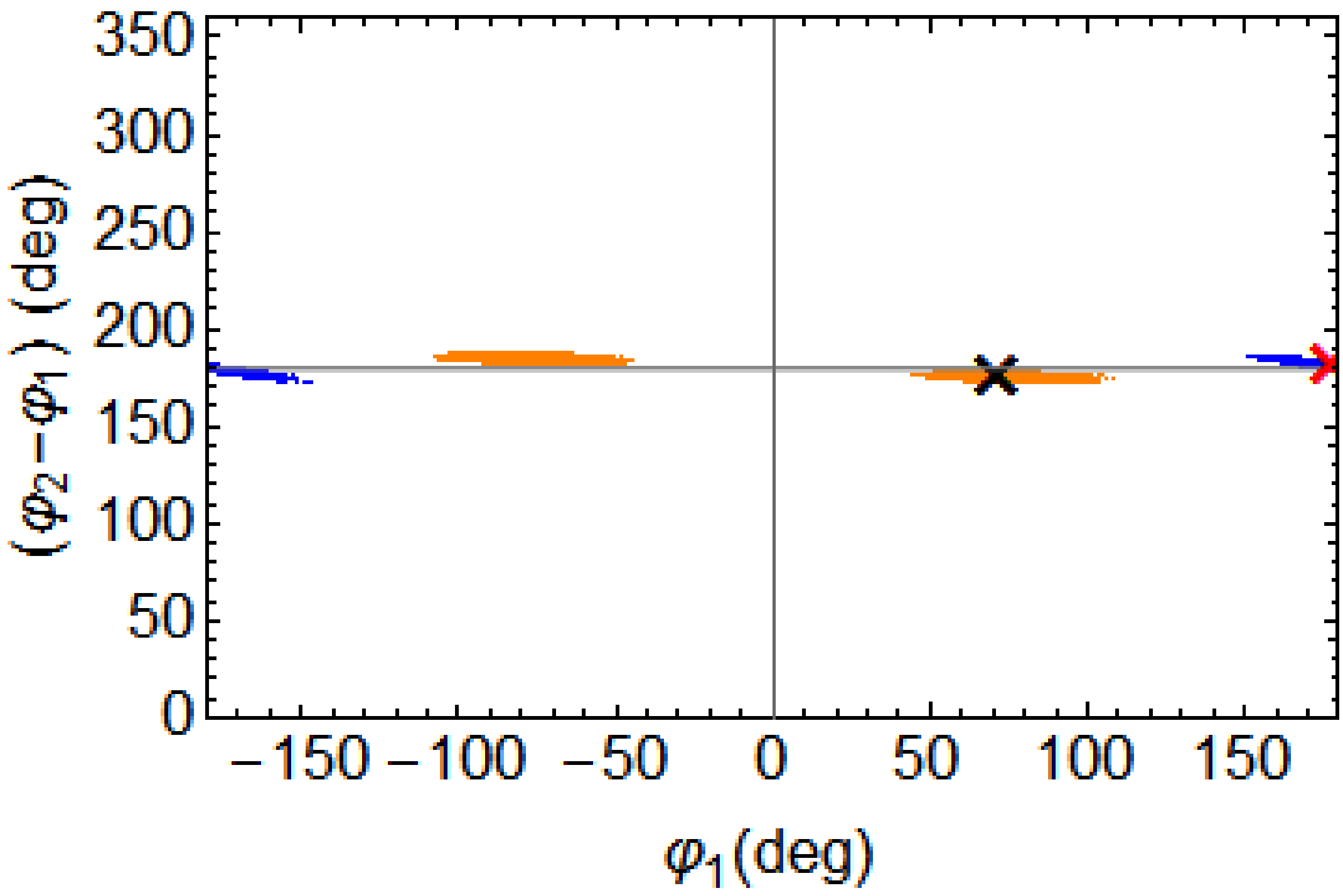}        }%
\end{center}
\vspace{-0.5cm}
\caption{Predictions for the low-energy neutrino parameters in the case of BM for both NH (yellow points) and IH (blue points). The black solid lines in the first two plots denote the bounds obtained from Eq.~(\ref{eq:t13_d23_bound}), and the dashed contours represent the $3\sigma$ ranges given in Ref.~\cite{Gonzalez-Garcia:2014bfa}.}
\label{fg:BM_LE}
\end{figure}

Similar to the case of TBM, the large pull on $\theta_{12}^{}$ originates from the fact that the correction to $\theta_{12}^{}$ is always positive at leading order. Thus, in order to reduce the value of $\theta_{12}^{}$ during the RG running, one has to suppress the leading-order contribution in the first place, by requiring a difference of $180^\circ$ between two Majorana phases, and then take account of the higher-order corrections. Although it is difficult to analytically examine these higher-order corrections, we can still investigate them numerically. In Fig.~\ref{fg:BM_LE}, we present the low-energy predictions of neutrino mixing angles and CP-violating phases, for which only the points leading to $\chi^2 < 230$ in NH and $\chi^2 < 190$ in IH are shown. As one can observe, the maximal correction to $\theta_{12}^{}$ is about $1^\circ$ for NH and $4^\circ$ for IH. Therefore, $\theta_{12}^{}$ is always outside the $3\sigma$ range in both NH and IH cases. Moreover, it is also confirmed that the difference between two Majorana phases is about $\pi$ in both cases.

In the end, we comment on the large pull of $\theta_{23}^{}$ for BM. It is also mainly caused by the previously identified correlation between $\theta_{13}^{}$ and $\theta_{23}^{}$, although now it is slightly modified by higher-order corrections, as can be seen from Fig.~\ref{fg:BM_LE}. Nevertheless, in both NH and IH cases, $\theta_{23}^{}$ is far away from its desired low energy best-fit value from Ref.~\cite{Gonzalez-Garcia:2014bfa}.
%%%%%%%%%%%%%%%%%%%%%%%%%%%%%%%%%%%%% Table 4 %%%%%%%%%%%%%%%%%%%%%%%%%%%%%%
\begin{table}[!t]
\centering
\begin{tabular}{c c c c c}
\hline
\multirow{2}{*}{Parameter} & \multicolumn{2}{c}{BM, NH} & \multicolumn{2}{c}{BM, IH} \\
\cline{2-3} \cline{4-5}
& best-fit & pull & best-fit & pull\\
\hline
$\overline{\eta}_b$ & 0.48 & - & -0.34 & - \\
$\tan\overline{\beta}$ & 26.3 & - & 49.4 & - \\
$\Lambda/ \text{GeV}$ & $7.59\times 10^{13}$ & - & $6.94\times 10^{13}$ & - \\
$\varphi_{1}^\Lambda / \text{deg}$ & 109.0 & - & -82.8 & - \\
$\varphi_{2}^\Lambda / \text{deg}$ & -79.8 & - & 85.5 & - \\
$m_{\text{lightest}}^\Lambda /\text{eV}$ & 0.197 & - & 0.075 & - \\
$(\Delta m_{21}^2)^\Lambda /10^{-5}~\text{eV}^2$ & 9.22 & - & 0.19 & - \\
$|\Delta m_{32}^2|^\Lambda /10^{-3}~\text{eV}^2$ & 2.68 & - & 2.89 & - \\
\hline
$\sin^2\theta_{12}$ & 0.482 & 14.2 & 0.435 & 10.5 \\
$\sin^2\theta_{13}$ & 0.0225 & 0.74 & 0.0240 & 1.96 \\
$\sin^2\theta_{23}$ & 0.640 & 4.69 & 0.324 & -8.23 \\
$\Delta m_{21}^2 /10^{-5}~\text{eV}^2$ & 7.43 & -0.368 & 7.36 & -0.79 \\
$|\Delta m_{3i}^2| /10^{-3}~\text{eV}^2$ & 2.478 & 0.444 & 2.450 & 1.006 \\
$m_{\text{lightest}} /\text{eV}$ & 0.194 & - & 0.185 & - \\
$\delta / \text{deg}$ & -109.0 & - & 179.5 & - \\
$\varphi_{1} / \text{deg}$ & 72.5 & - & 179.3 & - \\
$\varphi_{2} / \text{deg}$ & -112.3 & - & -0.4 & - \\
$m^{}_{ee} / \text{eV}$ & 0.015 & - & 0.007 & - \\
$m^{}_\beta / \text{eV}$ & 0.19 & - & 0.078 & - \\
\hline
$\chi^2_{\rm min}$ & & $\mathbf{224.8}$ & & $\mathbf{183.2}$ \\
\hline
\end{tabular}
\caption{The best-fit points of low-energy observables for the BM mixing pattern in both cases of NH and IH, where the input parameters at the high-energy scale are also given.}
\label{tb:BM}
\end{table}
%%%%%%%%%%%%%%%%%%%%%%%%%%%%%%%%%%%%%%%%%%%%%%%%%%%%%%%%%%%%%%%%%%%%%%%%%%%%%
%%%%%%%%%%%%%%%%%%%%%%%%%%%%%%%%%%%% Fig. 5 %%%%%%%%%%%%%%%%%%%%%%%%%%%%%%%%
%%%%%%%%%%%%%%%%%%%%%%%%%%%%%%%%%%%%%%%%%%%%%%%%%%%%%%%%%%%%%%%%%%%%%%%%%%%%%

\section{Conclusions}

The dynamics for neutrino mass generation and leptonic flavor mixing remains one of the most mysterious problems in particle physics. One promising solution is to introduce a discrete flavor symmetry in the neutrino mass model at a superhigh-energy scale $\Lambda$, where a simple pattern of leptonic mixing (e.g., TBM, GR and BM) can be derived. As all these constant mixing patterns predict a maximal mixing angle $\theta^{}_{23} = \pi/4$ and a vanishing angle $\theta^{}_{13}$ at $\Lambda$, we are motivated to investigate if the RG running effects from the superhigh-energy scale to a low-energy scale are significant enough to generate a nonzero $\theta^{}_{13}$ while keeping other mixing parameters consistent with neutrino oscillation data.

In the framework of MSSM, in order to thoroughly study the radiative corrections to the leptonic mixing, we have scanned over a wide range of each free model parameter at $\Lambda$, including the SUSY threshold parameter, absolute neutrino masses, CP-violating phases, the values of $\tan \beta$ and the high-energy scale $\Lambda$ itself. Using the latest global-fit results of neutrino mixing angles and mass-squared differences, we construct a $\chi^2$ function to quantify the agreement between theoretical predictions and experimental data. The main results from our analytical and numerical studies can be summarized as follows:
\begin{itemize}
\item Via the RG running, it is not difficult to produce a relatively large value of $\theta^{}_{13}$ if neutrino masses are nearly degenerate and the value of $\tan \beta$ is large. With the help of one-loop RGEs, we have derived the radiative corrections to three neutrino mixing angles in an analytical and approximate way. It has been found that $\theta^{}_{12}$ always increases when running toward the low-energy scale. Moreover, there exists a lower bound on the radiative correction to $\theta^{}_{23}$, namely, $\delta \theta^{}_{23} \gtrsim \theta^{}_{13} \tan \theta^\Lambda_{12}$.

\item For TBM and GR, the prediction for $\theta^\Lambda_{12}$ is already close to the best-fit value of $\theta^{}_{12}$, so the difference between two Majorana phases should be around $\pi$ such that the radiative correction to $\theta^{}_{12}$ is small. This also applies to the BM case, in which the increase of $\theta^{}_{12}$ should be kept as small as possible.

\item The correlation $\delta \theta^{}_{23} \gtrsim \theta^{}_{13} \tan \theta^\Lambda_{12}$ induces a severe tension with the observed $\theta^{}_{23}$. As a consequence, both TBM and GR can be compatible with neutrino oscillation data at the $3\sigma$ level, but not at the $1\sigma$ level. Furthermore, if the cosmological upper bound on the sum of neutrino masses is taken into account, these two mixing patterns will be excluded at the $3\sigma$ level. The situation is even worse for BM, which is disfavored even when the cosmological bound is not imposed.
\end{itemize}
Without the cosmological bound on neutrino masses, the $\chi^2$ fit with five degrees of freedom to neutrino oscillation data is given for all three mixing patterns, and the smallest values of $\chi^2_{\rm min}$ have been found in the GR case both for NH (i.e., $\chi^2_{\rm min} = 12.2$) and IH (i.e., $\chi^2_{\rm min} = 30.3$).

It is worthwhile to mention that the RG running of neutrino parameters is solved here within the framework of effective theories, which are valid below a superhigh-energy scale $\Lambda$. In a complete theory above $\Lambda$, such as a neutrino mass model with discrete flavor symmetries, a different set of RGEs should be considered and the decoupling of heavy particles are to be performed explicitly. Nevertheless, our studies have already conveyed some important messages for the model building of neutrino masses and leptonic flavor mixing at high-energy scales.

\section*{Acknowledgements}

The authors thank Zhen-hua Zhao and Ye-Ling Zhou for helpful discussions on the perturbative diagonalization of the neutrino mass matrix. One of them (S.Z.) is grateful to the Mainz Institute for Theoretical Physics (MITP) for its hospitality and partial support during the completion of this work, which has also been supported in part by the National Recruitment Program for Young Professionals and by the CAS Center for Excellence in Particle Physics (CCEPP).

\appendix

\section{Conventions and RGEs}

In Appendix A, we introduce the conventions that are used in this paper, and list the RGEs of all three phases in the MNSP matrix within the MSSM. First of all, the neutrino mass matrix can be diagonalized as follows
\begin{eqnarray}
U_\nu^{\rm T}  M^{}_\nu  U^{}_\nu = D^{}_\nu \; ,
\end{eqnarray}
where $D_\nu^{}=\text{Diag}\{m_1^{}, m_2^{}, m_3^{}\}$ with $m_i^{}$ (for $i = 1, 2, 3$) being real and positive neutrino mass eigenvalues. In the basis where the charged-lepton Yukawa coupling matrix $Y_l^{}$ is diagonal, the unitary matrix $U_\nu^{}$ is simply the MNSP matrix, which is conventionally parametrized as
\begin{eqnarray}
U_\nu = U_{\text{MNSP}} = V(\theta^{}_{12}, \theta^{}_{13}, \theta^{}_{23}, \delta) \cdot \text{Diag}\{e^{-{\rm i}\varphi^{}_1/2}, e^{-{\rm i}\varphi^{}_2/2}, 1\} \; ,
\end{eqnarray}
where
\begin{eqnarray}
V = \begin{pmatrix}
c^{}_{12}c^{}_{13} & c^{}_{13} s^{}_{12} &  s^{}_{13}e^{-{\rm i}\delta} \\
-c^{}_{23}s^{}_{12} - c^{}_{12}s^{}_{13}s^{}_{23}e^{{\rm i}\delta} & c^{}_{12}c^{}_{23}-s^{}_{12}s^{}_{23}s^{}_{13}e^{{\rm i}\delta} & c^{}_{13}s^{}_{23} \\
s^{}_{12}s^{}_{23}-c^{}_{12}c^{}_{23}s^{}_{13}e^{{\rm i}\delta} & -c^{}_{12}s^{}_{23}-c^{}_{23}s^{}_{12}s^{}_{13}e^{{\rm i}\delta} & c^{}_{13}c^{}_{23}
\end{pmatrix} \; ,
\end{eqnarray}
with $s_{ij}^{} \equiv \sin\theta_{ij}^{}$ and $c_{ij}^{} = \cos\theta_{ij}^{}$ have been defined. Furthermore, we have two neutrino mass-squared differences $\Delta m_{21}^2 \equiv m_2^2 - m_1^2$ and $\Delta m_{32}^2 \equiv m_3^2 - m_2^2$, and introduce their ratio $\zeta = \Delta m_{21}^2/\Delta m_{32}^2$.

Following the above conventions, one can obtain the RGEs of all three CP-violating phases. The RGE for the Dirac phase is given by~\cite{Antusch:2003kp}
\begin{eqnarray}
\frac{\text{d} \delta}{\text{d} t} = \frac{y_\tau^2}{32\pi^2} \frac{\delta^{(-1)}}{\theta_{13}} + \frac{y_\tau^2}{8\pi^2} \delta^{(0)} + \mathcal{O}(\theta_{13}) \; ,
\end{eqnarray}
where
\begin{eqnarray}
&~& \delta^{(-1)} = \sin 2\theta_{12} \sin 2\theta_{23} \frac{m_3}{\Delta m_{31}^2 (1+\zeta)}  [m_1 \sin(\varphi_1 - \delta) - (1+\zeta) m_2 \sin (\varphi_2 -\delta) + \zeta m_3 \sin \delta] \; , \\
&~& \delta^{(0)} = \frac{m_1 m_2}{\Delta m_{21}^2} s_{23}^2 \sin(\varphi_1 - \varphi_2) + \frac{m^{}_3}{\Delta m^2_{32}}  \cos 2\theta^{}_{23} \left[\frac{m^{}_1 s^2_{12}}{1 + \zeta}\sin \varphi^{}_1 + m^{}_2 c^2_{12} \sin \varphi^{}_2\right] \nonumber \\
&~& \hspace{4.95cm} + \frac{m^{}_3}{\Delta m^2_{32}}  c^2_{23} \left[\frac{m^{}_1 c^2_{12} }{1+\zeta} \sin(2\delta - \varphi^{}_1) +m^{}_2 s^2_{12} \sin(2\delta - \varphi^{}_2)\right] \; .
\end{eqnarray}
The RGEs of two Majorana phases read~\cite{Antusch:2003kp}
\begin{eqnarray}
\frac{\text{d} \varphi_1}{\text{d} t} &=& \frac{y_\tau^2}{4\pi^2} \left \{ \frac{m_1 m_2}{\Delta m_{21}^2} c_{12}^2 s_{23}^2 \sin(\varphi_1 - \varphi_2) + \frac{m^{}_3 \cos 2\theta^{}_{23}}{\Delta m^2_{32}} \left[\frac{m^{}_1 s_{12}^2}{1+\zeta} \sin \varphi^{}_1 + m^{}_2 c_{12}^2 \sin \varphi^{}_2 \right]\right \} + \mathcal{O}(\theta_{13}) \; , ~~\\
\frac{\text{d} \varphi_2}{\text{d} t} &=& \frac{y_\tau^2}{4\pi^2} \left \{ \frac{m_1 m_2}{\Delta m_{21}^2} s_{12}^2 s_{23}^2 \sin(\varphi_1 - \varphi_2) + \frac{m^{}_3 \cos 2\theta^{}_{23}}{\Delta m^2_{32}} \left[\frac{m^{}_1 s_{12}^2}{1+\zeta} \sin \varphi^{}_1 + m^{}_2 c_{12}^2 \sin \varphi^{}_2 \right] \right \} + \mathcal{O}(\theta_{13}) \; . ~~
\end{eqnarray}
The RGEs of neutrino masses and three mixing angles can also be found in the literature~\cite{Antusch:2003kp,RGE}. The formulas collected in this appendix are useful in the analytical discussions in Sec. 2. In our numerical calculations, the exact RGEs of neutrino parameters have been used.

\section{Perturbative diagonalization}

In Appendix B, we show how to derive the radiative corrections to neutrino masses and flavor mixing angles by perturbatively diagonalizing $M_\nu$ in Eq.~(\ref{eq:mnulambda}). Since the mixing angles $\theta_{ij}^\Lambda$'s at the high-energy scale $\Lambda$ are still good approximations to $\theta_{ij}$ at the low-energy scale $\lambda$ at the leading order, we first rotate $M^{}_\nu/I^{}_0$ by a unitary matrix $V \equiv U_\nu^\Lambda|_{\varphi_{1,2}^{\Lambda}=0}^{}$, namely, $V^{\rm T} \left (M^{}_\nu/I^{}_0\right) V \equiv \widehat{M}^{}_\nu$, and obtain

\begin{eqnarray}
\widehat{M}_\nu = \begin{pmatrix}
m_1^\Lambda e^{\mathrm{i}\varphi_1^\Lambda} (1- \epsilon \sin^2\theta_{12}^\Lambda) &  \displaystyle \frac{\epsilon}{4}(m_1^\Lambda e^{\mathrm{i}\varphi_1^\Lambda} + m_2^\Lambda e^{\mathrm{i}\varphi_2^\Lambda})\sin 2\theta_{12}^\Lambda &  \displaystyle -\frac{\epsilon}{2} (m_1^\Lambda e^{\mathrm{i}\varphi_1^\Lambda} + m_3^\Lambda) \sin\theta_{12}^\Lambda \\
\displaystyle \frac{\epsilon}{4}(m_1^\Lambda e^{\mathrm{i}\varphi_1^\Lambda} +  m_2^\Lambda e^{\mathrm{i}\varphi_2^\Lambda})\sin 2\theta_{12}^\Lambda &  m_2^\Lambda e^{\mathrm{i}\varphi_2^\Lambda} (1-\epsilon\cos^2\theta_{12}^\Lambda) & \displaystyle \frac{\epsilon}{2}(m_2^\Lambda e^{\mathrm{i}\varphi_2^\Lambda} + m_3^\Lambda)\cos\theta_{12}^\Lambda \\
\displaystyle -\frac{\epsilon}{2} (m_1^\Lambda e^{\mathrm{i}\varphi_1^\Lambda} + m_3^\Lambda) \sin\theta_{12}^\Lambda & \displaystyle \frac{\epsilon}{2} (m_2^\Lambda e^{\mathrm{i}\varphi_2^\Lambda} + m_3^\Lambda)\cos\theta_{12}^\Lambda & m_3^\Lambda(1-\epsilon)
\end{pmatrix} \; . ~~
\end{eqnarray}
Next, we adopt the standard approach and construct $\widehat{M}^\dagger_\nu \widehat{M}^{}_\nu$, which can be diagonalized by a unitary transformation. Expanding it in terms of the small parameter $\epsilon$, we arrive at
\begin{eqnarray}
\widehat{M}^\dagger_\nu \widehat{M}^{}_\nu = \begin{pmatrix}
(m_1^\Lambda)^2 & 0 & 0 \\
 0 & (m_2^\Lambda)^2 & 0 \\
0 & 0 & (m_3^\Lambda)^2
\end{pmatrix} + \epsilon \begin{pmatrix}
 -2(m_1^\Lambda)^2\sin^2\theta_{12}^\Lambda &  \displaystyle \frac{Y_{21}^{}}{4} \sin 2\theta_{12}^\Lambda & \displaystyle -\frac{Y_{31}}{2} \sin \theta^\Lambda_{12} \\
\displaystyle \frac{Y_{21}^*}{4} \sin 2\theta_{12}^\Lambda &  -2 (m_2^\Lambda )^2 \cos^2\theta_{12}^\Lambda & \displaystyle \frac{Y_{32}}{2} \cos \theta_{12}^\Lambda \\
\displaystyle -\frac{Y_{31}^*}{2} \sin \theta^\Lambda_{12} & \displaystyle \frac{Y_{32}^*}{2} \cos \theta_{12}^\Lambda & -2(m_3^\Lambda)^2
\end{pmatrix} + \mathcal{O}(\epsilon^2) \; ,~~~~
\end{eqnarray}
where $Y_{ij}^{} \equiv (m_i^\Lambda)^2 + 2 m_i^\Lambda m_j^\Lambda e^{\mathrm{i}(\varphi_i^\Lambda-\varphi_j^\Lambda)} + (m_j^\Lambda)^2$
for $ij = 21, 31, 32$ and $\varphi_3^\Lambda = 0$ should be understood.

According to the standard perturbation theory (see, e.g., Ref.~\cite{Sakurai:2011zz}), we require the perturbations not to alter the spectrum of eigenvalues at the leading order (i.e., no level-crossing theorem). In our case, this requirement means
\begin{eqnarray}
\epsilon |Y_{21}^{}| \sin \theta_{12}^\Lambda  \cos \theta^\Lambda_{12} &< & |(m_2^\Lambda)^2 - (m_1^\Lambda)^2| \; , \\
\epsilon |Y_{31}^{}|  \sin \theta_{12}^\Lambda &<&  |(m_3^\Lambda)^2 - (m_1^\Lambda)^2| \; , \\
\epsilon |Y_{32}^{}|  \cos \theta_{12}^\Lambda &<& |(m_3^\Lambda)^2 - (m_2^\Lambda)^2| \; .
\end{eqnarray}
The above inequalities do not hold \emph{a priori} for generic neutrino parameters at $\Lambda$. However, they are in fact satisfied \emph{a posteriori} according to our numerical results. The first inequality is fulfilled, since $\varphi^\Lambda_2 - \varphi^\Lambda_1 \approx \pi$ and $m^\Lambda_1 \approx m^\Lambda_2$ lead to $|Y^{}_{21}| \approx (m^\Lambda_2 - m^\Lambda_1)^2 < |(m_2^\Lambda)^2 - (m_1^\Lambda)^2|$. Moreover, taking typical values of $\epsilon \sim 0.01$ and $m_1^\Lambda \sim m_3^\Lambda \sim 0.2~\text{eV}$ from the numerical results indicates that the quantities on the left-hand sides of the last two inequalities should be $\lesssim 8\times 10^{-4}~\text{eV}^2$, compared to those $\sim 10^{-3}~\text{eV}^2$ on the right-hand sides.

As the validity of perturbation theory is justified, we proceed with the perturbative diagonalization
\begin{eqnarray}
\widehat{U} \widehat{M}^\dagger_\nu \widehat{M}^{}_\nu \widehat{U}^\dagger = {\rm Diag}\{m^2_1, m^2_2, m^2_3\}/I^2_0 \; ,
\end{eqnarray}
where the unitary matrix $\widehat{U}$ at the leading order is found to be
\begin{eqnarray}
\widehat{U} = \begin{pmatrix}
1 & Z_{21} (\sin 2\theta_{12}^\Lambda) \epsilon/4 & - Z_{31}(\sin \theta_{12}^\Lambda) \epsilon/2 \\
-Z_{21}(\sin 2\theta_{12}^\Lambda) \epsilon/4 & 1 & Z_{32}(\cos \theta_{12}^\Lambda) \epsilon/2 \\
Z_{31}(\sin \theta_{12}^\Lambda) \epsilon/2 & -Z_{32}(\cos \theta_{12}^\Lambda) \epsilon/2 & 1
\end{pmatrix} \; .
\end{eqnarray}
Note that the definitions of $Z^{}_{ij}$ for $ij = 21, 31, 32$ have been given in Eq.~(14). The unitary matrix $U^{}_\nu$ that diagonalizes $M^{}_\nu$ is then given by $U^{}_\nu \approx V \widehat{U}$, from which neutrino mixing angles $\theta^{}_{ij}$ can be extracted.

It is worthwhile to mention that such a diagonalization cannot provide any information about two Majorana phases, whose RG running effects have been studied numerically.


\begin{thebibliography}{99}
%\cite{Agashe:2014kda}
\bibitem{Agashe:2014kda}
  K.~A.~Olive {\it et al.}  [Particle Data Group Collaboration],
  %``Review of Particle Physics,''
  Chin.\ Phys.\ C {\bf 38}, 090001 (2014).
  %%CITATION = CHPHD,C38,090001;%%
  %943 citations counted in INSPIRE as of 24 Apr 2015

%
\bibitem{TBM}
  P.~F.~Harrison, D.~H.~Perkins and W.~G.~Scott,
  %``Tri-bimaximal mixing and the neutrino oscillation data,''
  Phys.\ Lett.\ B {\bf 530}, 167 (2002)
  [hep-ph/0202074];
  %%CITATION = HEP-PH/0202074;%%
  %1132 citations counted in INSPIRE as of 24 Apr 2015
  P.~F.~Harrison and W.~G.~Scott,
  %``Symmetries and generalizations of tri - bimaximal neutrino mixing,''
  Phys.\ Lett.\ B {\bf 535}, 163 (2002)
  [hep-ph/0203209];
  %%CITATION = HEP-PH/0203209;%%
  %509 citations counted in INSPIRE as of 24 Apr 2015
  Z.~z.~Xing,
  %``Nearly tri bimaximal neutrino mixing and CP violation,''
  Phys.\ Lett.\ B {\bf 533}, 85 (2002)
  [hep-ph/0204049];
  %%CITATION = HEP-PH/0204049;%%
  %492 citations counted in INSPIRE as of 24 Apr 2015
%\cite{He:2003rm}
%\bibitem{He:2003rm}
  X.~G.~He and A.~Zee,
  %``Some simple mixing and mass matrices for neutrinos,''
  Phys.\ Lett.\ B {\bf 560}, 87 (2003)
  %doi:10.1016/S0370-2693(03)00390-3
  [hep-ph/0301092].
  %%CITATION = doi:10.1016/S0370-2693(03)00390-3;%%
  %312 citations counted in INSPIRE as of 07 Jun 2016


%\cite{BM
\bibitem{BM}
  F.~Vissani,
  %``A Study of the scenario with nearly degenerate Majorana neutrinos,''
  hep-ph/9708483;
  %%CITATION = HEP-PH/9708483;%%
  %255 citations counted in INSPIRE as of 24 Apr 2015
   V.~D.~Barger, S.~Pakvasa, T.~J.~Weiler and K.~Whisnant,
  %``Bimaximal mixing of three neutrinos,''
  Phys.\ Lett.\ B {\bf 437}, 107 (1998)
  [hep-ph/9806387];
  %%CITATION = HEP-PH/9806387;%%
  %467 citations counted in INSPIRE as of 24 Apr 2015
  A.~J.~Baltz, A.~S.~Goldhaber and M.~Goldhaber,
  %``The Solar neutrino puzzle: An Oscillation solution with maximal neutrino mixing,''
  Phys.\ Rev.\ Lett.\  {\bf 81}, 5730 (1998)
  [hep-ph/9806540].
  %%CITATION = HEP-PH/9806540;%%
  %258 citations counted in INSPIRE as of 24 Apr 2015

%\cite{GR}
\bibitem{GR}
  A.~Datta, F.~S.~Ling and P.~Ramond,
  %``Correlated hierarchy, Dirac masses and large mixing angles,''
  Nucl.\ Phys.\ B {\bf 671}, 383 (2003)
  [hep-ph/0306002];
  %%CITATION = HEP-PH/0306002;%%
  %73 citations counted in INSPIRE as of 24 Apr 2015
  Y.~Kajiyama, M.~Raidal and A.~Strumia,
  %``The Golden ratio prediction for the solar neutrino mixing,''
  Phys.\ Rev.\ D {\bf 76}, 117301 (2007)
  [arXiv:0705.4559];
  %%CITATION = ARXIV:0705.4559;%%
  %93 citations counted in INSPIRE as of 24 Apr 2015
  L.~L.~Everett and A.~J.~Stuart,
  %``Icosahedral (A(5)) Family Symmetry and the Golden Ratio Prediction for Solar Neutrino Mixing,''
  Phys.\ Rev.\ D {\bf 79}, 085005 (2009)
  [arXiv:0812.1057].
  %%CITATION = ARXIV:0812.1057;%%
  %91 citations counted in INSPIRE as of 24 Apr 2015

\bibitem{GR_B}
  W.~Rodejohann,
  %``Unified Parametrization for Quark and Lepton Mixing Angles,''
  Phys.\ Lett.\ B {\bf 671}, 267 (2009)
  %doi:10.1016/j.physletb.2008.12.010
  [arXiv:0810.5239];
  A.~Adulpravitchai, A.~Blum and W.~Rodejohann,
  %``Golden Ratio Prediction for Solar Neutrino Mixing,''
  New J.\ Phys.\  {\bf 11}, 063026 (2009)
  %doi:10.1088/1367-2630/11/6/063026
  [arXiv:0903.0531].

%\cite{flavor}
\bibitem{flavor}
  G.~Altarelli and F.~Feruglio,
  %``Discrete Flavor Symmetries and Models of Neutrino Mixing,''
  Rev.\ Mod.\ Phys.\  {\bf 82}, 2701 (2010)
  [arXiv:1002.0211];
  %%CITATION = ARXIV:1002.0211;%%
  %374 citations counted in INSPIRE as of 24 Apr 2015
  H.~Ishimori, T.~Kobayashi, H.~Ohki, Y.~Shimizu, H.~Okada and M.~Tanimoto,
  %``Non-Abelian Discrete Symmetries in Particle Physics,''
  Prog.\ Theor.\ Phys.\ Suppl.\  {\bf 183}, 1 (2010)
  [arXiv:1003.3552];
  %%CITATION = ARXIV:1003.3552;%%
  %325 citations counted in INSPIRE as of 24 Apr 2015
  S.~F.~King and C.~Luhn,
  %``Neutrino Mass and Mixing with Discrete Symmetry,''
  Rept.\ Prog.\ Phys.\  {\bf 76}, 056201 (2013)
  [arXiv:1301.1340].
  %%CITATION = ARXIV:1301.1340;%%
  %182 citations counted in INSPIRE as of 24 Apr 2015
%\cite{King:2014nza}
%\bibitem{King:2014nza}
  S.~F.~King, A.~Merle, S.~Morisi, Y.~Shimizu and M.~Tanimoto,
  %``Neutrino Mass and Mixing: from Theory to Experiment,''
  New J.\ Phys.\  {\bf 16}, 045018 (2014)
  %doi:10.1088/1367-2630/16/4/045018
  [arXiv:1402.4271].
  %%CITATION = doi:10.1088/1367-2630/16/4/045018;%%
  %123 citations counted in INSPIRE as of 07 Jun 2016
%\cite{dayabay}
\bibitem{dayabay}
  F.~P.~An {\it et al.}  [Daya Bay Collaboration],
  %``Observation of electron-antineutrino disappearance at Daya Bay,''
  Phys.\ Rev.\ Lett.\  {\bf 108}, 171803 (2012)
  [arXiv:1203.1669],
  %%CITATION = ARXIV:1203.1669;%%
  %1168 citations counted in INSPIRE as of 24 Apr 2015
  F.~P.~An {\it et al.}  [Daya Bay Collaboration],
  %``Improved Measurement of Electron Antineutrino Disappearance at Daya Bay,''
  Chin.\ Phys.\ C {\bf 37}, 011001 (2013)
  [arXiv:1210.6327],
  %%CITATION = ARXIV:1210.6327;%%
  %231 citations counted in INSPIRE as of 24 Apr 2015
  F.~P.~An {\it et al.}  [Daya Bay Collaboration],
  %``Spectral measurement of electron antineutrino oscillation amplitude and frequency at Daya Bay,''
  Phys.\ Rev.\ Lett.\  {\bf 112}, 061801 (2014)
  [arXiv:1310.6732].
  %%CITATION = ARXIV:1310.6732;%%
  %99 citations counted in INSPIRE as of 24 Apr 2015

%\cite{reno}
\bibitem{reno}
  J.~K.~Ahn {\it et al.}  [RENO Collaboration],
  %``Observation of Reactor Electron Antineutrino Disappearance in the RENO Experiment,''
  Phys.\ Rev.\ Lett.\  {\bf 108}, 191802 (2012)
  [arXiv:1204.0626].
  %%CITATION = ARXIV:1204.0626;%%
  %1012 citations counted in INSPIRE as of 24 Apr 2015

%\cite{chooz}
\bibitem{chooz}
  Y.~Abe {\it et al.}  [Double Chooz Collaboration],
  %``Indication for the disappearance of reactor electron antineutrinos in the Double Chooz experiment,''
  Phys.\ Rev.\ Lett.\  {\bf 108}, 131801 (2012)
  [arXiv:1112.6353].
  %%CITATION = ARXIV:1112.6353;%%
  %668 citations counted in INSPIRE as of 24 Apr 2015

%\cite{modifications}
\bibitem{modifications}
  Z.~Z.~Xing,
  %``The T2K Indication of Relatively Large $\theta_{13}$ and a Natural Perturbation to the Democratic Neutrino Mixing Pattern,''
  Chin.\ Phys.\ C {\bf 36}, 101 (2012)
  %doi:10.1088/1674-1137/36/2/001
  [arXiv:1106.3244];
  %%CITATION = doi:10.1088/1674-1137/36/2/001;%%
  %55 citations counted in INSPIRE as of 07 Jun 2016
  X.~G.~He and A.~Zee,
  %``Minimal Modification to Tri-bimaximal Mixing,''
  Phys.\ Rev.\ D {\bf 84}, 053004 (2011)
  [arXiv:1106.4359];
  %%CITATION = ARXIV:1106.4359;%%
  %80 citations counted in INSPIRE as of 24 Apr 2015
  S.~Zhou,
  %``Relatively large theta13 and nearly maximal theta23 from the approximate S3 symmetry of lepton mass matrices,''
  Phys.\ Lett.\ B {\bf 704}, 291 (2011)
  [arXiv:1106.4808];
  %%CITATION = ARXIV:1106.4808;%%
  %61 citations counted in INSPIRE as of 24 Apr 2015
  T.~Araki,
  %``Getting at large $\theta_{13}$ with almost maximal $\theta_{23}$ from tri-bimaximal mixing,''
  Phys.\ Rev.\ D {\bf 84}, 037301 (2011)
  [arXiv:1106.5211];
  %%CITATION = ARXIV:1106.5211;%%
  %49 citations counted in INSPIRE as of 24 Apr 2015
  W.~Chao and Y.~j.~Zheng,
  %``Relatively Large Theta13 from Modification to the Tri-bimaximal, Bimaximal and Democratic Neutrino Mixing Matrices,''
  JHEP {\bf 1302}, 044 (2013)
  [arXiv:1107.0738];
  %%CITATION = ARXIV:1107.0738;%%
  %56 citations counted in INSPIRE as of 24 Apr 2015
  D.~Marzocca, S.~T.~Petcov, A.~Romanino and M.~Spinrath,
  %``Sizeable $\theta_{13}$ from the Charged Lepton Sector in SU(5), (Tri-)Bimaximal Neutrino Mixing and Dirac CP Violation,''
  JHEP {\bf 1111}, 009 (2011)
  [arXiv:1108.0614];
  %%CITATION = ARXIV:1108.0614;%%
  %73 citations counted in INSPIRE as of 24 Apr 2015
  %\cite{Ge:2011qn}
  S.~F.~Ge, D.~A.~Dicus and W.~W.~Repko,
  %``Residual Symmetries for Neutrino Mixing with a Large $\theta_{13}$ and Nearly Maximal $\delta_D$,''
  Phys.\ Rev.\ Lett.\  {\bf 108}, 041801 (2012)
  %doi:10.1103/PhysRevLett.108.041801
  [arXiv:1108.0964];
  %%CITATION = doi:10.1103/PhysRevLett.108.041801;%%
  %135 citations counted in INSPIRE as of 25 Jul 2016
  S.~F.~King and C.~Luhn,
  %``A4 models of tri-bimaximal-reactor mixing,''
  JHEP {\bf 1203}, 036 (2012)
  [arXiv:1112.1959];
  %%CITATION = ARXIV:1112.1959;%%
  %41 citations counted in INSPIRE as of 24 Apr 2015
  S.~Gupta, A.~S.~Joshipura and K.~M.~Patel,
  %``Minimal extension of tri-bimaximal mixing and generalized Z_2 X Z_2 symmetries,''
  Phys.\ Rev.\ D {\bf 85}, 031903 (2012)
  [arXiv:1112.6113];
  %%CITATION = ARXIV:1112.6113;%%
  %42 citations counted in INSPIRE as of 24 Apr 2015
  D.~Marzocca, S.~T.~Petcov, A.~Romanino and M.~C.~Sevilla,
  %``Nonzero |U_e3| from Charged Lepton Corrections and the Atmospheric Neutrino Mixing Angle,''
  JHEP {\bf 1305}, 073 (2013)
  [arXiv:1302.0423];
  %%CITATION = ARXIV:1302.0423;%%
  %26 citations counted in INSPIRE as of 24 Apr 2015
  S.~K.~Garg and S.~Gupta,
  %``Corrections for tribimaximal, bimaximal and democratic neutrino mixing matrices,''
  JHEP {\bf 1310}, 128 (2013)
  [arXiv:1308.3054];
  %%CITATION = ARXIV:1308.3054;%%
  %11 citations counted in INSPIRE as of 24 Apr 2015
  %\cite{Hanlon:2013ska}
  A.~D.~Hanlon, S.~F.~Ge and W.~W.~Repko,
  %``Phenomenological consequences of residual $ \mathbb{Z}^s_2$ and $ \overline {\mathbb{Z}}^s_2$ symmetries,''
  Phys.\ Lett.\ B {\bf 729}, 185 (2014)
  %doi:10.1016/j.physletb.2013.12.063
  [arXiv:1308.6522];
  %%CITATION = doi:10.1016/j.physletb.2013.12.063;%%
  %18 citations counted in INSPIRE as of 25 Jul 2016
  J.~Kile, M.~J.~P\'{e}rez, P.~Ramond and J.~Zhang,
  %``$_{13}$ and the flavor ring,''
  Phys.\ Rev.\ D {\bf 90}, no. 1, 013004 (2014)
  [arXiv:1403.6136];
  %%CITATION = ARXIV:1403.6136;%%
  %2 citations counted in INSPIRE as of 24 Apr 2015
  Z.~h.~Zhao,
  %``Minimal modifications to the Tri-Bimaximal neutrino mixing,''
  JHEP {\bf 1411}, 143 (2014)
  [arXiv:1405.3022].
  %%CITATION = ARXIV:1405.3022;%%
  %1 citations counted in INSPIRE as of 24 Apr 2015

%\cite{group_scan}
\bibitem{group_scan}
  M.~Holthausen, K.~S.~Lim and M.~Lindner,
  %``Lepton Mixing Patterns from a Scan of Finite Discrete Groups,''
  Phys.\ Lett.\ B {\bf 721}, 61 (2013)
  [arXiv:1212.2411];
  %%CITATION = ARXIV:1212.2411;%%
  %48 citations counted in INSPIRE as of 24 Apr 2015
  J.~Talbert,
  %``[Re]constructing Finite Flavour Groups: Horizontal Symmetry Scans from the Bottom-Up,''
  JHEP {\bf 1412}, 058 (2014)
  [arXiv:1409.7310];
  %%CITATION = ARXIV:1409.7310;%%
  %1 citations counted in INSPIRE as of 24 Apr 2015
  C.~Y.~Yao and G.~J.~Ding,
  %``CP Symmetry and Lepton Mixing from a Scan of Finite Discrete Groups,''
  [arXiv:1606.05610].
  %%CITATION = ARXIV:1606.05610;%%


%\cite{Zhang:2011aw}
\bibitem{Zhang:2011aw}
  H.~Zhang and S.~Zhou,
  %``Radiative corrections and explicit perturbations to the tetra-maximal neutrino mixing with large $\theta_{13}$,''
  Phys.\ Lett.\ B {\bf 704}, 296 (2011)
  %doi:10.1016/j.physletb.2011.09.033
  [arXiv:1107.1097];
  %%CITATION = doi:10.1016/j.physletb.2011.09.033;%%
  %41 citations counted in INSPIRE as of 30 Jun 2016
%\cite{Rodejohann:2011uz}
%\bibitem{Rodejohann:2011uz}
  W.~Rodejohann, H.~Zhang and S.~Zhou,
  %``Systematic search for successful lepton mixing patterns with nonzero $\theta_{13}$,''
  Nucl.\ Phys.\ B {\bf 855}, 592 (2012)
  [arXiv:1107.3970].
  %%CITATION = ARXIV:1107.3970;%%
  %33 citations counted in INSPIRE as of 24 Apr 2015


%\cite{seesaw_threshold}
\bibitem{seesaw_threshold}
  J.~w.~Mei and Z.~z.~Xing,
  %``Radiative generation of theta(13) with the seesaw threshold effect,''
  Phys.\ Rev.\ D {\bf 70}, 053002 (2004)
  [hep-ph/0404081];
  %%CITATION = HEP-PH/0404081;%%
  %31 citations counted in INSPIRE as of 24 Apr 2015
  S.~Antusch, J.~Kersten, M.~Lindner, M.~Ratz and M.~A.~Schmidt,
  %``Running neutrino mass parameters in see-saw scenarios,''
  JHEP {\bf 0503}, 024 (2005)
  [hep-ph/0501272];
  %%CITATION = HEP-PH/0501272;%%
  %239 citations counted in INSPIRE as of 24 Apr 2015
  S.~Gupta, S.~K.~Kang and C.~S.~Kim,
  %``Renormalization Group Evolution of Neutrino Parameters in Presence of Seesaw Threshold Effects and Majorana Phases,''
  Nucl.\ Phys.\ B {\bf 893}, 89 (2015)
  [arXiv:1406.7476].
  %%CITATION = ARXIV:1406.7476;%%
  %2 citations counted in INSPIRE as of 24 Apr 2015


%\cite{Antusch:2003kp}
\bibitem{Antusch:2003kp}
  S.~Antusch, J.~Kersten, M.~Lindner and M.~Ratz,
  %``Running neutrino masses, mixings and CP phases: Analytical results and phenomenological consequences,''
  Nucl.\ Phys.\ B {\bf 674}, 401 (2003)
  [hep-ph/0305273].
  %%CITATION = HEP-PH/0305273;%%
  %264 citations counted in INSPIRE as of 24 Apr 2015

%\cite{Luo:2005fc}
\bibitem{Luo:2005fc}
  S.~Luo and Z.~z.~Xing,
  %``Generalized tri-bimaximal neutrino mixing and its sensitivity to radiative corrections,''
  Phys.\ Lett.\ B {\bf 632}, 341 (2006)
  [hep-ph/0509065].
  %%CITATION = HEP-PH/0509065;%%
  %87 citations counted in INSPIRE as of 24 Apr 2015


%\cite{Dighe:2006sr}
\bibitem{Dighe:2006sr}
  A.~Dighe, S.~Goswami and W.~Rodejohann,
  %``Corrections to Tri-bimaximal Neutrino Mixing: Renormalization and Planck Scale Effects,''
  Phys.\ Rev.\ D {\bf 75}, 073023 (2007)
  [hep-ph/0612328].
  %%CITATION = HEP-PH/0612328;%%
  %56 citations counted in INSPIRE as of 24 Apr 2015

%\cite{Dighe:2007ksa}
\bibitem{Dighe:2007ksa}
  A.~Dighe, S.~Goswami and P.~Roy,
  %``Radiatively broken symmetries of nonhierarchical neutrinos,''
  Phys.\ Rev.\ D {\bf 76}, 096005 (2007)
  [arXiv:0704.3735].
  %%CITATION = ARXIV:0704.3735;%%
  %38 citations counted in INSPIRE as of 24 Apr 2015

%\cite{Dighe:2008wn}
\bibitem{Dighe:2008wn}
  A.~Dighe, S.~Goswami and S.~Ray,
  %``Renormalization group evolution of neutrino mixing parameters near $\theta_{13}$ = 0 and models with vanishing $\theta_{13}$ at the high scale,''
  Phys.\ Rev.\ D {\bf 79}, 076006 (2009)
  [arXiv:0810.5680].
  %%CITATION = ARXIV:0810.5680;%%
  %12 citations counted in INSPIRE as of 24 Apr 2015

%\cite{Goswami:2009yy}
\bibitem{Goswami:2009yy}
  S.~Goswami, S.~T.~Petcov, S.~Ray and W.~Rodejohann,
  %``Large |U(e3)| and Tri-bimaximal Mixing,''
  Phys.\ Rev.\ D {\bf 80}, 053013 (2009)
  [arXiv:0907.2869].
  %%CITATION = ARXIV:0907.2869;%%
  %44 citations counted in INSPIRE as of 24 Apr 2015

%\cite{Luo:2012ce}
\bibitem{Luo:2012ce}
  S.~Luo and Z.~z.~Xing,
  %``Impacts of the observed $\theta_{13}$ on the running behaviors of Dirac and Majorana neutrino mixing angles and CP-violating phases,''
  Phys.\ Rev.\ D {\bf 86}, 073003 (2012)
  [arXiv:1203.3118].
  %%CITATION = ARXIV:1203.3118;%%
  %23 citations counted in INSPIRE as of 24 Apr 2015


%\cite{Ade:2015xua}
\bibitem{Ade:2015xua}
  P.~A.~R.~Ade {\it et al.}  [Planck Collaboration],
  %``Planck 2015 results. XIII. Cosmological parameters,''
  arXiv:1502.01589.
  %%CITATION = ARXIV:1502.01589;%%
  %207 citations counted in INSPIRE as of 24 Apr 2015


%\cite{Gonzalez-Garcia:2014bfa}
\bibitem{Gonzalez-Garcia:2014bfa}
  M.~C.~Gonzalez-Garcia, M.~Maltoni and T.~Schwetz,
  %``Updated fit to three neutrino mixing: status of leptonic CP violation,''
  JHEP {\bf 1411}, 052 (2014)
  [arXiv:1409.5439].
  %%CITATION = ARXIV:1409.5439;%%
  %88 citations counted in INSPIRE as of 24 Apr 2015

%\cite{kapparunning}
\bibitem{kapparunning}
  P.~H.~Chankowski and Z.~Pluciennik,
  %``Renormalization group equations for seesaw neutrino masses,''
  Phys.\ Lett.\ B {\bf 316}, 312 (1993)
  [hep-ph/9306333];
  %%CITATION = HEP-PH/9306333;%%
  %254 citations counted in INSPIRE as of 24 Apr 2015
  K.~S.~Babu, C.~N.~Leung and J.~T.~Pantaleone,
  %``Renormalization of the neutrino mass operator,''
  Phys.\ Lett.\ B {\bf 319}, 191 (1993)
  [hep-ph/9309223];
  %%CITATION = HEP-PH/9309223;%%
  %313 citations counted in INSPIRE as of 24 Apr 2015
  S.~Antusch, M.~Drees, J.~Kersten, M.~Lindner and M.~Ratz,
  %``Neutrino mass operator renormalization revisited,''
  Phys.\ Lett.\ B {\bf 519}, 238 (2001)
  [hep-ph/0108005].
  %%CITATION = HEP-PH/0108005;%%
  %174 citations counted in INSPIRE as of 24 Apr 2015

%\cite{RGE}
\bibitem{RGE}
 S.~Antusch, J.~Kersten, M.~Lindner and M.~Ratz,
  %``Running neutrino masses, mixings and CP phases: Analytical results and phenomenological consequences,''
  Nucl.\ Phys.\ B {\bf 674}, 401 (2003)
  [hep-ph/0305273];
  %%CITATION = HEP-PH/0305273;%%
  %264 citations counted in INSPIRE as of 24 Apr 2015
  S.~Antusch, J.~Kersten, M.~Lindner, M.~Ratz and M.~A.~Schmidt,
  %``Running neutrino mass parameters in see-saw scenarios,''
  JHEP {\bf 0503}, 024 (2005)
  [hep-ph/0501272];
  %%CITATION = HEP-PH/0501272;%%
  %239 citations counted in INSPIRE as of 24 Apr 2015
  J.~w.~Mei,
  %``Running neutrino masses, leptonic mixing angles and CP-violating phases: From M(Z) to Lambda(GUT),''
  Phys.\ Rev.\ D {\bf 71}, 073012 (2005)
  [hep-ph/0502015].
  %%CITATION = HEP-PH/0502015;%%
  %67 citations counted in INSPIRE as of 24 Apr 2015

%\cite{Ohlsson:2013xva}
\bibitem{Ohlsson:2013xva}
  T.~Ohlsson and S.~Zhou,
  %``Renormalization group running of neutrino parameters,''
  Nature Commun.\  {\bf 5}, 5153 (2014)
  %doi:10.1038/ncomms6153
  [arXiv:1311.3846].
  %%CITATION = doi:10.1038/ncomms6153;%%
  %17 citations counted in INSPIRE as of 30 Jun 2016

%\cite{Dighe:2006zk}
\bibitem{Dighe:2006zk}
  A.~Dighe, S.~Goswami and P.~Roy,
  %``Quark-lepton complementarity with quasidegenerate Majorana neutrinos,''
  Phys.\ Rev.\ D {\bf 73}, 071301 (2006)
  [hep-ph/0602062].
  %%CITATION = HEP-PH/0602062;%%
  %55 citations counted in INSPIRE as of 24 Apr 2015

%\cite{Antusch:2013jca}
\bibitem{Antusch:2013jca}
  S.~Antusch and V.~Maurer,
  %``Running quark and lepton parameters at various scales,''
  JHEP {\bf 1311}, 115 (2013)
  [arXiv:1306.6879].
  %%CITATION = ARXIV:1306.6879;%%
  %20 citations counted in INSPIRE as of 24 Apr 2015
  %\cite{Xing:2011aa}
%\bibitem{Xing:2011aa}
  Z.~z.~Xing, H.~Zhang and S.~Zhou,
  %``Impacts of the Higgs mass on vacuum stability, running fermion masses and two-body Higgs decays,''
  Phys.\ Rev.\ D {\bf 86}, 013013 (2012)
 % doi:10.1103/PhysRevD.86.013013
  [arXiv:1112.3112];
  %%CITATION = doi:10.1103/PhysRevD.86.013013;%%
  %84 citations counted in INSPIRE as of 25 Mar 2016
%\cite{Xing:2007fb}
%\bibitem{Xing:2007fb}
%  Z.~z.~Xing, H.~Zhang and S.~Zhou,
  %``Updated Values of Running Quark and Lepton Masses,''
  Phys.\ Rev.\ D {\bf 77}, 113016 (2008)
 % doi:10.1103/PhysRevD.77.113016
  [arXiv:0712.1419];
  %%CITATION = doi:10.1103/PhysRevD.77.113016;%%
  %270 citations counted in INSPIRE as of 25 Mar 2016
%\cite{Fusaoka:1998vc}
%\bibitem{Fusaoka:1998vc}
  H.~Fusaoka and Y.~Koide,
  %``Updated estimate of running quark masses,''
  Phys.\ Rev.\ D {\bf 57}, 3986 (1998)
  %doi:10.1103/PhysRevD.57.3986
  [hep-ph/9712201].
  %%CITATION = doi:10.1103/PhysRevD.57.3986;%%
  %312 citations counted in INSPIRE as of 25 Mar 2016

%\cite{SUSY}
\bibitem{SUSY}
  L.~J.~Hall, R.~Rattazzi and U.~Sarid,
  %``The Top quark mass in supersymmetric SO(10) unification,''
  Phys.\ Rev.\ D {\bf 50}, 7048 (1994)
  [hep-ph/9306309, hep-ph/9306309];
  %%CITATION = HEP-PH/9306309,;%%
  %804 citations counted in INSPIRE as of 24 Apr 2015
  M.~Carena, M.~Olechowski, S.~Pokorski and C.~E.~M.~Wagner,
  %``Electroweak symmetry breaking and bottom - top Yukawa unification,''
  Nucl.\ Phys.\ B {\bf 426}, 269 (1994)
  [hep-ph/9402253];
  %%CITATION = HEP-PH/9402253;%%
  %707 citations counted in INSPIRE as of 24 Apr 2015
  R.~Hempfling,
  %``Yukawa coupling unification with supersymmetric threshold corrections,''
  Phys.\ Rev.\ D {\bf 49}, 6168 (1994);
  %%CITATION = PHRVA,D49,6168;%%
  %394 citations counted in INSPIRE as of 24 Apr 2015
  T.~Blazek, S.~Raby and S.~Pokorski,
  %``Finite supersymmetric threshold corrections to CKM matrix elements in the large tan Beta regime,''
  Phys.\ Rev.\ D {\bf 52}, 4151 (1995)
  [hep-ph/9504364];
  %%CITATION = HEP-PH/9504364;%%
  %195 citations counted in INSPIRE as of 24 Apr 2015
  S.~Antusch and M.~Spinrath,
  %``Quark and lepton masses at the GUT scale including SUSY threshold corrections,''
  Phys.\ Rev.\ D {\bf 78}, 075020 (2008)
  [arXiv:0804.0717];
  %%CITATION = ARXIV:0804.0717;%%
  %60 citations counted in INSPIRE as of 24 Apr 2015
  A.~Crivellin and C.~Greub,
  %``Two-loop supersymmetric QCD corrections to Higgs-quark-quark couplings in the generic MSSM,''
  Phys.\ Rev.\ D {\bf 87}, 015013 (2013)
  [Phys.\ Rev.\ D {\bf 87}, 079901 (2013)]
  [arXiv:1210.7453].
  %%CITATION = ARXIV:1210.7453;%%
  %17 citations counted in INSPIRE as of 24 Apr 2015

%\cite{multinest}
\bibitem{multinest}
  F.~Feroz and M.~P.~Hobson,
  %``Multimodal nested sampling: an efficient and robust alternative to MCMC methods for astronomical data analysis,''
  Mon.\ Not.\ Roy.\ Astron.\ Soc.\  {\bf 384}, 449 (2008)
  [arXiv:0704.3704];
  %%CITATION = ARXIV:0704.3704;%%
  %273 citations counted in INSPIRE as of 24 Apr 2015
  F.~Feroz, M.~P.~Hobson and M.~Bridges,
  %``MultiNest: an efficient and robust Bayesian inference tool for cosmology and particle physics,''
  Mon.\ Not.\ Roy.\ Astron.\ Soc.\  {\bf 398}, 1601 (2009)
  [arXiv:0809.3437];
  %%CITATION = ARXIV:0809.3437;%%
  %361 citations counted in INSPIRE as of 24 Apr 2015
  F.~Feroz, M.~P.~Hobson, E.~Cameron and A.~N.~Pettitt,
  %``Importance Nested Sampling and the MultiNest Algorithm,''
  arXiv:1306.2144.
  %%CITATION = ARXIV:1306.2144;%%
  %52 citations counted in INSPIRE as of 24 Apr 2015

%\cite{Luo:2006tb}
\bibitem{Luo:2006tb}
  S.~Luo and Z.~Z.~Xing,
  %``On the quasi-fixed point in the running of CP-violating phases of Majorana neutrinos,''
  Phys.\ Lett.\ B {\bf 637}, 279 (2006)
  [hep-ph/0603091].
  %%CITATION = HEP-PH/0603091;%%
  %14 citations counted in INSPIRE as of 20 May 2015

%\cite{deGouvea:2013onf}
\bibitem{deGouvea:2013onf}
  A.~de Gouvea {\it et al.}  [Intensity Frontier Neutrino Working Group Collaboration],
  %``Working Group Report: Neutrinos,''
  arXiv:1310.4340.
  %%CITATION = ARXIV:1310.4340;%%
  %48 citations counted in INSPIRE as of 03 Feb 2015


%\cite{Rodejohann:2012xd}
\bibitem{Rodejohann:2012xd}
  W.~Rodejohann,
  %``Neutrinoless double beta decay and neutrino physics,''
  J.\ Phys.\ G {\bf 39}, 124008 (2012)
  [arXiv:1206.2560].
  %%CITATION = ARXIV:1206.2560;%%
  %67 citations counted in INSPIRE as of 03 Feb 2015


%\cite{Osipowicz:2001sq}
\bibitem{Osipowicz:2001sq}
  A.~Osipowicz {\it et al.}  [KATRIN Collaboration],
  %``KATRIN: A Next generation tritium beta decay experiment with sub-eV sensitivity for the electron neutrino mass. Letter of intent,''
  hep-ex/0109033.
  %%CITATION = HEP-EX/0109033;%%
  %435 citations counted in INSPIRE as of 03 Feb 2015

%\cite{fortheKATRIN:2013saa}
\bibitem{fortheKATRIN:2013saa}
  R.~G.~H.~Robertson [KATRIN Collaboration],
  %``KATRIN: an experiment to determine the neutrino mass from the beta decay of tritium,''
  arXiv:1307.5486.
  %%CITATION = ARXIV:1307.5486;%%
  %10 citations counted in INSPIRE as of 03 Feb 2015

%\cite{Luo:2014upa}
\bibitem{Luo:2014upa}
  S.~Luo and Z.~z.~Xing,
  %``Resolving the octant of $\theta_{23}$ via radiative $\mu-\tau$ symmetry breaking,''
  Phys.\ Rev.\ D {\bf 90}, no. 7, 073005 (2014)
 % doi:10.1103/PhysRevD.90.073005
  [arXiv:1408.5005];
  %%CITATION = doi:10.1103/PhysRevD.90.073005;%%
  %11 citations counted in INSPIRE as of 29 Jun 2016
%\cite{Zhou:2014sya}
%\bibitem{Zhou:2014sya}
  Y.~L.~Zhou,
  %``$\mu$-$\tau$ reflection symmetry and radiative corrections,''
  arXiv:1409.8600.
  %%CITATION = ARXIV:1409.8600;%%
  %1 citations counted in INSPIRE as of 24 Apr 2015

%\cite{Xing:2015fdg}
\bibitem{Xing:2015fdg}
  Z.~z.~Xing and Z.~h.~Zhao,
  %``A review of ¦Ì-¦Ó flavor symmetry in neutrino physics,''
  Rept.\ Prog.\ Phys.\  {\bf 79}, no. 7, 076201 (2016)
  %doi:10.1088/0034-4885/79/7/076201
  [arXiv:1512.04207];
  %%CITATION = doi:10.1088/0034-4885/79/7/076201;%%
  %5 citations counted in INSPIRE as of 29 Jun 2016
  %\cite{Zhao:2016orh}
%\bibitem{Zhao:2016orh}
  Z.~h.~Zhao,
  %``On the breaking of mu-tau flavor symmetry,''
  arXiv:1605.04498.
  %%CITATION = ARXIV:1605.04498;%%

%\cite{Zhang:2016djh}
\bibitem{Zhang:2016djh}
  J.~Zhang and S.~Zhou,
  %``Radiative corrections to the solar lepton mixing sum rule,''
  JHEP {\bf 1608}, 024 (2016)
  %doi:10.1007/JHEP08(2016)024
  [arXiv:1604.03039].
  %%CITATION = doi:10.1007/JHEP08(2016)024;%%
  %4 citations counted in INSPIRE as of 08 Sep 2016

%\cite{Sakurai:2011zz}
\bibitem{Sakurai:2011zz}
  J.~J.~Sakurai and J.~Napolitano,
  ``Modern quantum physics,''
  Boston, USA: Addison-Wesley (2011)
  %12 citations counted in INSPIRE as of 24 Apr 2015
\end{thebibliography}
\end{document}